\documentclass[%
 aip,
 jmp,%
 amsmath,amssymb,
 reprint,%
]{revtex4-1}

\usepackage{graphicx}
\graphicspath{{Figures_pdf/}}
\usepackage{dcolumn}
\usepackage{bm}
\usepackage{subfigure}

\usepackage{amsmath,amssymb,amsfonts}
\usepackage{graphics}
\usepackage{color}
\usepackage{xcolor}
\definecolor{rev1}{rgb}{0,0,0}

\usepackage{comment}
\usepackage{hyperref}
\usepackage{lipsum}
\usepackage{tabularx}
\usepackage[linesnumbered,lined,ruled,longend]{algorithm2e}
\usepackage{mathrsfs}
\usepackage{stackengine}
\DeclareMathOperator{\argmin}{arg\,min} 
\newcommand{\overbar}[1]{\mkern 1.5mu\overline{\mkern-1.5mu#1\mkern-1.5mu}\mkern 1.5mu}

\begin{document}
\newcommand*{\vertbar}{\rule[-1ex]{0.5pt}{2.5ex}}
\newcommand*{\horzbar}{\rule[.5ex]{2.5ex}{0.5pt}}



 
\title{Feature engineering and symbolic regression methods for detecting hidden physics from sparse \textcolor{rev1}{sensor observation data}}

\author{Harsha Vaddireddy}
\affiliation{ 
School of Mechanical \& Aerospace Engineering, Oklahoma State University, Stillwater, OK 74078, USA.
}%

\author{Adil Rasheed}%
\affiliation{ 
	Department of Engineering Cybernetics, Norwegian University of Science and Technology, N-7465, Trondheim, Norway.
}%

\author{Anne E Staples}%
\affiliation{ 
	Department of Biomedical Engineering and Mechanics, Virginia Tech, Blacksburg, VA 24061, USA.
}%

\author{Omer San}%
 \email{osan@okstate.edu}
\affiliation{ 
School of Mechanical \& Aerospace Engineering, Oklahoma State University, Stillwater, OK 74078, USA.
}%


\date{\today}

\begin{abstract}
We put forth a modular approach for distilling hidden flow physics from discrete and sparse observations. To address functional expressiblity, a key limitation of the black-box machine learning methods, we have exploited the use of symbolic regression as a principle for identifying relations and operators that are related to the underlying processes. This approach combines evolutionary computation with feature engineering to provide a tool for discovering hidden parameterizations embedded in the trajectory of fluid flows in the Eulerian frame of reference. Our approach in this study mainly involves gene expression programming (GEP) and sequential threshold ridge regression (STRidge) algorithms. We demonstrate our results in three different applications: (i) equation discovery, (ii) truncation error analysis, and (iii) hidden physics discovery, for which we include both predicting unknown source terms from a set of sparse observations and discovering subgrid scale closure models. We illustrate that both GEP and STRidge algorithms are able to distill the Smagorinsky model from an array of tailored features in solving the Kraichnan turbulence problem. Our results demonstrate the huge potential of these techniques in complex physics problems, and reveal the importance of feature selection and feature engineering in model discovery approaches.

\end{abstract}


\keywords{Symbolic regression, gene expression programming, compressive sensing,  model discovery, modified equation analysis, hidden physics discovery.} 
\maketitle


\section{Introduction}
\label{sec:intro}
Since the dawn of mathematical modelling of complex physical processes, scientists have been attempting to formulate predictive models to infer current and future states. These first principle models are generally conceptualized from conservation laws, sound physical arguments, and empirical heuristics drawn from either conducting experiments or hypothesis made by an insightful researcher. However, there are many complex systems (some being climate science, weather forecasting, and disease control modelling) with their governing equations known partially and their hidden physics await to be modelled. In the last decade, there have been rapid advances in machine learning\cite{ml1, ml2} and easy access to rich data, thanks to the plummeting costs of sensors and high performance computers. 
 
This paradigm shift in data driven techniques can be readily exploited to distill new or improved physical models for nonlinear dynamical systems. Extracting predictive models based on observing complex patterns from vast multimodal data can be loosely termed as reverse engineering nature. This approach is not particularly new, for example, Kepler used planets' positional data to approximate their elliptic orbits. The reverse engineering approach is most appropriate in the modern age as we can leverage computers to directly infer physical laws from data collected from omnipresent sensors that otherwise might not be comprehensible to humans. Symbolic regression methods are a class of data driven algorithms that aim to find a mathematical model that can describe and predict hidden physics from observed input-response data. Some of the popular machine learning techniques that are adapted for the task of  symbolic regression are neural networks\cite{ann1, ann2}, compressive sensing/sparse optimization\cite{candes2008enhancing, cs4}, and  evolutionary algorithms\cite{koza,ferreira2001gene}. 

Symbolic regression (SR) approaches based on evolutionary computation\cite{koza, ferreira2006gene} are a class of frameworks that are capable of finding analytically tractable functions.  Traditional deterministic regression algorithms assume a mathematical form and only find parameters that best fit the data. On the other hand, evolutionary SR approaches aim to simultaneously find parameters and also learn the best-fit functional form of the model from input-response data. Evolutionary algorithms search for functional abstractions with a preselected set of mathematical operators and operands while minimizing the error metrics. Furthermore, the optimal model is selected from Pareto front analysis with respect to minimizing accuracy versus model complexity. Genetic programming (GP)\cite{koza} is a popular choice leveraged by most of the SR frameworks. GP is an extended and improved version of a genetic algorithm (GA)\cite{mitchell1998introduction, holland1992adaptation} which is inspired by Darwin's theory of natural evolution. Seminal work was done in identifying hidden physical laws\cite{gp1,gp2} from the input-output response using the GP approach.  GP has been applied in the context of the SR approach in digital signal processing\cite{yang2005force}, nonlinear system identification\cite{ferariu2009multiobjective} and aerodynamic parametric estimation\cite{luo2015adaptive}. Furthermore, GP as an SR tool was applied to identify complex closed-loop feedback control laws for turbulent separated flows\cite{brunton2015closed,gautier2015closed,duriez2015feedback,debien2016closed}. Hidden physical laws of the evolution of a harmonic oscillator based on sensor measurements and the real world prediction of solar power production at a site were identified using GP as an SR approach\cite{quade2016prediction}.

Improved versions of GP focus on better representation of the chromosome, which helps in the free evolution of the chromosome with constraints on the complexity of its growth, and faster searches for the best chromosome. Some of these improved versions of GP are gene expression programming (GEP)\citep{ferreira2001gene}, parse matrix evolution (PME)\cite{luo2012parse}, and linear genetic programming (LGP)\citep{brameier2007linear}. GEP takes advantage of the linear coded chromosome approach from GA and the parse tree evolution of GP to alleviate the disadvantages of both GA and GP. GEP was applied to diverse applications as an SR tool to recover nonlinear dynamical systems \cite{faradonbeh2017prediction,faradonbeh2017roadheader, hoseinian2017semi, ccanakci2009prediction}. Recently, GEP was modified for tensor regression, termed as multi-GEP, and has been applied to recover functional models approximating the nonlinear behavior of the stress tensor in the Reynolds-averaged Navier-Stokes (RANS) equations\cite{weatheritt2016novel}. Furthermore, this novel algorithm was extended to identify closure models in a combustion setting for large eddy simulations (LES)\cite{schoepplein2018application}. Similarly, a new damping function has been discovered using the GEP algorithm for the hybrid RANS/LES methodology\cite{weatheritt2017hybrid}. Generally, evolutionary based SR approaches can identify models with complex nonlinear compositions given enough computational time.

Compressive sensing (CS)\cite{candes2008enhancing,cs4} is predominately applied to signal processing in seeking the sparsest solution (i.e., a solution with the fewest number of features). Basis pursuit algorithms\cite{rauhut2010compressive}, also identified as sparsity promoting optimization techniques\cite{tibshirani1996regression, sp3}, play a fundamental role in CS. Ordinary least squares (OLS) optimization generally results in identifying models with large complexity which are prone to overfitting. In sparse optimization, the OLS objective function is regularized by an additional constraint on the coefficient vector. This regularization helps in taming and shrinking large coefficients and thereby promoting sparsity in feature selection and avoiding overfitted solutions. The least absolute shrinkage and selection operator (LASSO)\cite{tibshirani1996regression, hastie2015statistical} is one of the most popular regularized least squares (LS) regression methods. In LASSO, an L$_1$ penalty is added to the LS objective function to recover sparse solutions\cite{cs5}. In Bayesian terms, LASSO is a maximum a posteriori estimate (MAP) of LS with Laplacian priors. LASSO performs feature selection and simultaneously shrinks large coefficients which may manifest to overfit the training data.  Ridge regression\cite{ridge} is another regularized variant where an L$_2$ penalty is added to the LS objective function. Ridge regression is also defined as a MAP estimate of LS with a Gaussian prior. The L$_2$ penalty helps in grouping multiple correlated basis functions and increases robustness and convergence stability for ill-conditioned systems. The elastic net approach\cite{zou2005regularization, friedman2010regularization} is a hybrid of the LASSO and ridge approaches combining the strengths of both algorithms. 

Derived from these advances, a seminal work was done in employing sparse regression to identify the physical laws of nonlinear dynamical systems\cite{sindy}. This work leverages the structure of sparse physical laws, i.e., only a few terms represent the dynamics. The authors have constructed a large feature library of potential basis functions that has the expressive power to define the dynamics and then seek to find a sparse feature set from this overdetermined system. To achieve this, a sequential threshold least squares (STLS) algorithm\cite{sindy} has been introduced in such a way that a hard threshold on OLS coefficients is performed recursively to obtain sparse solutions. This algorithm was leveraged to form a framework called sparse identification of nonlinear dynamics (SINDy)\cite{sindy} to extract the physical laws of nonlinear dynamical systems represented by ordinary differential equations (ODEs). This work re-envisioned model discovery from the perspective of sparse optimization and compressive sensing. The SINDy framework recovered various benchmark dynamical systems such as the chaotic Lorenz system and vortex shedding behind a cylinder. However, STLS regression finds it challenging to discover physical laws that are represented by spatio-temporal data or high-dimensional measurements and have highly correlated features in the basis library. This limitation was addressed using a regularized variant of STLS called the sequential threshold ridge regression (STRidge) algorithm\cite{pdefind}. This algorithm was intended to discover unknown governing equations that are represented by partial differential equations (PDEs), hence forming a framework termed as PDE-functional identification of nonlinear dynamics (PDE-FIND)\cite{pdefind}. PDE-FIND was applied to recover canonical PDEs representing various nonlinear dynamics. This framework also performs reasonably well under the addition of noise to data and measurements. These sparse optimization frameworks generally have a free parameter associated with the regularization term that is tuned by the user to recover models ranging from highly complex to parsimonious.

In a similar direction of discovering governing equations using sparse regression techniques,  L$_1$ regularized LS minimization was used to recover various nonlinear PDEs\cite{schaeffer2013sparse, schaeffer2017learning} using both high fidelity and distorted (noisy) data. Additionally, limited and distorted data samples were used to recover chaotic and high-dimensional nonlinear dynamical systems\cite{tran2017exact, schaeffer2018extracting}. To automatically filter models with respect to model complexity (number of terms in the model) versus test accuracy, Bayes information criteria were used to rank the most informative models\cite{mangan2017model}. Furthermore, SINDy coupled with model information criteria is used to infer canonical biological models\cite{mangan2016inferring} and introduce a reduced order modelling (ROM) framework\cite{loiseau2018sparse}. STRidge\cite{pdefind} was applied as a deterministic SR method to derive algebraic Reynolds-stress models for the RANS equations\cite{schmelzer2018data}. Recently, various sparse regression algorithms like LASSO\cite{tibshirani1996regression}, STRidge\cite{pdefind}, sparse relaxed regularized regression\cite{zheng2018unified}, and the forward-backward greedy algorithm\cite{zhang2009adaptive} were investigated to recover truncation error terms of various modified differential equations (MDEs) coming from canonical PDEs\cite{thaler2019108851}. The frameworks discussed above assume that the structure of the model to be recovered is sparse in nature; that is, only a small number of terms govern the dynamics of the system. This assumption holds for many physical systems in science and engineering.

Fast function extraction (FFX)\cite{mcconaghy2011ffx} is another deterministic SR approach based on pathwise regularized learning that is also called the elastic net algorithm \cite{zou2005regularization}. The resulting models of FFX are selected through non-dominated filtering concerning accuracy and model complexity, similar to evolutionary computations. FFX is influenced by both GP and CS to better distill physical models from data. FFX has been applied to recover hidden physical laws\cite{quade2016prediction}, canonical governing equations\cite{vaddireddy2019equation} and Reynolds stress models for the RANS equations\cite{schmelzer2019machine}. Some other potential algorithms for deterministic SR are elite bases regression (EBR)\cite{chen2017elite} and  prioritized grammar enumeration (PGE)\cite{worm2013prioritized}. EBR uses only elite features in the search space selected by measuring correlation coefficients of features for the target model. PGE is another deterministic approach that aims for the substantial reduction of the search space where the genetic operators and random numbers from GP are replaced with grammar production rules and systematic choices.

An artificial neural network (ANN), also referred to as deep learning if multiple hidden layers are used, is a machine learning technique that transforms input features through nonlinear interactions and maps to output target features\cite{ann1, ann2}. ANNs attracted attention in recent times due to their exemplary performance in modelling complex nonlinear interactions across a wide range of applications including image processing\cite{ann3}, video classification\cite{ann4} and autonomous driving \cite{ann5}. ANNs produce black-box models that are not quite open to physical inference or interpretability. Recently, physics-informed neural networks (PINNs)\cite{raissi2019physics} were proposed in the flavor of SR that is capable of identifying scalar parameters for known physical models. PINNs use a loss function in symbolic form to help ANNs adhere to the physical structure of the system. Along similar directions, a Gaussian process regression (GPR) has been also investigated for the discovery of coefficients by recasting unknown coefficients as GPR kernel hyper-parameters for various time dependent PDEs\cite{raissi2018numerical, raissi2018hidden}. \textcolor{rev1}{As a nonlinear system identification tool, the GPR approach provides a powerful framework to model dynamical systems\cite{kocijan2005dynamic,gregorvcivc2008nonlinear}. State calibration with the four dimensional variational data assimilation (4D VAR)\cite{cordier2013identification} and deep learning techniques such as long short-term memory (LSTM)\cite{wang2018model} have been used for model identification in ROM settings.} Convolutional neural networks (CNNs) are constructed to produce hidden physical laws from using the insight of establishing direct connections between filters and finite difference approximations of differential operators\cite{cai2012image, dong2017image}. This approach has been demonstrated to discover underlying PDEs from learning the filters by minimizing the loss functions\cite{pmlr-v80-long18a, long2019pde}.

\textcolor{rev1}{In this paper, we have exploited the use of SR in three different applications,  equation discovery,  truncation error analysis, and  hidden physics discovery.} We demonstrate the use of the evolutionary computation algorithm, GEP, and the sparse regression algorithm, STRidge, in the context of the SR approach to discover various physical laws represented by linear and nonlinear PDEs from observing input-response data. We begin by demonstrating the identification of canonical linear and nonlinear PDEs that are up to fifth order in space. For identifying one particular PDE, we demonstrate the natural feature extraction ability of GEP and the limits in the expressive and predictive power of using a feature library when dealing with STRidge in discovering physical laws. We then demonstrate the discovery of highly nonlinear truncation error terms of the Burgers MDE using both GEP and STRidge. \textcolor{rev1}{We highlight that the analysis of truncation errors is very important in the implicit large eddy simulation as a way to determine inherent turbulence models. This analysis is usually very tedious and elaborate, and our study provides a clear example of how SR tools are suitable in such research.} Following truncation error terms identification, we apply GEP using sparse data to recover hidden source terms represented by complex function compositions for a one-dimensional (1D) advection-diffusion process and a two-dimensional (2D) vortex-merger problem. Furthermore, both GEP and STRidge are used to demonstrate the identification of the eddy viscosity kernel along with its ad-hoc modelling coefficient closing LES equations simulating the 2D decaying turbulence problem. \textcolor{rev1}{An important result is the ability of the proposed methodology to distill the Smagorinsky model from an array of tailored features in solving the Kraichnan turbulence problem.}   
   
The rest of the paper is organized as follows. Section~\ref{sec:meth} gives a brief description of the GEP and STRidge algorithms. In Section~\ref{sec:eq}, GEP, and STRidge are tested on identifying different canonical PDEs. Section~\ref{sec:trunc} deals with the identification of nonlinear truncation terms of the Burgers MDE using both STRidge and GEP. In Section~\ref{sec:hidden} we exploit GEP for identification of hidden source terms in a 1D advection-diffusion process and a 2D vortex-merger problem. We additionally demonstrate recovery of the eddy viscosity kernel and its modelling coefficient by both GEP and STRidge for closing the LES equations simulating the 2D decaying turbulence problem in the same section. Finally, Section~\ref{sec:conc} draws our conclusions and highlights some ideas for future extensions of this work. 


\section{Methodology}
\label{sec:meth}
We recover various physical models from data using two symbolic regression tools namely, GEP, an evolutionary computing algorithm, and STRidge, which is a deterministic algorithm that draws its influences from compressive sensing and sparse optimization. 
We take the example of the equation discovery problem that is discussed in Section~\ref{sec:eq} to elaborate on the methodology of applying GEP and STRidge for recovering various physical models. 
We restrict the PDEs to be recovered to quadratic nonlinear and up to the fifth order in space. The general nonlinear PDE to be recovered is in the form of,
\begin{align}\label{goveq1}
\begin{split}
     u_t &= \mathscr{F}(\sigma, u, u^2, u_{x}, u_x^2, uu_x, u_{2x}, \ldots, u_{5x}^2),\\
\end{split}
\end{align}   
where subscripts denote order of partial differentiation and $\sigma$ is an arbitrary parameter. For  example, consider the problem of identifying the viscous Burgers equation as shown below,
\begin{align}\label{goveq2}
\begin{split}
     u_{t} + uu_{x} &= \nu u_{2x},\\
\end{split}
\end{align} 
where $u(x,t) \in \mathbb{R}^{m \times n}$ is the velocity field and $\nu$ is the kinematic viscosity. In our study, $m$ is the number of time snapshots and $n$ is the number of spatial locations. The solution field $u(x,t)$ is generally obtained by solving Eq.~\ref{goveq2} analytically or numerically. The solution field might also be obtained from sensor measurements that can be arranged as shown below,

\begin{align}\label{goveq3}
\begin{split}
\textbf{u} &= \left.\left[ 
                  \vphantom{\begin{array}{c}1\\1\\1\\1\\1\end{array}}
                  \smash{\overbrace{
  \begin{array}{cccc}
    u_1(t_{1}) &  u_2(t_{1}) &    \ldots    & u_n(t_{1}) \\
    u_1(t_{2})    & u_2(t_{2})    & \ldots & u_n(t_{2})    \\
    \vdots &  \vdots & \ddots       & \vdots \\
    u_1(t_{m})    & u_2(t_{m})    & \ldots & u_n(t_{m}) \\
                      \end{array}
                      }^{\mbox{\text{spatial locations}}}}
              \right]\right\}
              \,\text{time snapshots}
\end{split}
\end{align} 
For recovering PDEs, we need to construct a library of basis functions called as feature library that contains higher order derivatives of the solution field $u(x,t)$. Higher order spatial and temporal partial derivative terms can be approximated using any numerical scheme once the recording of the discrete data set given by Eq.~\ref{goveq3} is available. In our current setup, we use the leapfrog scheme for approximating the temporal derivatives and central difference schemes for spatial derivatives as follows,
\begin{equation}\label{goveq4}
\left.\begin{aligned}
u_t    &= \dfrac{u^{p+1}_{j}-u^{p-1}_{j}}{2dt}\\
u_{2t} &= \dfrac{u^{p+1}_j-2u^p_j+u^{p-1}_j}{dt^2} \\
u_x    &= \dfrac{u^p_{j+1}-u^p_{j-1}}{2dx}\\
u_{2x} &= \dfrac{u^p_{j+1}-2u^p_j+u^p_{j-1}}{dx^2} \\
u_{3x} &= \dfrac{u^p_{j+2}-2u^p_{j+1}+2u^p_{j-1}-u^p_{j-2}}{2dx^3} \\
u_{4x} &= \dfrac{u^p_{j+2}-4u^p_{j+1}+6u^p_{j}+-4u^p_{j-1}-u^p_{j-2}}{dx^4} \\
u_{5x} &= \dfrac{u^p_{j+3}-4u^p_{j+2}+5u^p_{j+1}-5u^p_{j-1}+4u^p_{j-2}-u^p_{j-3}}{2dx^5} \\
\end{aligned}
\right\},
\end{equation}
where temporal and spatial steps are given by $dt$ and $dx$, respectively. Within the expressions presented in Eq.~\ref{goveq4}, the spatial location is denoted using subscript index $j$, and the temporal instant using superscript index $p$.  

We note that other approaches such as automatic differentiation or spectral differentiation for periodic domains can easily be adopted within our study. Both GEP and STRidge take the input library consisting of features (basis functions) that are built using Eq.~\ref{goveq2} and Eq.~\ref{goveq3}. This core library, used for the equation discovery problem in Section~\ref{sec:eq}, is shown below,
\begin{equation}\label{goveq5}
\left.\begin{aligned}
\textbf{V(t)}  &=  
\left[
  \begin{array}{c}    
     \textbf{U$_t$}
  \end{array}
\right]\\
 \mathbf{\widetilde \Theta(U)}  &=  
\left[
  \begin{array}{cccccccc}    
    \textbf{U}  &  \textbf{U$_x$} & \textbf{U$_{2x}$} & \textbf{U$_{3x}$} & \textbf{U$_{4x}$} & \textbf{U$_{5x}$}
\end{array}
\right]
\end{aligned}
\right\}.
\end{equation}

The solution $u(x,t)$ and its spatial and temporal derivatives are arranged with size $m \cdot n \times 1$ in each column of Eq.~\ref{goveq5},. For example, the features (basis functions) $\textbf{U}$ and $\textbf{U$_{2x}$}$  are arranged as follows,
\begin{equation}\label{goveq6}
\textbf{U}  =  
\left[
  \begin{array}{c}
    u(x_0, t_0)\\
    u(x_0, t_1)\\
    \vertbar   \\
    u(x_j, t_p)\\
    \vertbar\\
 u(x_n, t_m)\\   
  \end{array}
\right], \hspace{2mm} \textbf{U$_{2x}$}  =  
\left[
  \begin{array}{c}
    u_{2x}(x_0, t_0)\\
    u_{2x}(x_0, t_1)\\
    \vertbar   \\
    u_{2x}(x_j, t_p)\\
    \vertbar\\
 u_{2x}(x_n, t_m)\\   
  \end{array}
\right],
\end{equation}
where subscript $j$ denotes the spatial location and subscript $p$ denotes the time snapshot.
The features (basis functions) in the core library  $\mathbf{\widetilde \Theta(U)}$ is expanded to include interacting  features limited to quadratic nonlinearity and also a constant term. The final expanded library is given as,
\begin{equation}\label{goveq7}
 \mathbf{\Theta(U)}  =  
\left[
  \begin{array}{cccccccc}    
    \textbf{1}   & \textbf{U} & \textbf{U$^{2}$}  &  \textbf{U$_x$} &  \textbf{UU$_x$} & \textbf{U$^2_x$} & \ldots  & \textbf{U$^2_{5x}$}
  \end{array}
\right],
\end{equation}
where the size of the library is $\mathbf{\Theta(U)}$  $\in$ $\mathbb{R}^{m \cdot n \times N_{\beta}}$ and  $N_{\beta}$ is number of features (basis functions) i.e., $N_{\beta}=28$ for our setup. For example, if we have 501 spatial points and 101 time snapshots with 28 bases, then  $\mathbf{\Theta(U)}$ (Eq.~\ref{goveq7}) contains $501 \times 101$ rows and 28 columns.

Note that the core feature library $\mathbf{\widetilde \Theta(U)}$  in Eq.~\ref{goveq5} is given as an input to GEP to recover PDEs and the algorithm extracts higher degree nonlinear interactions of core features in  $\mathbf{\widetilde \Theta(U)}$ automatically. However, for sparse optimization techniques such as STRidge, explicit input of all possible combinations of core features in Eq.~\ref{goveq5} are required. Therefore,  $\mathbf{\Theta(U)}$ in Eq.~\ref{goveq7} forms the input to STRidge algorithm for equation identification. This forms the fundamental difference in terms of feature building for both algorithms. The following Subsection~\ref{sec:gep} gives a brief introduction to GEP and its specific hyper-parameters that control the efficacy of the algorithm in identifying  physical models from observing data. Furthermore, the Subsection~\ref{sec:str} describes how to form linear system representations in terms of $\mathbf{V(t)}$ and  $\mathbf{\Theta(U)}$ and briefly describe STRidge optimization approach to identifying sparse features and thereby building parsimonious models using spatio-temporal data.


\subsection{Gene Expression Programming}
\label{sec:gep}
Gene expression programming (GEP)\cite{ferreira2001gene, ferreira2002gene} is a genotype-phenotype evolutionary optimization algorithm which takes advantage of simple chromosome representation of genetic algorithm (GA)\cite{mitchell1998introduction} and the free expansion of complex chromosomes of genetic programming (GP)\cite{koza}. As in most evolutionary algorithms, this technique also starts with generating initial random populations, iteratively selecting candidate solutions according to a fitness function,    and improving candidate solutions by modifying through genetic variations using one or more genetic operators. The main difference between GP and GEP is how both techniques define the nature of their individuals. In GP, the individuals are nonlinear entities of different sizes and shapes represented as parse trees and in GEP the individuals are encoded as linear strings of fixed length called genome and chromosome, similar to GA representation of individual and later expressed as nonlinear entities of different size and shape called phenotype or expression trees (ET). GEP is used for a very broad range of applications, but here it is introduced as a symbolic regression tool to extract constraint free solutions from  input-response data.

\begin{figure}[!ht]
	\centering
	\includegraphics[width=0.5\textwidth,trim={1.5cm 0 0 0},clip]{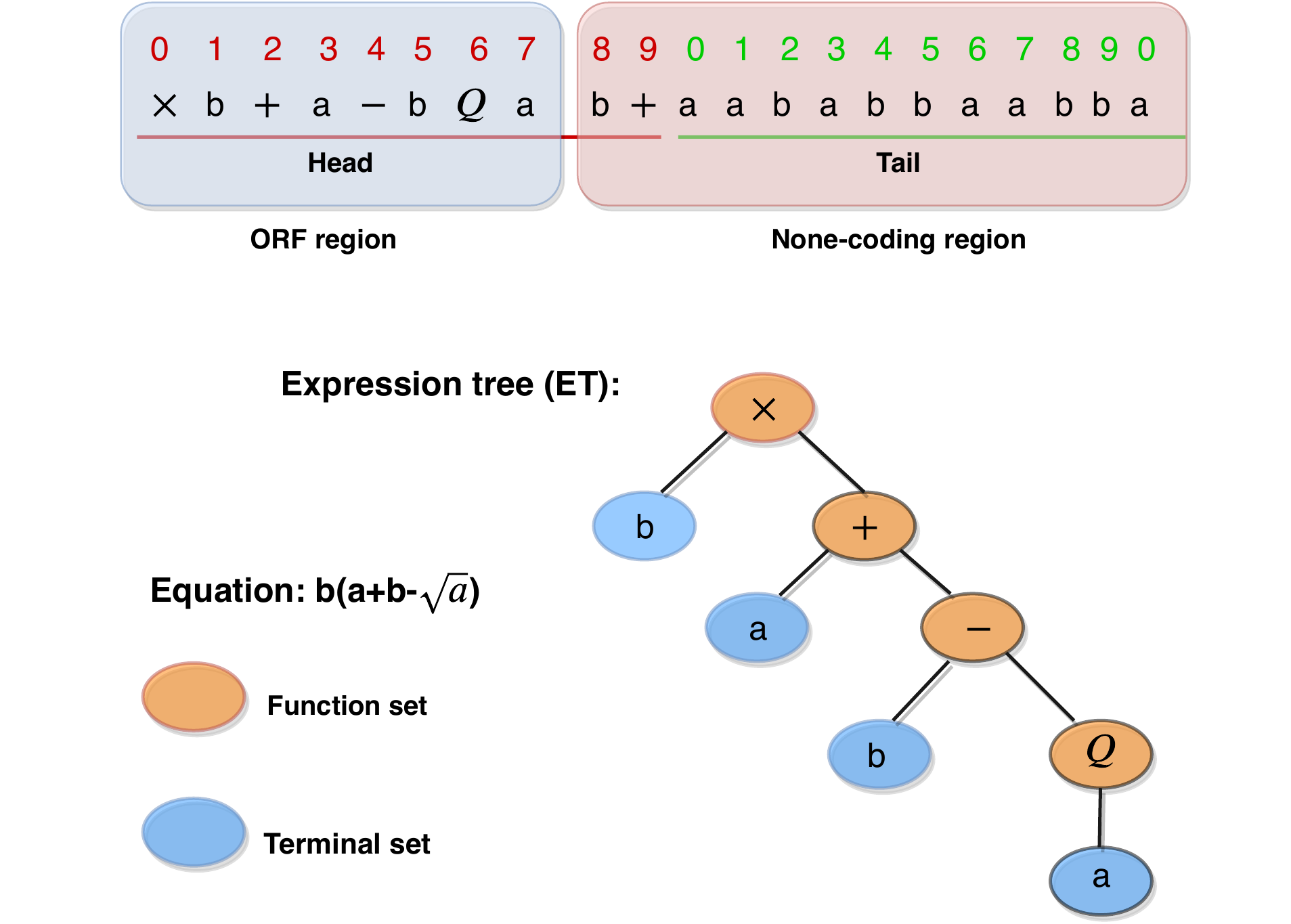}
	\caption{ET of a gene/chromosome with its structure in GEP. $Q$ represents the square root operator.}
	\label{fig:gep0}
\end{figure}

The arrangement of a typical gene/chromosome in GEP is shown in  Fig.~\ref{fig:gep0}. The GEP gene is composed of head and tail regions as illustrated in Fig.~\ref{fig:gep0}. The head of a gene consists of both symbolic terms from functions (elements from a function set $F$) and terminals (elements from a terminal set $T$) whereas the tail consists of only terminals. The function set $F$ may contain arithmetic mathematical operators (e.g., {$+, \times, -, /$}), nonlinear functions (e.g., {sin, cos, tan, arctan, sqrt, exp}), or Boolean operators (e.g., {Not , Nor , Or , And})  and the terminal set contains the symbolic variables. The gene always starts with a randomly generated mathematical operator from the function set $F$. The head length is one of the important hyper-parameters of GEP, and it is  determined using trial and error  as there is no definite method to assign it. Once the head length is determined, the size of the tail is computed as a function of the  head length and the maximum arity of a mathematical operator  in the function set $F$\cite{ferreira2006gene}. It can be calculated by the following equation,
\begin{align}\label{tail_t}
 \textrm{tail length} = \textrm{head length}\times(a_{max} - 1) + 1,
\end{align}
where $a_{max}$ is the maximum argument of a function in $F$. The single gene can be extended to multigenic chromosomes where individual genes are linked using a linking function (eg., $+, \times, /, -$). The general rule of thumb is to have a larger head and higher number of genes when dealing with complex problems\cite{ferreira2006gene}. 

The structural organization of the GEP gene is arranged in terms of open reading frames (ORFs) inspired from biology where the coding sequence of a gene equivalent to an ORF begins with a start codon, continue with an amino acid codon and ends with a termination codon. In contrast to a gene in biology, the start site is always the first position of a gene in  GEP, but the termination point does not always coincide with the last position of a gene. These regions of the gene are termed non coding regions downstream of the termination point. Only the ORF region is expressed in the ET and can be clearly seen in Fig.~\ref{fig:gep0}.

Even though the none-coding regions in GEP genes do not participate in final solution, the power of  GEP evolvability lies in this region. The syntactically correct genes in GEP evolve after modification through diverse genetic operators due to this region chromosome. This is the paramount difference between GEP and GP implementations where in latter, many syntactically invalid individuals are produced and need to be discarded while evolving the solutions and additional special constraint are imposed on the depth/complexity of candidate solution to be evolved to avoid bloating problem \cite{duriez2015feedback}. 

Fig.~\ref{fig:gep1} displays the typical flowchart of the GEP algorithm. The process is described briefly below, 
\begin{enumerate}

    \item The optimization procedure starts with a random generation of chromosomes built upon combinations of functions and terminals. The size of the random population is a hyper-parameter and the larger the population size, better the probability of finding the best candidate solution.
    
    \item After the population is generated, the chromosomes are expressed as ETs, which is  converted to a numerical expression. This expression is then evaluated using a fitness function. In our setup, we employ the mean squared error between the best predicted model $f^{*}$ and the true model $f$ as the fitness function given by,
\begin{equation}\label{eq:mse}
    MSE = \frac{1}{N}\sum_{l=1}^{N}\left(f^{*}_{(lk)} - f_{(l)}\right)^2,
\end{equation}
where  $f^{*}_{lk}$ is the value predicted by the chromosome  $k$  for the fitness case  $l$ (out of $N$ samples cases) and $f_l$  is the true or measurement value for the $l^{th}$ fitness case.   

\begin{figure}[!ht]
	\centering
	\includegraphics[width=0.5\textwidth]{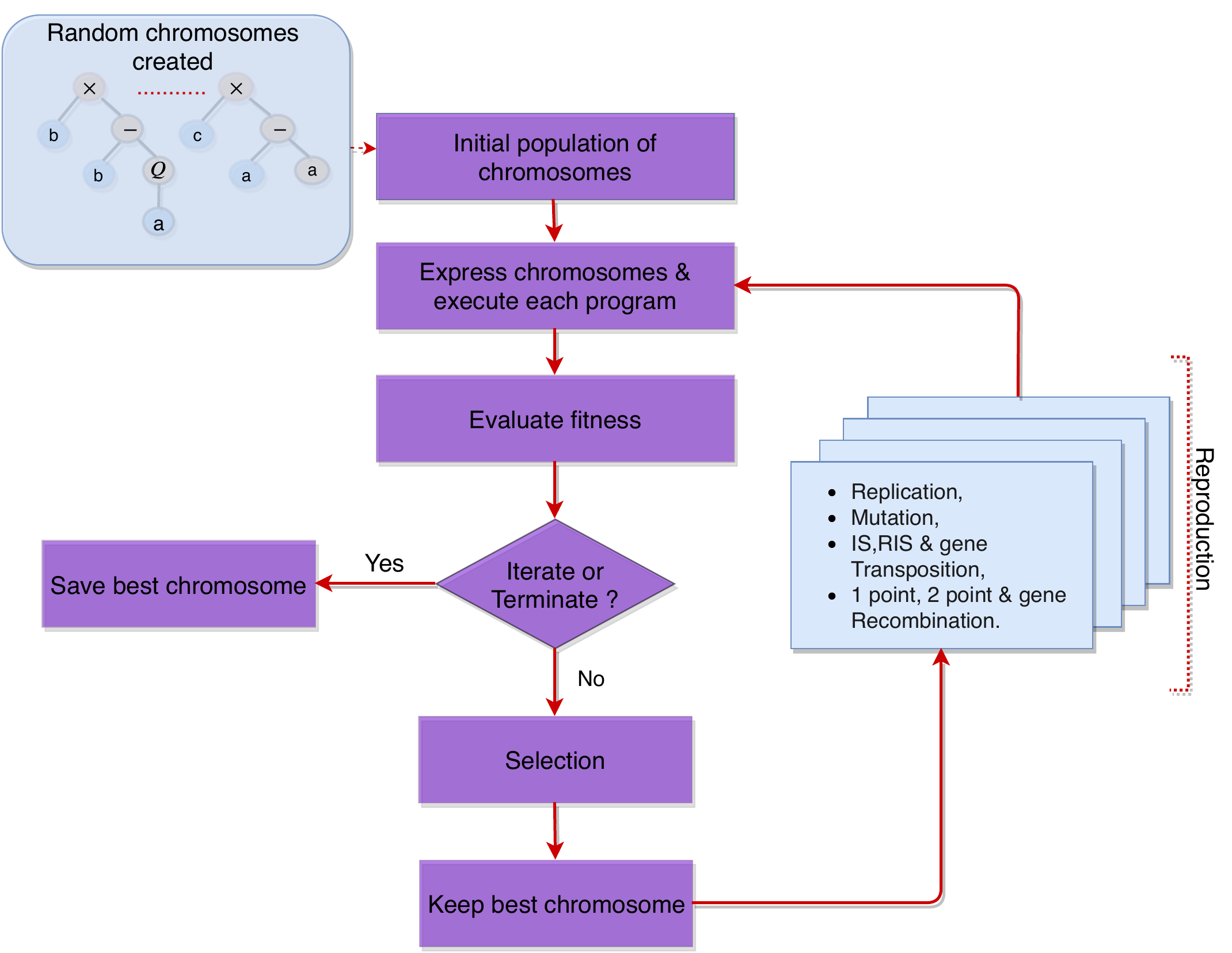}
	\caption{Flowchart of the gene expression programming.}
	\label{fig:gep1}
\end{figure}

 \item The termination criteria is checked after all fitness evaluations, to continue evolving or to save the best fitness chromosome as our final predicted model. In our current setup, we terminate after a specified number of generations. 
  
  \item The evolvability/reproduction of chromosome through genetic operators which is the core part of the GEP evolutionary algorithm executes if termination criteria is not met. Before the genetic operations on chromosome begins, the best chromosome according to fitness function is cloned to the next generations using a selection method. Popular selection methods include tournament selection with elitism and roulette-wheel selection with elitism.  In our current setup, we use tournament selection with elitism. 
  
  \item The four genetic operators that introduce variation in populations are mutation, inversion, transposition, and recombination. The GEP transposition operator is applied to the elements of the chromosome in three ways: insertion sequence (IS), root insertion sequence (RIS) and gene insertion sequence and similarly three kinds of recombination are applied namely one point, two point, and gene recombination. 
  
  \item The process is continued up to termination criteria is met, which is the number of generations in our current setup. 
  
\end{enumerate}

\begin{table}[htpb]
\caption{ GEP hyper-parameters for various genetic operators selected for all the test cases in this study.}
\label{tab:ghyp0}
\bgroup
\def\arraystretch{1.5}
\setlength{\tabcolsep}{0.05em}
\begin{tabular}{@{}lc}
\hline
\textbf{Hyper-parameters} & \textbf{Value}   \\ \hline
\noalign{\smallskip}
Selection                &   Tournament selection \\
Mutation rate            &   $0.05$\\
Inversion                &   $0.1$ \\
IS transposition rate    &   $0.1$ \\
RIS transposition rate   &   $0.1$ \\
Gene transposition rate  &   $0.1$ \\
One point recombination  &   $0.3$ \\
Two point recombination  &   $0.2$ \\
Gene recombination       &   $0.1$ \\
Dc specific mutation rate      &   $0.05$ \\
Dc specific inversion rate     &   $0.1$  \\
Dc specific transposition rate &   $0.1$  \\
Random constant mutation rate  &  $0.02$  \\
\hline
\end{tabular}
\egroup
\end{table}

Numerical constants occur in most mathematical models and, therefore, it is important to any symbolic regression tools to effectively integrate floating point constants in their optimization search. GP \cite{koza} handles numerical constants by introducing random numerical constants in a specified range to its parse trees. The random constants are moved around the parse trees using the crossover operator. GEP  handles the creation of random numerical constants (RNCs) by using an extra terminal `?' and a separate domain Dc composed of symbols chosen to represent random numerical constants \cite{ferreira2006gene}. This Dc specific domain starts from the end of the tail of the gene. 

For each gene, RNCs are generated during the creation of a random initial population and kept in an array. To maintain the genetic variations in the pool of RNCs, additional genetic operators are introduced to take effect on Dc specific regions. Hence in addition to the usual genetic operators such as mutation, inversion, transposition and recombination, the GEP-RNC algorithm has Dc specific inversion, transposition, and random constant mutation operators.  Hence, with these modifications to the algorithm, an appropriate diversity of random constants can be generated and evolved through operations of genetic operators. The values for each genetic operator selected for this study are listed in Table~\ref{tab:ghyp0}. These values are selected from various examples given by Ferreira\cite{ferreira2006gene} combined with the trial and error approach. Additionally, to simplify our study, we use the same parameters for all the test cases even though they may not be the best values for the test case under investigation.  

Once decent values of genetic operators that can explore the search space are selected, the size of the head length, population, and the number of genes form the most important hyper-parameters for GEP. Generally, larger head length and a greater number of genes are selected for identifying complex expressions. Larger population size helps in a diverse set of initial candidates which may help GEP in finding the best chromosome in less number of generations. However, computational overhead increases with an increase in the size of the population. Furthermore, the best chromosome can be identified in fewer generations with the right selection of the linking function between genes. GEP algorithm inherently performs poor in predicting the numerical constants that are ubiquitous in  physical laws. Hence, the GEP-RNC algorithm is used where a range of random constants are predefined to help GEP to find numerical constants. This also becomes important in GEP identifying the underlying expression in fewer generations. Finally, we note that due to the heuristic nature of evolutionary algorithms, any other combinations of hyper-parameters might work perfectly in identifying the symbolic expressions.  \textcolor{rev1}{In this study, we use geppy\cite{geppy}, an open source library for symbolic regression using GEP, which is built as an extension to distributed evolutionary algorithms in Python (DEAP) package\cite{DEAP_JMLR2012}. All codes used in this study are made available on Github (\href{https://github.com/sayin/SR}{https://github.com/sayin/SR}).}


\subsection{Sequential Threshold Ridge Regression }
\label{sec:str}
Compressive sensing/sparse optimization\cite{cs0, cs4} has been exploited for sparse feature selection from a large library of potential candidate features and recovering dynamical systems represented by ODEs and PDEs\cite{sindy, pdefind, mangan2017model} in a highly efficient computational manner. In our setup, we use this STRidge\cite{pdefind} algorithm to recover various hidden physical models from observed data. In continuation with the Section~\ref{sec:meth} where we define feature library $\mathbf{\Theta(U)}$ and target/output data $\mathbf{V(t)}$, this subsection briefly explains the formation of an overdetermined linear system for STRidge optimization to identify various physical models from data.

The Burgers PDE given in Eq.~\ref{goveq2} or any other PDE under consideration can be written in the form of linear system representation in terms of $\mathbf{\Theta(U)}$ and $\mathbf{V(t)}$,

\begin{equation}\label{goveq8}
\textbf{V$(t)$} = \mathbf{\Theta(U)} \cdot  \boldsymbol\beta,
\end{equation}
where $\boldsymbol{\beta} =$ [$\beta_1, \beta_2, \ldots, \beta_{N_{\beta}}$] is coefficient vector of size $\mathbb{R}^{N_{\beta}}$ where $N_{\beta}$ is number of features (basis functions) in library $\mathbf{\Theta(U)}$. Note that $\mathbf{\Theta(U)}$ is an over-complete library (the number of measurements is greater than the number of features) and having rich feature (column) space to represent the dynamics under consideration.  Thus, we form an overdetermined linear system in Eq.~\ref{goveq8}. The goal of STRidge is to find a sparse coefficient vector $\boldsymbol\beta$ that only consists of active features, which best represent the dynamics. The rest of the features are hard thresholded to zero. For example, in the Burgers equation given by Eq.~\ref{goveq2}, STRidge ideally has to find the coefficient vector $\boldsymbol\beta$ that corresponds to the features $uu_x$ and $u_{2x}$ and simultaneously it should set all other feature coefficients to zero.

\begin{figure}[!ht]
	\centering
	\includegraphics[width=0.5\textwidth]{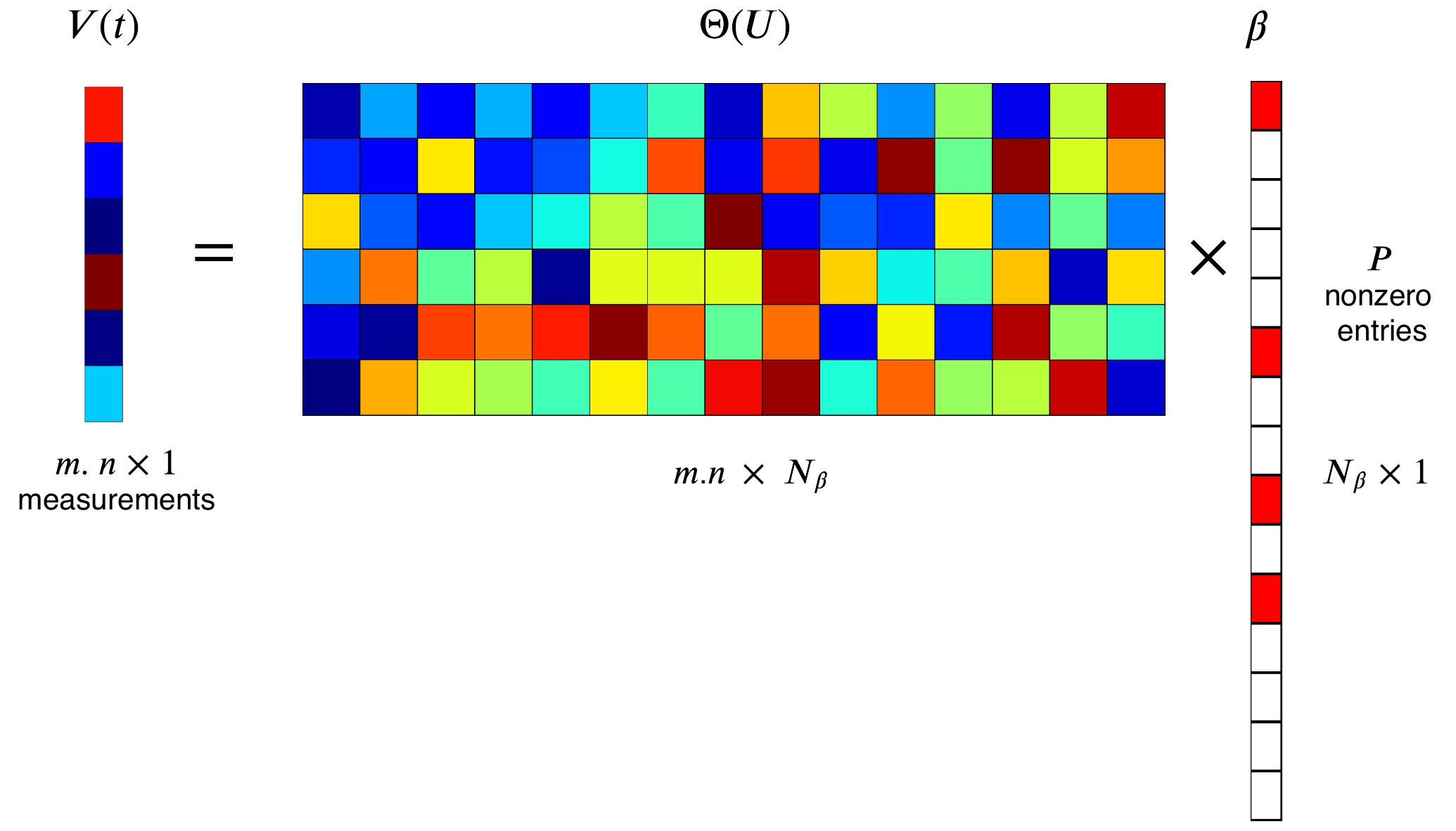}
	\caption{Structure of compressive matrices with sparse non zero entries in  coefficient vector $\boldsymbol{\beta}$. Red boxes in $\boldsymbol{\beta}$ vector correspond to active feature coefficients and all other coefficients being set to zero. }
	\label{fig:str0}
\end{figure}

The linear system defined in  Eq.~\ref{goveq8} can be solved for $\beta$ using the ordinary least squares (OLS) problem. But OLS minimization tries to form a functional relationship with all the features in $\mathbf{\Theta(U)}$  resulting in all non zero values in the coefficient vector $\boldsymbol\beta$. Thus solving Eq.~\ref{goveq8} using OLS infers radically complex functional form to represent the underlying PDE and generally results in overfitted models. Regularized least square minimization can be applied to constraint the coefficients and avoid overfitting.  Hence regularized LS optimization is preferred to identify the sparse features (basis functions) along with their coefficient estimation. Typical estimation of sparse coefficient vector with $P$ non zero entries in  $\boldsymbol{\beta}$ is shown in Fig.~\ref{fig:str0}.  General sparse regression objective function to approximate the solution of the coefficient vector $\boldsymbol{\beta}$ is given by,

\begin{align}
\label{goveq9}
{\beta}^\ast = \argmin_\beta ||\mathbf{\Theta}\cdot\beta -\mathbf{V(t)}||^2_2 + \lambda|| \beta||_0,
\end{align} 
where $\lambda$ is regularizing weight and $||\beta||_0$ corresponds to L$_0$  penalty which makes the problem $np$-hard.  Hence to arrive at convex optimization problem of Eq.~\ref{goveq9}, L$_1$ and L$_2$  penalty is generally used to approximate the solution of the coefficient vector $\boldsymbol{\beta}$. 

The addition of L$_1$ penalty to LS objective function which corresponds to maximum a posteriori estimate (MAP) of Laplacian prior and termed as least absolute shrinkage and selection operator (LASSO) in compressive sensing. It is defined by,

\begin{align}
\label{goveq9}
{\beta}^\ast = \argmin_\beta ||\mathbf{\Theta}\cdot\beta -\mathbf{V(t)}||^2_2 + \lambda|| \beta||_1.
\end{align} 
However, the performance of LASSO deteriorates when the feature space is correlated \cite{pdefind}. The sequential threshold least squares (STLS) algorithm was proposed to identify dynamical systems represented by ODEs \cite{sindy}. In STLS, a hard threshold is performed on least square estimates of regression coefficients and hard threshold is recursively performed on remaining non zero coefficients. However, the efficacy of STLS reduces when dealing with the identification of systems containing multiple correlated columns in $\mathbf{\Theta}$. Hence L$_2$ regularized least squares termed as ridge regression \cite{ridge},  which corresponds to the maximum a posteriori estimate using a Gaussian prior, is proposed to handle the identification of PDEs. Ridge regression is defined by,
 
\begin{align}
\label{goveq10}
{\beta}^\ast &= \argmin_\beta ||\mathbf{\Theta}\cdot\beta -\mathbf{V(t)}||^2_2 + \lambda|| \beta||_2, \nonumber \\
             &= (\mathbf{\Theta}^T\mathbf{\Theta} + \lambda^TI)\mathbf{\Theta}^T\mathbf{V(t)}.
\end{align} 

Ridge regression is substituted for ordinary least squares in STLS and the resulting algorithm as sequential threshold ridge regression (STRidge) \cite{pdefind}. The STRidge framework\cite{pdefind}  is illustrated in Algorithm~\ref{alg0} for the sake of completeness. Note that, if $\lambda = 0$,  STRidge becomes STLS procedure. For more elaborate details on updating tolerance ($tol$) to perform hard thresholding in Algorithm~\ref{alg0}, readers are encouraged to refer supplementary document of Rudy et al\cite{pdefind}.

\begin{algorithm}[h!]
\SetAlgoLined
\LinesNumberedHidden
  \KwIn{ $ \boldsymbol{\Theta}, \mathbf{V(t)}, \lambda, tol$, iters }
  \KwOut{$\beta^{\ast}$ }  
$ {\beta}^\ast = \argmin_\beta ||\mathbf{\Theta}\cdot\beta -\mathbf{V(t)}||^2_2 + \lambda|| \beta||^2_2 $ \\
 large = \{$p: |\beta^{\ast}_p|\geq tol$\} \\
 $\beta^{\ast}[~\textrm{large}] = 0$ \\
 $\beta^{\ast}[\textrm{large}] = \textrm{STRidge}(\boldsymbol\Theta[:,\textrm{large}], \mathbf{V(t)}, \lambda, tol, \textrm{iters}-1)$ \\
 return  $\beta^{\ast}$
 \caption{STRidge($\boldsymbol{\Theta}$, $\mathbf{V(t)}$, $\lambda$, $tol$, iters)\cite{pdefind}}
 \label{alg0}
\end{algorithm}

We use the framework provided by Rudy et al.\cite{pdefind} in our current study. The hyper-parameters in STRidge include the regularization weight $\lambda$ and tolerance level $tol$ which are to be tuned to identify appropriate physical models. In the present study, the sensitivity of feature coefficients for various values of $\lambda$ and the final value of  $\lambda$ where the best model is identified is showed. The following sections deal with various numerical experiments to test the GEP and STRidge frameworks.


\begin{table*}[!htpb]
\caption{Summary of canonical PDEs selected for recovery.}
\label{tab:eq dis}
\begin{tabular}{@{}llc@{\hskip 1.5mm}c@{}}
\hline\noalign{\smallskip}
\textbf{PDE} & \textbf{\begin{tabular}[c]{@{}l@{}}Exact solution \end{tabular}} & \textbf{\begin{tabular}[c]{@{}c@{}}Constant \\ parameters\end{tabular}} & \textbf{\begin{tabular}[c]{@{}c@{}}Discretization\\ $n$ (spatial) \\ $m$ (temporal)\end{tabular}} \\ \hline
\noalign{\smallskip}
\begin{tabular}[c]{@{}l@{}}Wave eq.\\ $u_t = - a u_x$\end{tabular} &  $ u(t,x) = \textrm{sin}(2\pi(x - at))$ & $a = 1.0$ & \begin{tabular}[c]{@{}c@{}} $x \in [0,1]~(n = 101),$\\  $t \in [0,1]~(m = 101) $ \end{tabular} \\[20pt]

\begin{tabular}[c]{@{}l@{}}Heat eq.\\$u_t = - \alpha u_{2x}$\end{tabular} & $u(t,x) = -\textrm{sin}(x)\textrm{exp}(- \alpha t)$  & $\alpha = 1.0$ & \begin{tabular}[c]{@{}c@{}}$x \in [-\pi,\pi]~(n = 201),$\\  $t \in [0,1]]~(m = 101)$\end{tabular}   \\[20pt]

\begin{tabular}[c]{@{}l@{}}Burgers eq. (i) \\ $u_t = - uu_{x} + \nu u_{2x}$\end{tabular} & $u(t,x) =\dfrac{x}{(t+1)\left( 1 + (\sqrt{t + 1}) \textrm{exp}({\frac{1}{16\nu}}\frac{4x^2-t-1}{t+1})\right)}$ & $\nu = 0.01$ & \begin{tabular}[c]{@{}c@{}}$x \in [0,1]~ (n = 101),$\\$t \in [0,1]~ (m = 101)$\end{tabular}  \\[20pt] \\

\begin{tabular}[c]{@{}l@{}}Burgers eq. (ii) \\ $u_t = - uu_{x} + \nu u_{2x}$\end{tabular} & $u(t,x) = \dfrac{2\nu \pi \textrm{exp}(-\pi^2\nu t) \textrm{sin}(\pi x)}{a+ \textrm{exp}(-\pi^2\nu t) \textrm{cos}(\pi x)}$ & \begin{tabular}[c]{@{}c@{}} $\nu = 0.01,$ \\ $a=5/4$ \end{tabular}   & \begin{tabular}[c]{@{}c@{}}$x \in [0,1]~(n = 101),$\\  $t \in [0,100]~(m = 101)$\end{tabular}  \\[20pt] \\

\begin{tabular}[c]{@{}l@{}}Korteweg-de Vries eq.\\ $u_t = -\alpha uu_{x} - \beta u_{3x}$\end{tabular} & $u(t,x) = 12\left(\dfrac{4\textrm{cosh}(2x -8t)+ \textrm{cosh}(4x - 64t) + 3}{(3\textrm{cosh}(x - 28t) + \textrm{cosh}(3x - 36t))^2}\right)$ & \begin{tabular}[c]{@{}c@{}} $\alpha = 6.0,$ \\ $\beta = 1.0$ \end{tabular} & \begin{tabular}[c]{@{}c@{}}$x \in [-10,10]~(n = 501),$\\  $t \in [0,1]~(m = 201)$\end{tabular}  \\[20pt]

\begin{tabular}[c]{@{}l@{}}Kawahara eq.\\ $u_t = -uu_{x} -  \alpha u_{3x} - \beta u_{5x}$\end{tabular} & $u(t,x) =\dfrac{105}{169}\textrm{sech}\left(\dfrac{1}{2\sqrt{13}}\left(x - a t\right)\right)^4$ & \begin{tabular}[c]{@{}c@{}}$\alpha = 1.0,$\\ $\beta = 1.0,$ \\ $a=36/169$ \end{tabular} & \begin{tabular}[c]{@{}c@{}}$x \in [-20,20]~(n = 401),$\\  $t \in [0,1]~( m = 101)$\end{tabular}  \\[20pt]

\begin{tabular}[c]{@{}l@{}}Newell-Whitehead-Segel eq.\\ $u_t = \kappa u_{2x}+  \alpha u - \beta u^{q}$\end{tabular} & $u(t,x) = \dfrac{1}{\left(1+ \textrm{exp}(\dfrac{x}{\sqrt{6}} - \dfrac{5t}{6}) \right)^2} $ & \begin{tabular}[c]{@{}c@{}}$\kappa = 1.0,$\\ $\alpha = 1.0,$\\ $\beta = 1.0,$\\ $q = 2$\end{tabular} & \begin{tabular}[c]{@{}c@{}}$x \in [-40,40]~(n = 401),$\\  $t \in [0,2]~(m = 201)$\end{tabular}  \\[20pt] \\

\begin{tabular}[c]{@{}l@{}}Sine-Gordon eq.\\ $u_{2t} = \kappa u_{2x} - \alpha\textrm{sin}(u)$\end{tabular} & $u(t,x) = 4 \textrm{tan}^{-1}(\textrm{sech}(x)t)$ & \begin{tabular}[c]{@{}c@{}}$\kappa = 1.0,$\\ $\alpha = 1.0$\end{tabular} & \begin{tabular}[c]{@{}c@{}}$x \in [-2,2]~(n = 401),$\\  $t \in [0,1]~(m = 101) $\end{tabular}  \\
\hline\noalign{\smallskip}
\end{tabular}
\end{table*}

\begin{table*}[!htpb]
\caption{ GEP hyper-parameters selected for identification of various PDEs.}
\label{tab:hp1}
\bgroup
\def\arraystretch{1.5}
\setlength{\tabcolsep}{0.8em}
\begin{tabular}{@{}lcccc}
\hline
\textbf{Hyper-parameters}  & \textbf{Wave eq.} & \textbf{Heat eq.}  & \textbf{Burgers eq.~(i)}  & \textbf{Burgers eq.~(ii)}  \\ \hline
\noalign{\smallskip}
Head length              &   $2$   &  $2$      &  $4$    &    $2$    \\
Number of genes          &   $1$   &  $2$      &  $1$    &    $2$    \\
Population size          &   $25$  &  $25$     &  $20$   &    $50$   \\
Generations              &   $100$ &  $100$    &  $500$  &    $500$  \\
Length of RNC array      &   $10$  &  $10$     &  $30$   &    $5$    \\
Random constant minimum  &   $-10$  &  $-1$     &  $-1$   &    $-1$   \\
Random constant maximum  &   $10$  &  $1$      &  $1$    &    $1$    \\
\hline
\end{tabular}
\egroup
\end{table*}

\begin{table*}[!htpb]
\caption{ GEP hyper-parameters selected for identification of various PDEs.}
\label{tab:hp2}
\bgroup
\def\arraystretch{1.5}
\setlength{\tabcolsep}{0.8em}
\begin{tabular}{@{}lcccc}
\hline
\textbf{Hyper-parameters}  &  \textbf{KdV eq.}  & \textbf{Kawahara eq.}  & \textbf{NWS eq.}  &  \textbf{Sine-Gordon eq.} \\ \hline
\noalign{\smallskip}
Head length              &   $6$     &    $2$   &  $5$    &    $3$    \\
Number of genes          &   $5$     &    $1$   &   $3$   &    $2$    \\
Population size          &   $20$    &    $20$  &  $30$   &    $100$  \\
Generations              &   $500$   &    $100$ &  $100$  &    $500$  \\
Length of RNC array      &   $30$    &    $5$   &  $25$   &    $20$   \\
Random constant minimum  &   $1$     &   $-1$   &   $-10$ &    $-10$  \\
Random constant maximum  &   $10$    &    $1$   &   $10$  &    $10$   \\
\hline
\end{tabular}
\egroup
\end{table*}

\section{Equation Discovery}
\label{sec:eq}

Partial differential equations (PDEs) play a prominent role in all branches of science and engineering. They are generally derived from conservation laws, sound physical arguments, and empirical heuristic from an insightful researcher. The recent explosion of machine learning algorithms provides new ways to identify hidden physical laws represented by PDEs using only data. In this section, we demonstrate the identification of various linear and nonlinear canonical PDEs using the GEP and STRidge algorithms from using data alone. Analytical solutions of PDEs are used to form the data. Table~\ref{tab:eq dis}  summarizes various PDEs along with their analytical solutions $u(t,x)$ and domain discretization. Building a feature library and corresponding response data to identify PDEs is discussed in detail in Section~\ref{sec:meth}.  

\begin{table}[!htpb]
\caption{GEP functional and terminal set used for equation discovery. `?' is a random constant.}
\label{tab:gepset}
\bgroup
\def\arraystretch{1.5}
\setlength{\tabcolsep}{2.1em}
\begin{tabular}{@{}lll}
\hline
\textbf{Parameter} & \textbf{Value}    \\ \hline
\noalign{\smallskip}
Function set     &   $+, -, \times, /$, sin, cos  \\
Terminal set     &   $\mathbf{\widetilde \Theta(U)}$, $?$ \\
Linking function &   $+$  \\
\hline
\end{tabular}
\egroup
\end{table}

We reiterate the methodology for PDE identification in Section~\ref{sec:meth}. The analytical solution $u(t,x)$  is solved at discrete spatial and temporal locations resulting from the discretization of space and time domains as given in Table~\ref{tab:eq dis}. The discrete analytical solution is used as input data for calculating higher order spatial and temporal data using the finite difference approximations listed in Eq.~\ref{goveq4}. Furthermore, the feature library is built using discrete solution $u(t,x)$ and higher order derivative which is discussed in Section~\ref{sec:meth}. As GEP is a natural feature extractor, core feature library $\mathbf{\widetilde \Theta(U)}$ given in  Eq.~\ref{goveq5} is enough to form input data, i.e., GEP terminal set. Table~\ref{tab:gepset} shows the function set and terminal set used for equation identification and Table~\ref{tab:ghyp0} lists the hyper-parameter values for various genetic operators.   However, extended core feature library $\mathbf{\Theta(U)}$ which contains a higher degree interactions of features is used as input for STRidge as the expressive power of STRidge depends on exhaustive combinations of features in the input library. The temporal derivative of $u(t,x)$ is target or response data $\mathbf{V(t)}$ given in Eq.~\ref{goveq5} for both GEP and STRidge.

\subsection{Wave Equation}
\begin{figure}[!htpb]
	\centering
	\includegraphics[width=0.5\textwidth,trim={1.5cm 0 0 0},clip]{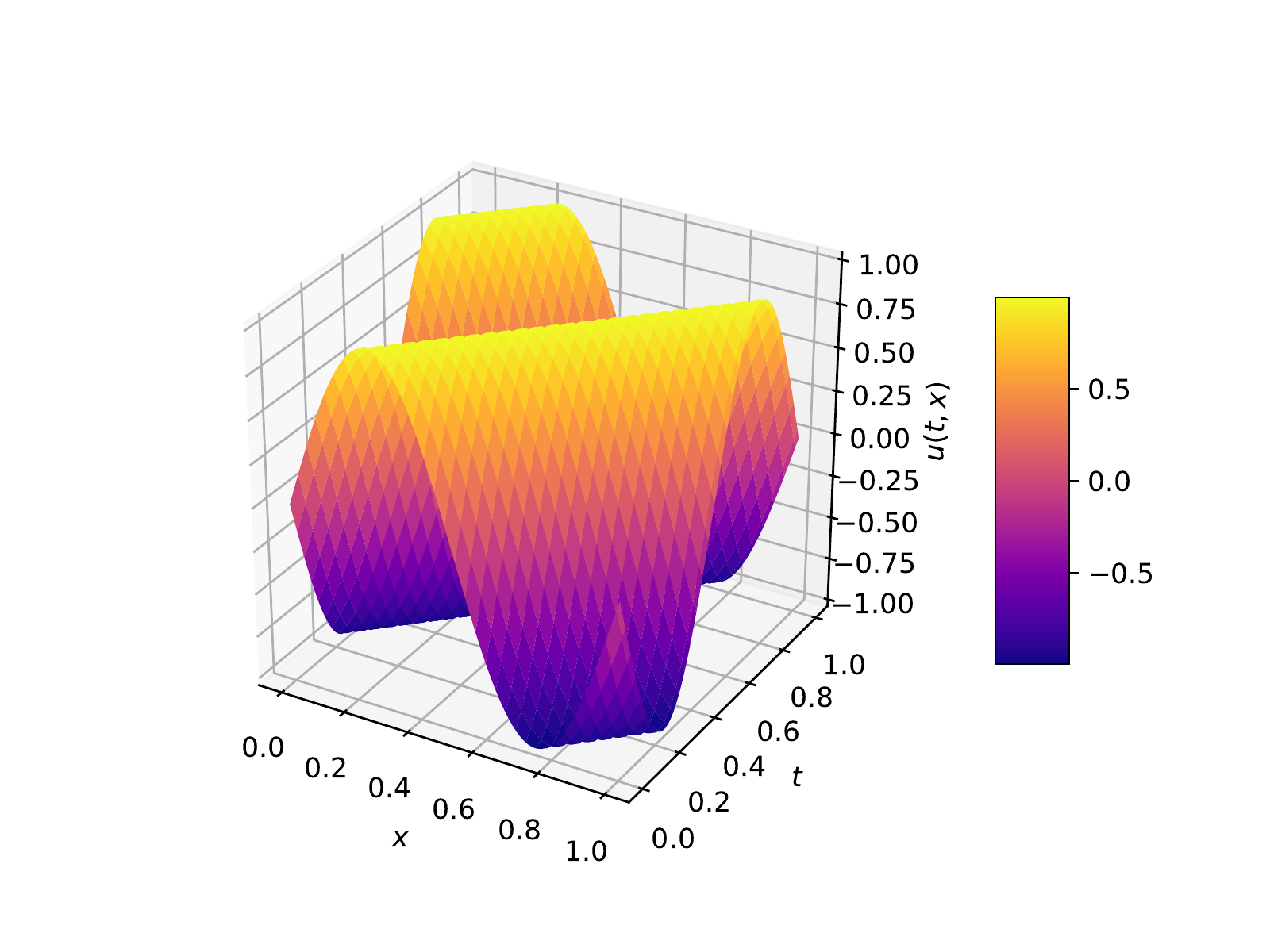}
	\caption{Analytical solution of the wave equation.}
	\label{fig:wave}
\end{figure}

\begin{figure}[!htpb]
	\centering
	\includegraphics[scale=1]{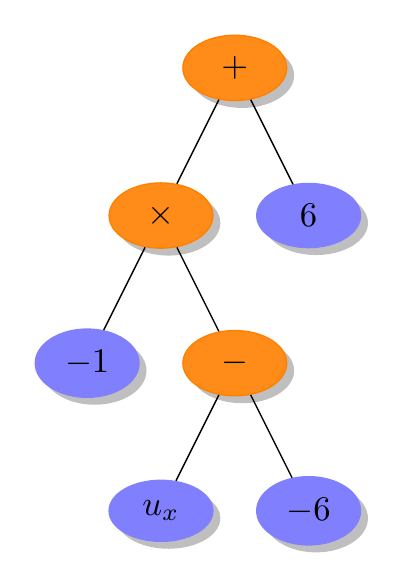}
	\caption{Wave equation in terms of ET identified by GEP.}
	\label{fig:wt}
\end{figure}

Our first test case is the wave equation which is a first order linear PDE. The PDE and its analytical solution are listed in Table~\ref{tab:eq dis}. We choose the constant wave speed $a=1.0$ for propagation of the solution $u(t,x)$.  Fig.~\ref{fig:wave} shows the analytical solution  $u(t,x)$ of the wave equation. The GEP hyper-parameters used for identification of the wave equation are listed in  Table~\ref{tab:hp1}. We use a smaller head length and a single gene for simple cases like a linear wave PDE. We note that any other combinations of hyper-parameters may identify the underlying PDE. Fig.~\ref{fig:wt} illustrates the identified PDE in the ET form. When the ET form is simplified, we can show that the resulting equation is the correct wave PDE, identified with its wave propagation speed parameter $a$. 

\begin{figure}[!htpb]
	\centering
	\includegraphics[width=0.5\textwidth]{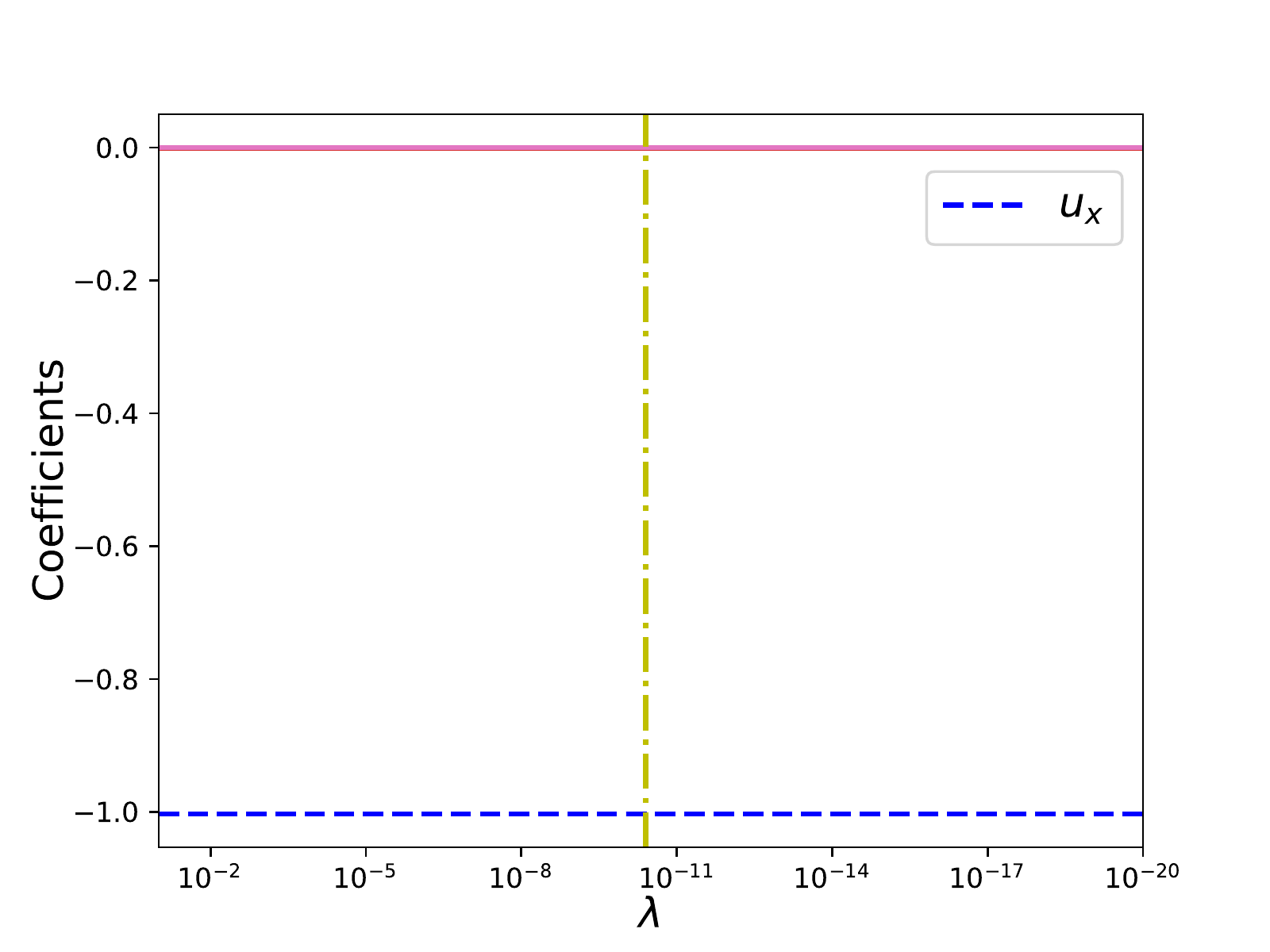}
	\caption{STRidge coefficients as a function of regularization parameter $\lambda$ for the wave equation.}
	\label{fig:wst}
\end{figure}

The regularization weight ($\lambda$) in STRidge is swept across various values as shown in Fig.~\ref{fig:wst}. The yellow line in Fig.~\ref{fig:wst} represents the value of $\lambda$ at which the best identified PDE is selected. Note that in this simple case STRidge was able to find the wave equation for almost all the values of $\lambda$'s that are selected. Table~\ref{tab:wr} shows the wave PDE recovered by both GEP and STRidge.

\begin{table}[!htpb]
\caption{Wave equation identified by GEP and STRidge.}
\label{tab:wr}
\bgroup
\def\arraystretch{1.5}
\setlength{\tabcolsep}{2.1em}
\begin{tabular}{@{}lll}
\hline
& \textbf{Recovered} & \textbf{Test error}   \\ \hline
\noalign{\smallskip}
True      &   $u_t = - 1.00~ u_x$ &   \\
GEP       &   $u_t = - 1.00~ u_x$ &  $1.72\times10^{-28}$ \\
STRidge   &   $u_t = - 1.00~ u_x$ &  $9.01\times10^{-29}$\\
\hline
\end{tabular}
\egroup
\end{table}

\subsection{Heat Equation}
\begin{figure}[!htpb]
	\centering
	\includegraphics[width=0.5\textwidth,trim={1.5cm 0 0 0},clip]{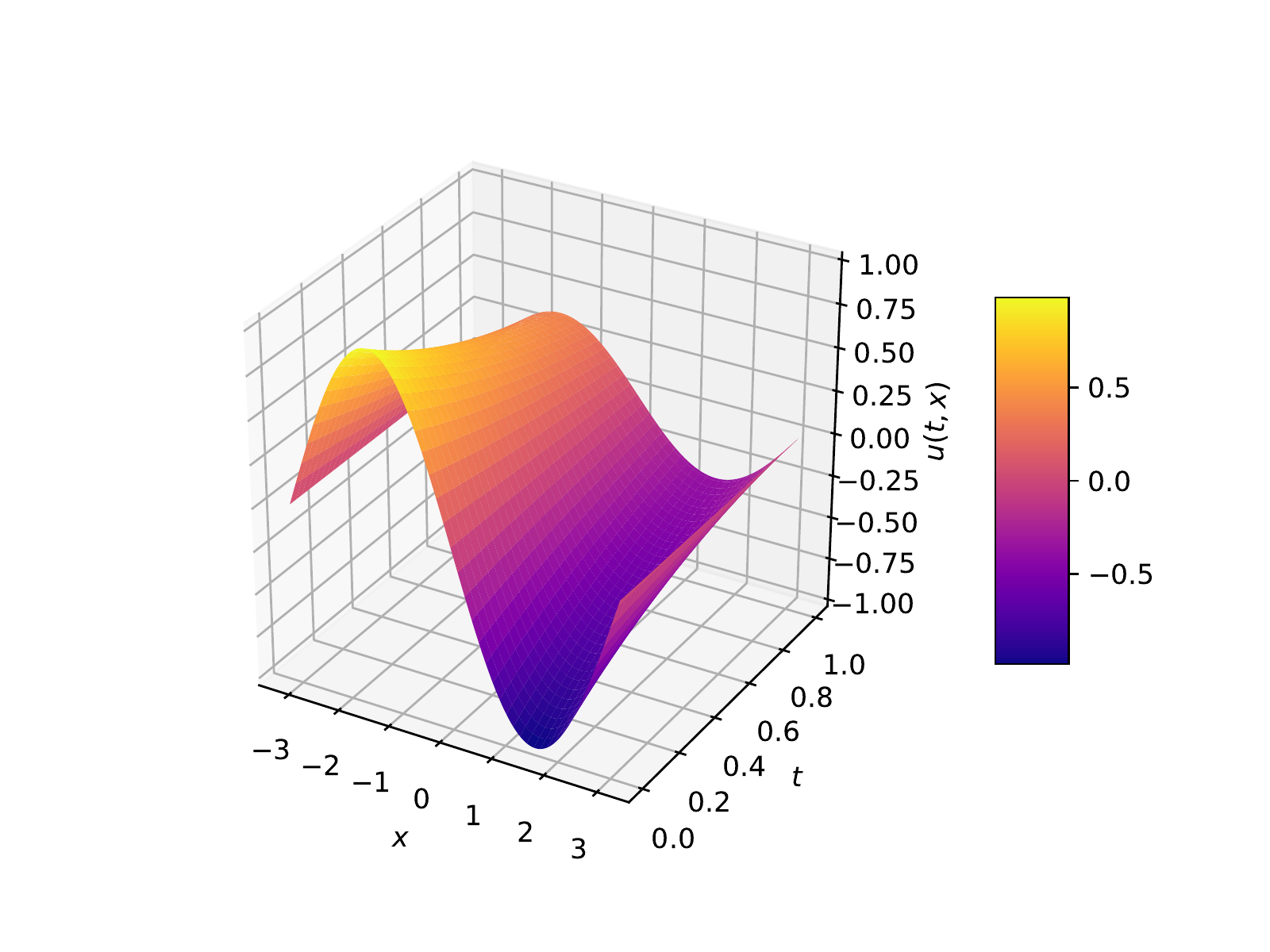}
	\caption{Analytical solution of the heat equation.}
	\label{fig:heat}
\end{figure}

\begin{figure}[!htpb]
	\centering
	\includegraphics[scale=1]{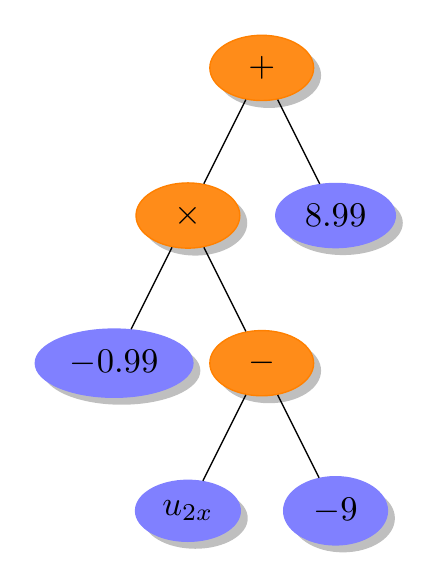}
	\caption{Heat equation in terms of ET identified by GEP.}
	\label{fig:ht}
\end{figure}

We use the heat equation which is a second order linear PDE to test both SR approaches. The PDE and its analytical solution is listed in Table~\ref{tab:eq dis}. The physical parameter $\alpha=1.0$ may represent thermal conductivity.  Fig.~\ref{fig:heat} displays the analytical solution  $u(t,x)$ of the heat equation. Table~\ref{tab:hp1} lists the GEP hyper-parameters used for identification of the heat equation. Fig.~\ref{fig:ht} shows the identified PDE in the form of an ET. When the ET form is simplified, we can show that the resulting model is the heat equation identified with its coefficient $\alpha$.  

\begin{figure}[!htpb]
	\centering
	\includegraphics[width=0.5\textwidth]{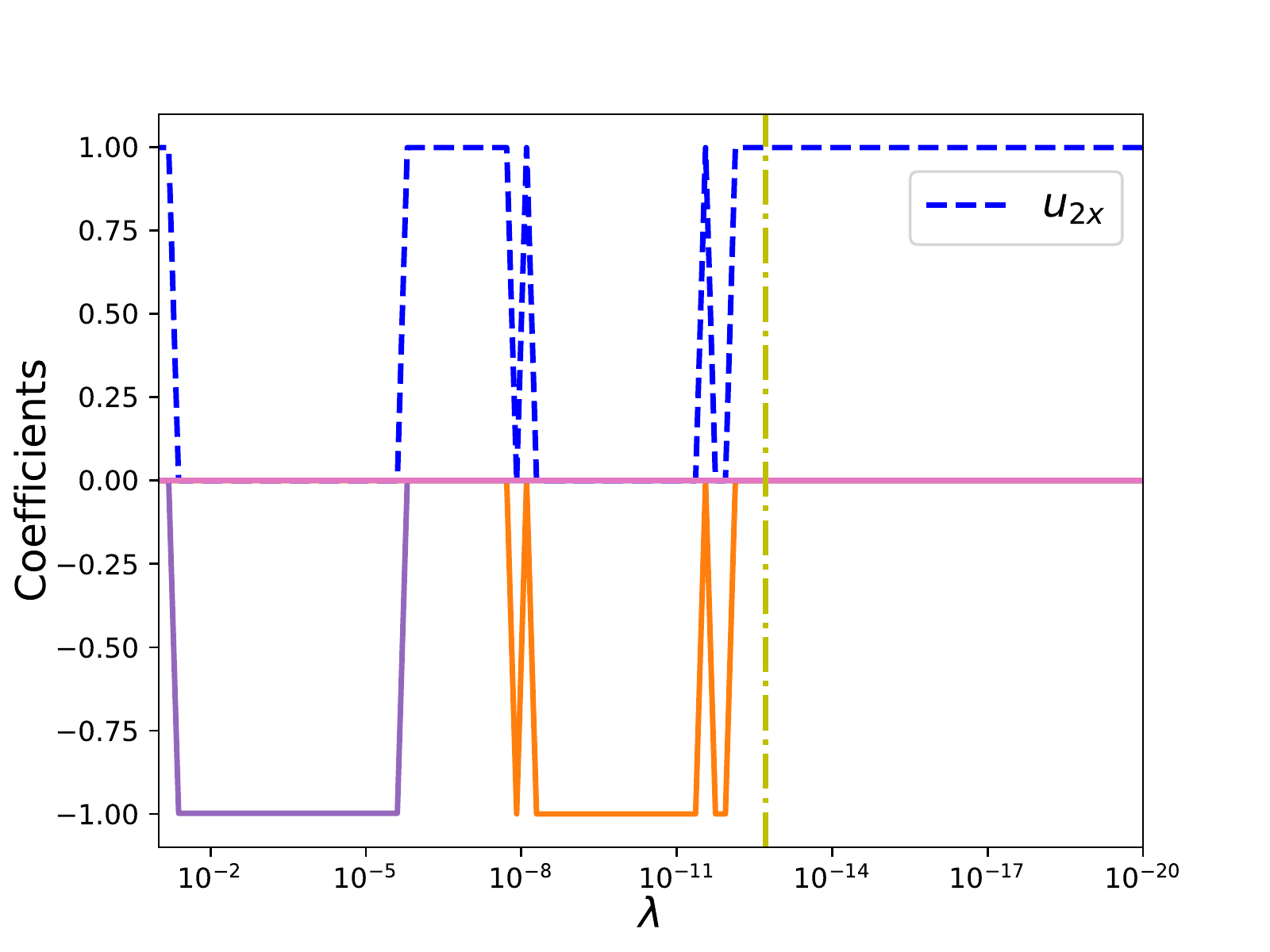}
	\caption{STRidge coefficients as a function of regularization parameter $\lambda$ for the heat equation.}
	\label{fig:hst}
\end{figure}

The regularization weight ($\lambda$) in STRidge is swept across various values as shown Fig.~\ref{fig:hst}. The yellow line in Fig.~\ref{fig:hst} represents the value of $\lambda$ selected at which STRidge finds the heat equation accurately.  Note that STRidge was able to find the heat equation for low values of the regularization weight $\lambda$ as shown in Fig.~\ref{fig:hst}. Table~\ref{tab:hr} shows the heat equation recovered by both GEP and STRidge.  STRidge was able to find a more accurate coefficient ($\alpha$) value than GEP. Furthermore, a small constant value is also identified along with the heat equation by GEP.

\begin{table}[!htpb]
\caption{Heat equation identified by GEP and STRidge.}
\label{tab:hr}
\bgroup
\def\arraystretch{1.5}
\setlength{\tabcolsep}{0.6em}
\begin{tabular}{@{}lll}
\hline
& \textbf{Recovered} & \textbf{Test error}   \\ \hline
\noalign{\smallskip}
True      &   $u_t = - 1.00~u_{2x}$ &   \\
GEP       &   $u_t = - 0.99~u_{2x} - 5.33\times10^{-15}$ &  $5.55\times10^{-24}$ \\
STRidge   &   $u_t = - 1.00~u_{2x}$ &  $4.09\times10^{-30}$\\
\hline
\end{tabular}
\egroup
\end{table}

\subsection{Burgers Equation (i)}
\begin{figure}[!htpb]
	\centering
	\includegraphics[width=0.5\textwidth,trim={1.5cm 0 0 0},clip]{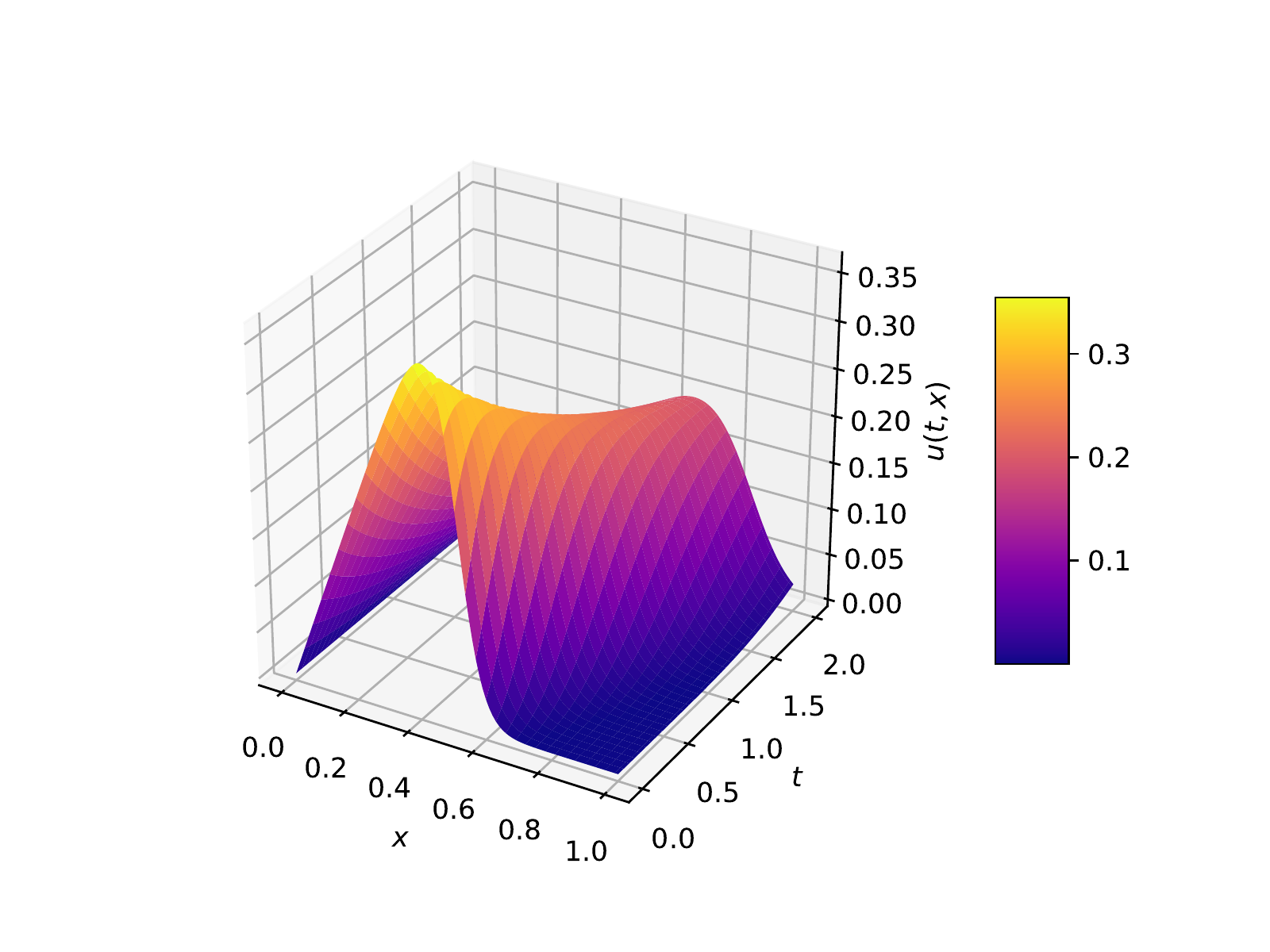}
	\caption{Analytical solution of the Burgers equation (i).}
	\label{fig:bs}
\end{figure}

\begin{figure}[!htpb]
	\centering
	\includegraphics[scale=1]{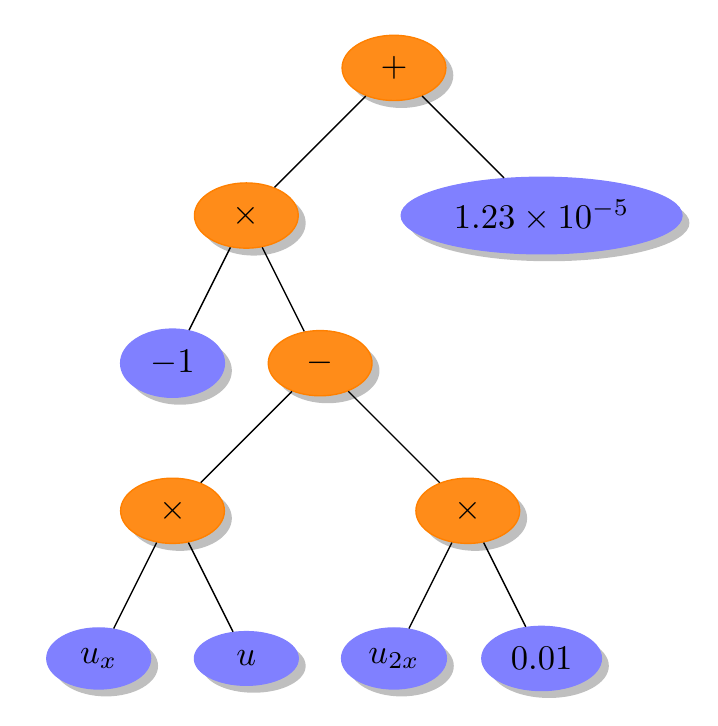}
	\caption{Burgers equation (i) in terms of ET identified by GEP.}
	\label{fig:bst}
\end{figure}

Burgers equation is a fundamental nonlinear PDE occurring in various areas such as fluid mechanics, nonlinear acoustics, gas dynamics and traffic flow  \cite{bateman1915some, whitham2011linear}. The interest in the Burgers equation arises due to the non linear term $uu_x$ and presents a challenge to both GEP and STRidge in the identification of its PDE using data. The form of the Burgers PDE and its analytical solution\cite{maleewong2011line} is listed in Table~\ref{tab:eq dis}. The physical parameter $\nu=0.01$ can be considered as the kinematic viscosity in fluid flows.  Fig.~\ref{fig:bs} shows the analytical solution  $u(t,x)$ of the Burgers equation. Table~\ref{tab:hp1} shows the GEP hyper-parameters used for identification of the Burgers equation. Fig.~\ref{fig:bst} shows the identified PDE in the form of the ET. When ET form is simplified, we can show that the resulting model is the Burgers equation identified along with the coefficient of the nonlinear term and  the kinematic viscosity.  GEP uses more generations for identifying the Burgers PDE due to its nonlinear behavior along with the identification of feature interaction term  $uu_x$. 

\begin{figure}[!htpb]
	\centering
	\includegraphics[width=0.5\textwidth]{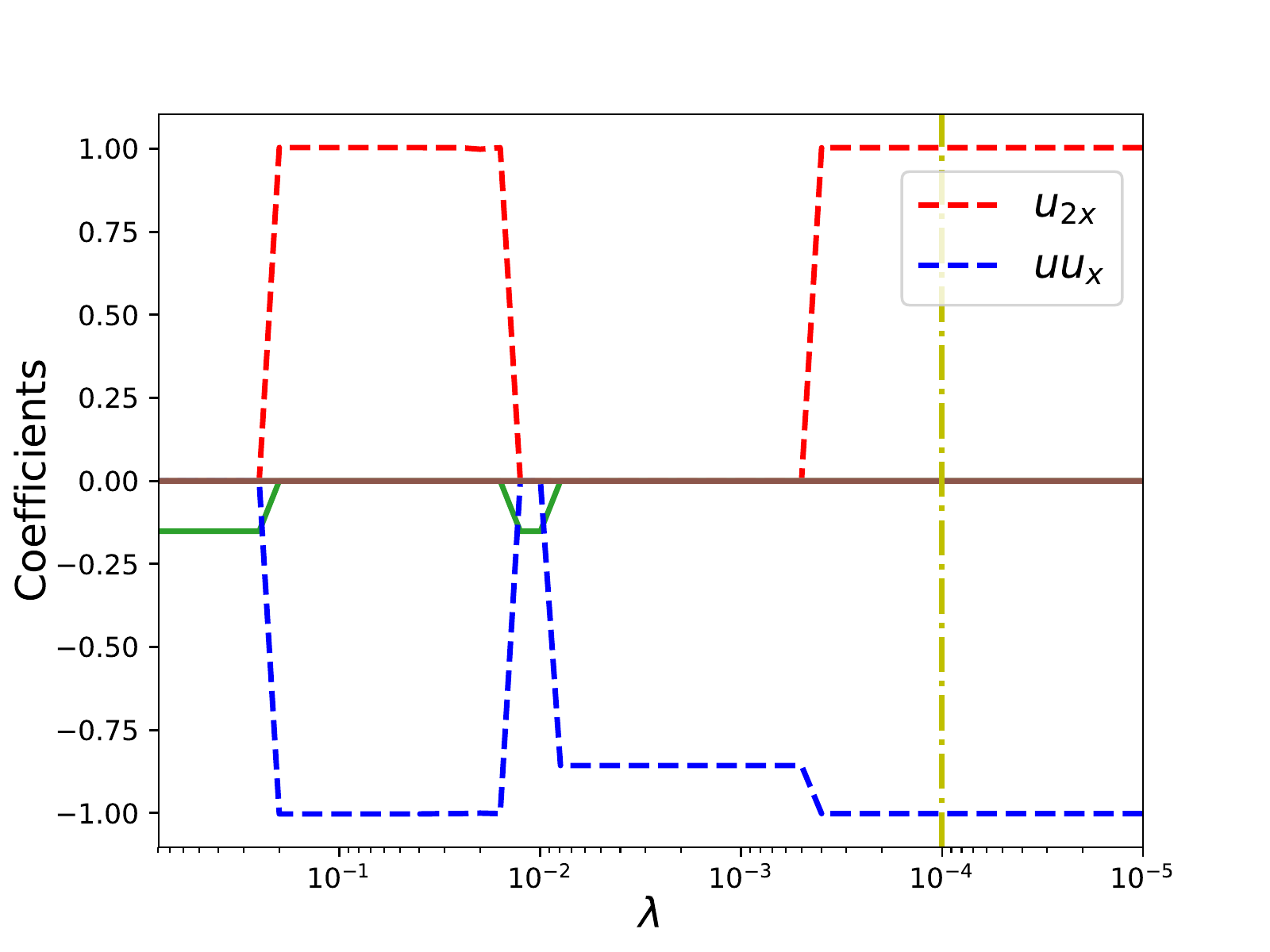}
	\caption{STRidge coefficients as a function of regularization parameter $\lambda$ for the Burgers equation (i).}
	\label{fig:bsst}
\end{figure}

The regularization weight ($\lambda$) in STRidge is swept across various values as shown in Fig.~\ref{fig:bsst}. The yellow line in Fig.~\ref{fig:bsst} represents the value of $\lambda$ at which the best identified PDE is selected. Note that the STRidge algorithm was able to find the Burgers equation at multiple values of regularization weights $\lambda$. Table~\ref{tab:br1} shows the Burgers PDE recovered by both GEP and STRidge. There is an additional constant coefficient term recovered by GEP. Furthermore, the recovery of the nonlinear term using a limited set of input features shows the usefulness of GEP. 

\begin{table}[!htpb]
\caption{Burgers equation (i) identified by GEP and STRidge.}
\label{tab:br1}
\bgroup
\def\arraystretch{1.5}
\setlength{\tabcolsep}{0.3em}
\begin{tabular}{@{}lll}
\hline
& \textbf{Recovered} & \textbf{Test error}   \\ \hline
\noalign{\smallskip}
True      &   $u_t = - uu_{x} + 0.01~ u_{2x}$ &   \\
GEP       &    $u_t = - uu_{x} + 0.01~ u_{2x} - 1.23\times10^{-5}$ &  $6.10\times10^{-08}$ \\
STRidge   &   $u_t = - uu_{x} + 0.01~ u_{2x}$ &  $5.19\times10^{-08}$\\
\hline
\end{tabular}
\egroup
\end{table}

\subsection{Burgers Equation (ii)}
\begin{figure}[!htpb]
	\centering
	\includegraphics[width=0.5\textwidth,trim={1.5cm 0 0 0},clip]{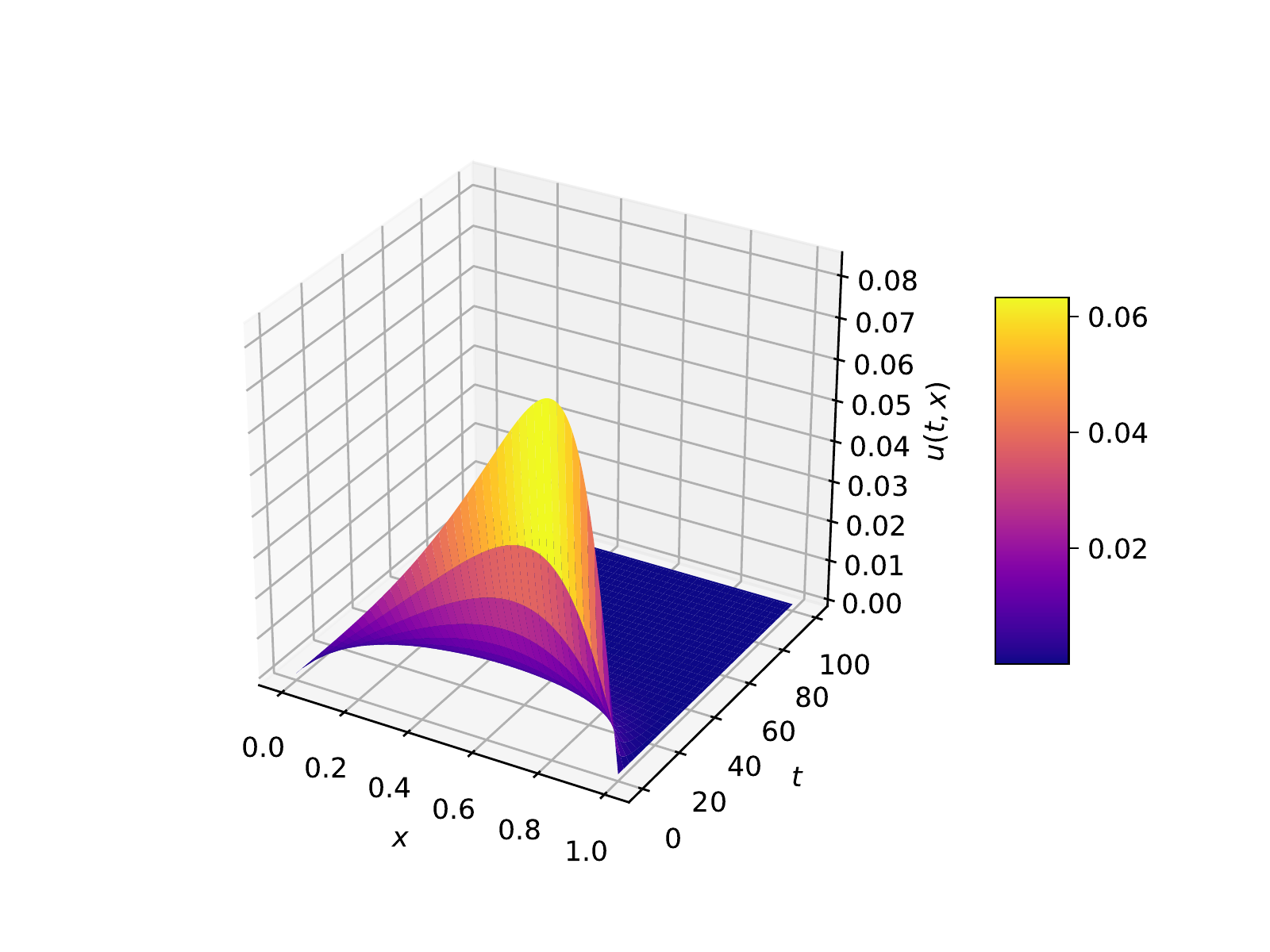}
	\caption{Analytical solution of the Burgers equation (ii).}
	\label{fig:bw}
\end{figure}

\begin{figure}[!htpb]
	\centering
	\includegraphics[width=0.4\textwidth]{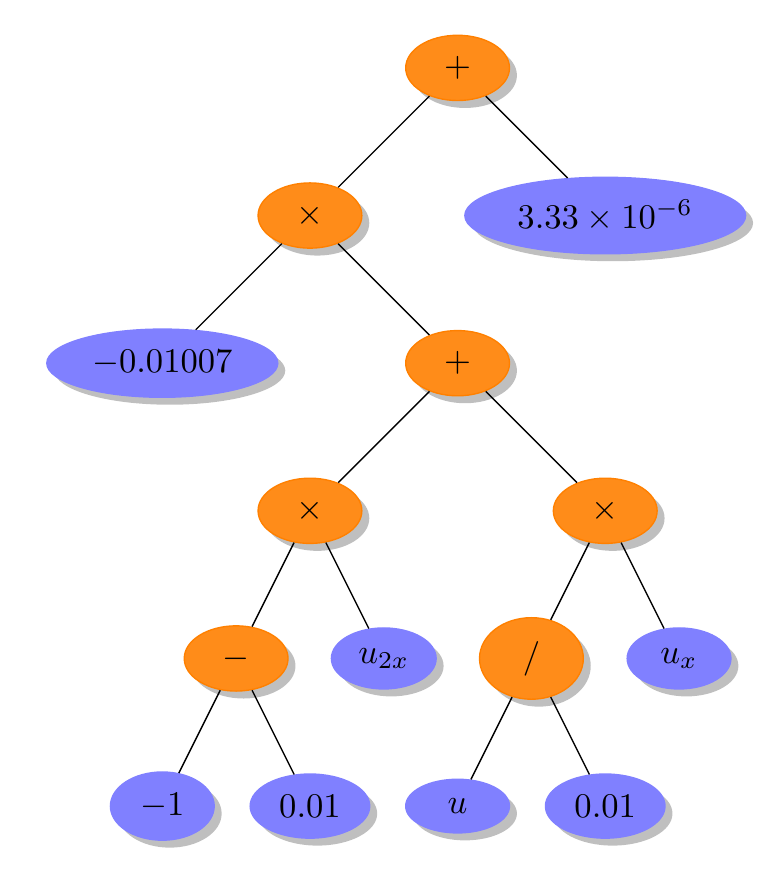}
	\caption{Burgers equation (ii) in terms of ET identified by GEP.}
	\label{fig:bwt}
\end{figure}

Burgers PDE with a different analytical solution is used to test the effectiveness of GEP and STRidge as the input data is changed but represented by the same physical law. The analytical solution of the Burgers equation (ii) is listed in Table~\ref{tab:eq dis}. The physical parameter $\nu=0.01$ is used to generate the data.  Fig.~\ref{fig:bw} shows the alternate analytical solution  $u(t,x)$ of the Burgers equation. Table~\ref{tab:hp1} shows the GEP hyper-parameters used for identification of the Burgers equation (ii). Fig.~\ref{fig:bwt} shows the identified PDE in the form of ET. When ET form is simplified, we can show that the resulting model is the Burgers equation identified along with the coefficient of nonlinear term and kinematic viscosity. With an alternate solution, GEP uses a larger head length,  more genes, and a larger population for identifying the same Burgers PDE.

\begin{figure}[!htpb]
	\centering
	\includegraphics[width=0.5\textwidth]{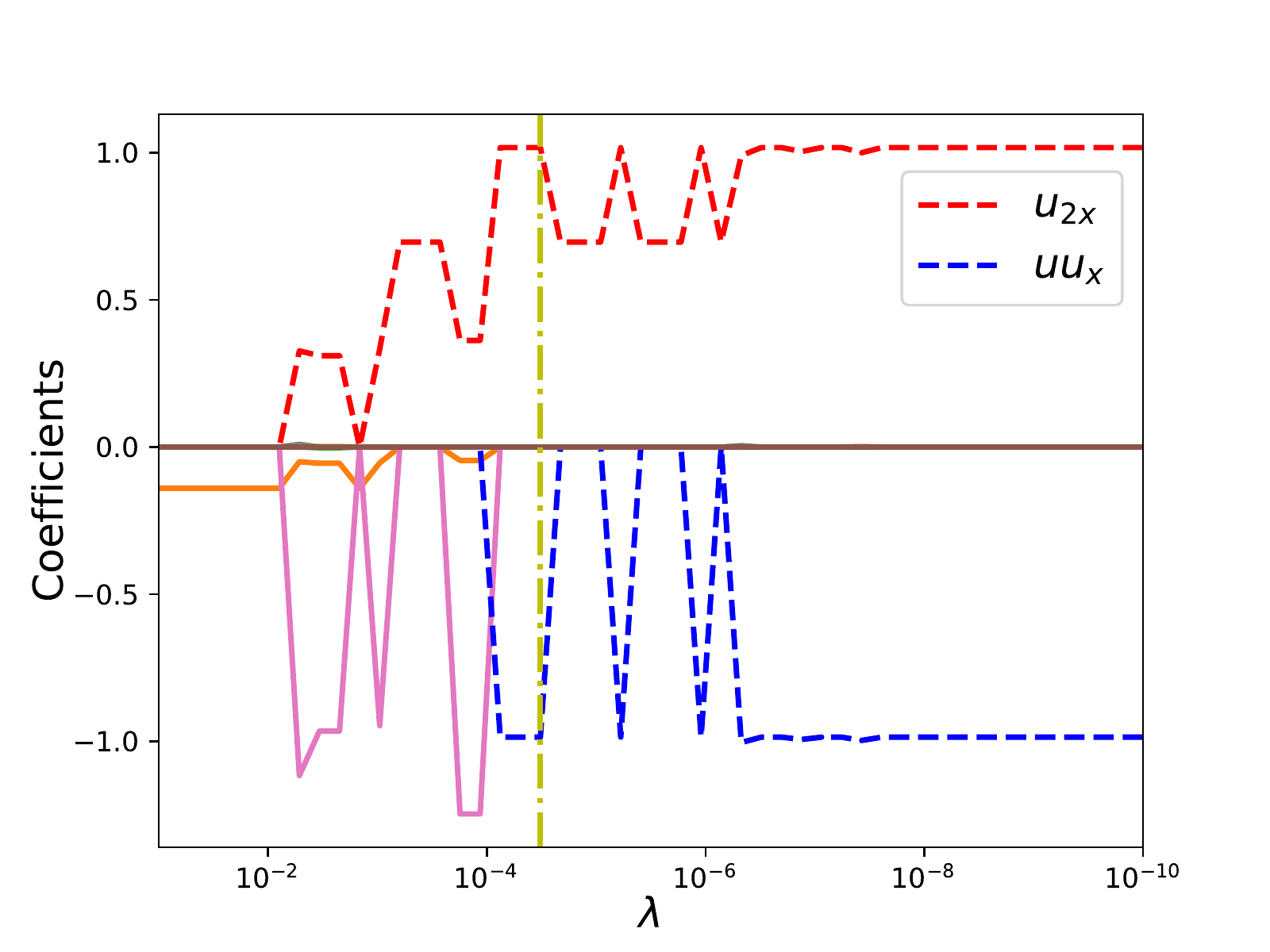}
	\caption{STRidge coefficients as a function of regularization parameter $\lambda$ for the Burgers equation (ii).}
	\label{fig:bwst}
\end{figure}

The regularization weight ($\lambda$) in STRidge is swept across various values as shown Fig.~\ref{fig:bwst}. The yellow line in Fig.~\ref{fig:bwst} represents the value of $\lambda$ at which the best identified PDE is selected. Note that STRidge was able to find the Burgers equation at various values of regularization weight $\lambda$. Table~\ref{tab:br2} shows the Burgers PDE recovered by both GEP and STRidge.

\begin{table}[!htpb]
\caption{Burgers equation (ii) identified by GEP and STRidge.}
\label{tab:br2}
\bgroup
\def\arraystretch{1.5}
\setlength{\tabcolsep}{0.2em}
\begin{tabular}{@{}lll}
\hline
& \textbf{Recovered} & \textbf{Test error}   \\ \hline
\noalign{\smallskip}
True      &   $u_t  = -1.00~ uu_{x} + 0.01~ u_{2x}$   &   \\
GEP       &   $u_t  = -1.01~ uu_{x} + 0.01~ u_{2x} - 3.33\times10^{-6}$  &  $1.94\times10^{-09}$ \\
STRidge   &   $u_t  = -0.99~ uu_{x} + 0.01~ u_{2x}$  &  $1.85\times10^{-08}$\\
\hline
\end{tabular}
\egroup
\end{table}

\subsection{Korteweg-de Vries (KdV) Equation}
\begin{figure}[!htpb]
	\centering
	\includegraphics[width=0.5\textwidth,trim={1.5cm 0 0 0},clip]{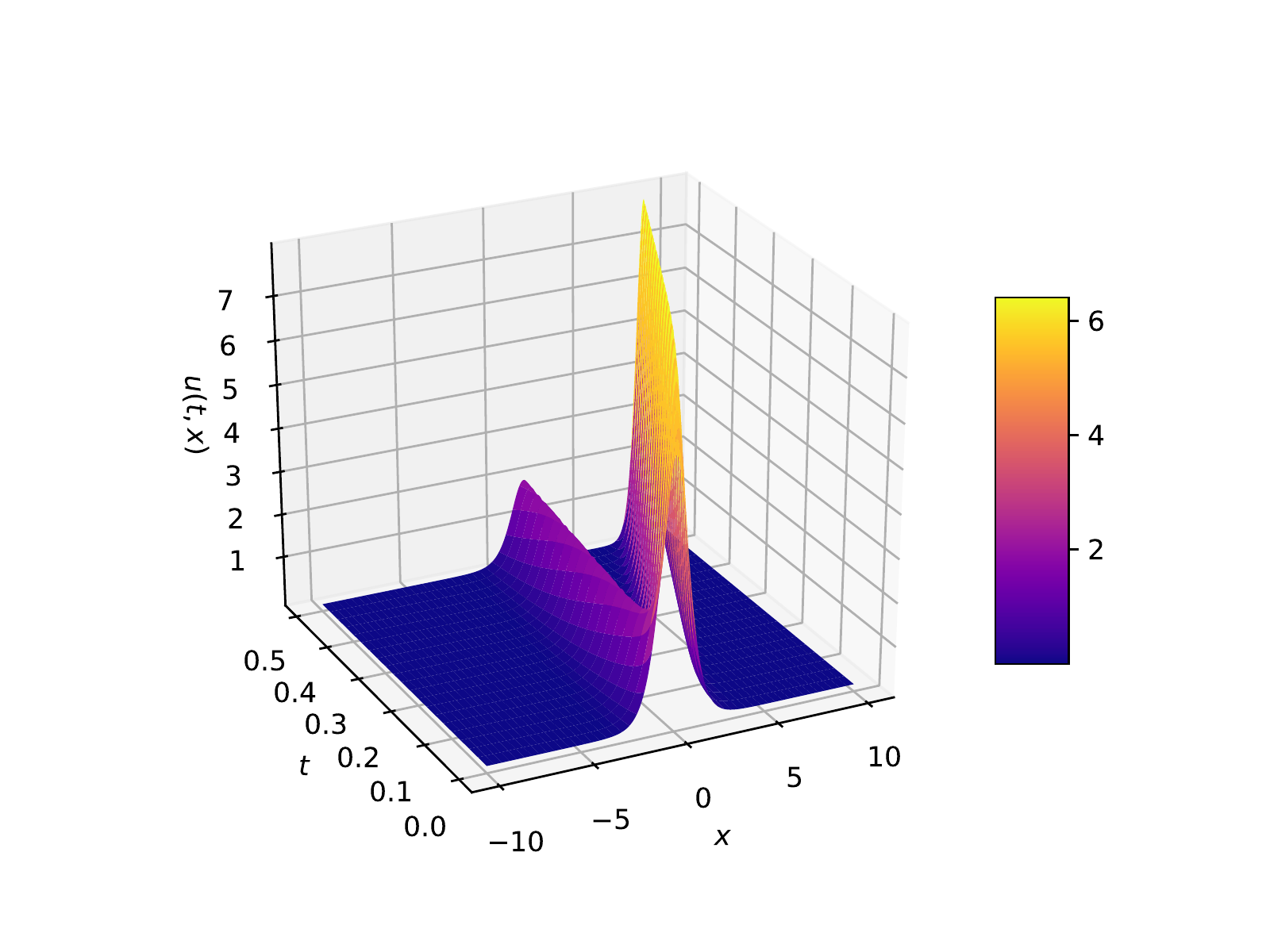}
	\caption{Analytical solution of the KdV equation.}
	\label{fig:kdv}
\end{figure}

\begin{figure}[!htpb]
	\centering
	\includegraphics[width=0.5\textwidth]{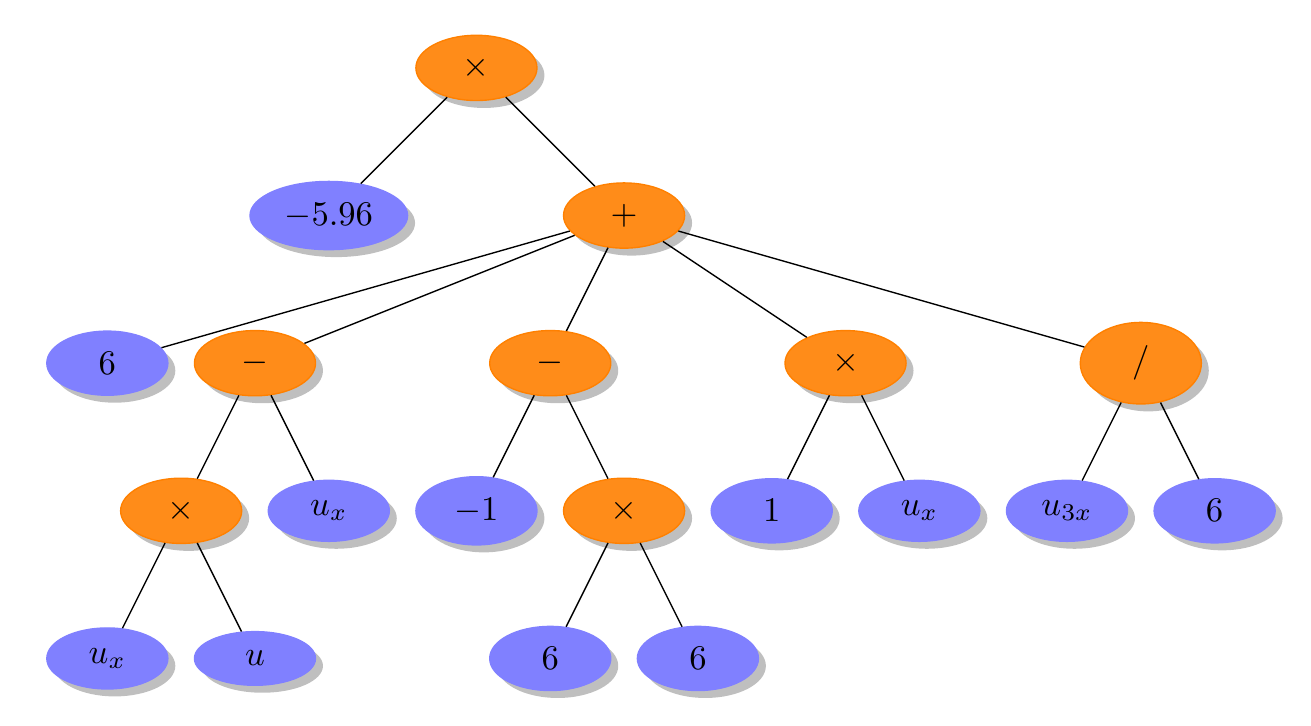}
	\caption{KdV equation in terms of ET identified by  GEP.}
	\label{fig:kdvt}
\end{figure}

Korteweg and de Vries derived the KdV equation to model Russell’s phenomenon of solitons \cite{kdvphi18, wazzan2009modified}. The KdV equation also appears when modelling the behavior of magneto-hydrodynamic waves in warm plasma's, acoustic waves in an inharmonic crystal and ion-acoustic waves \cite{ozics2006simple}. Many different forms of the KdV equation available in the literature but we use the form given in Table~\ref{tab:eq dis}. Fig.~\ref{fig:kdv} shows the  analytical solution  $u(t,x)$ of the KdV equation\cite{lamb1980elements}. It can be seen that this analytical solution refers to two solutions colliding together which forms good test case for SR techniques like GEP and STRidge. Table~\ref{tab:hp2} shows the GEP hyper-parameters used for identification of the KdV equation. Due to the higher nonlinear dynamics represented by higher order PDE, GEP requires large head length and genes compared to other test cases in equation discovery.  Fig.~\ref{fig:kdvt} shows the identified PDE in the form of the ET. When ET form is simplified, we can observe that the resulting model is the KdV equation identified along with its coefficients. 

\begin{figure}[!htpb]
	\centering
	\includegraphics[width=0.5\textwidth]{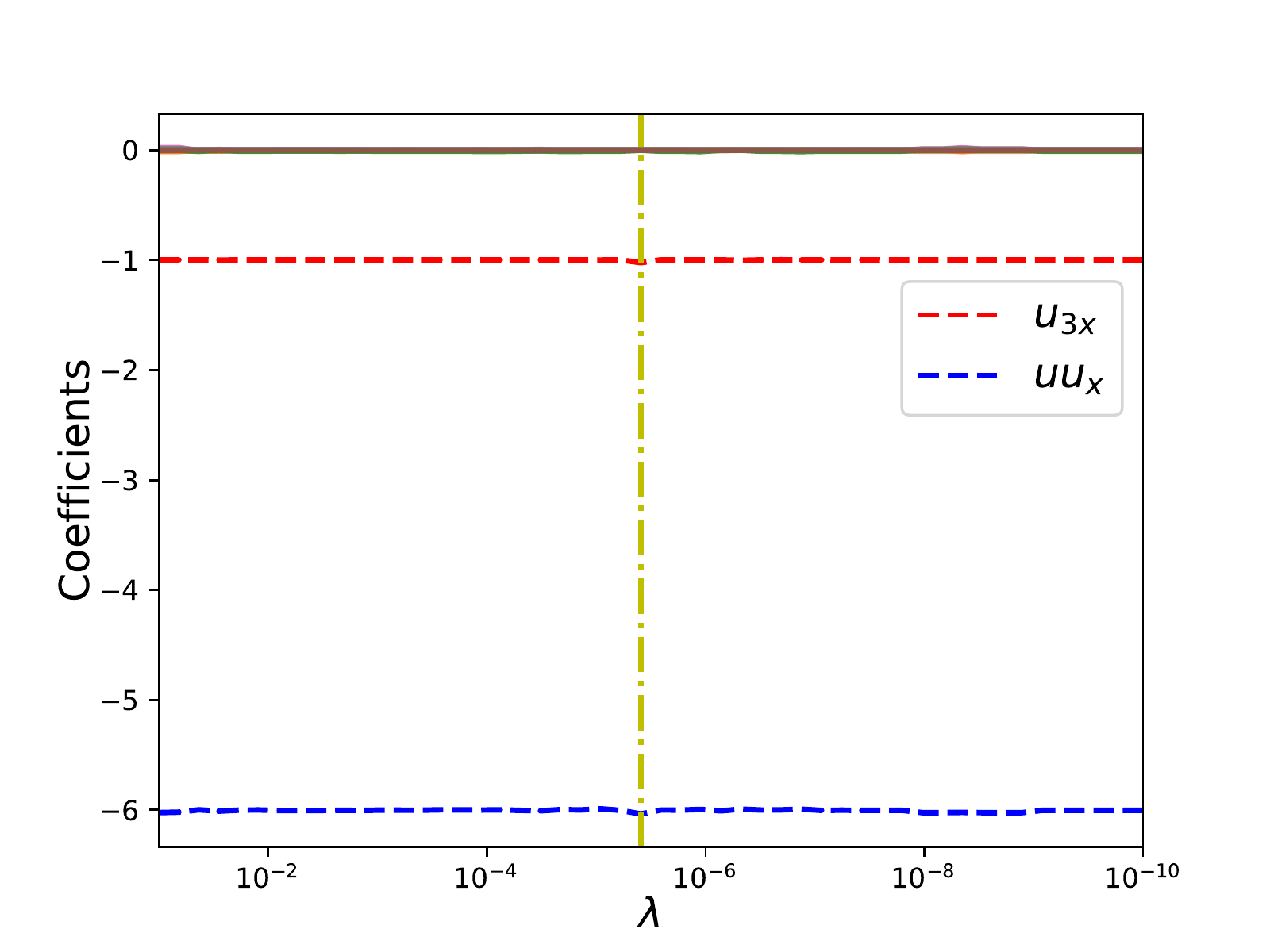}
	\caption{STRidge coefficients as a function of regularization parameter $\lambda$ for the KdV equation.}
	\label{fig:kdvst}
\end{figure}

The regularization weight ($\lambda$) in STRidge is swept across various values as shown Fig.~\ref{fig:kdvst}. The yellow line in Fig.~\ref{fig:kdvst} represents the value of $\lambda$ at which the best identified PDE is selected.  Note that STRidge was able to find the KdV equation at various values of the regularization weights ($\lambda$). Table~\ref{tab:kdr} shows the KdV equation recovered by both GEP and STRidge. The physical model identified by STRidge is more accurate to the true PDE than the model identified by GEP.

\begin{table}[!htpb]
\caption{KdV equation identified by GEP and STRidge.}
\label{tab:kdr}
\bgroup
\def\arraystretch{1.5}
\setlength{\tabcolsep}{0.13em}
\begin{tabular}{@{}llc}
\hline
& \textbf{Recovered} & \textbf{Test error}   \\ \hline
\noalign{\smallskip}
True      &   $u_t  = -6.00~ uu_{x} + 1.00~ u_{3x}$   &   \\
GEP       &   $u_t  = -5.96~ uu_{x} + 0.99~ u_{3x} - 5.84\times10^{-4}$  &  $0.29$ \\
STRidge   &   $u_t  = -6.04~ uu_{x} + 1.02~ u_{3x}$  &  $0.02$\\
\hline
\end{tabular}
\egroup
\end{table}

\subsection{Kawahara Equation}
\begin{figure}[!htpb]
	\centering
	\includegraphics[width=0.5\textwidth,trim={1.5cm 0 0 0},clip]{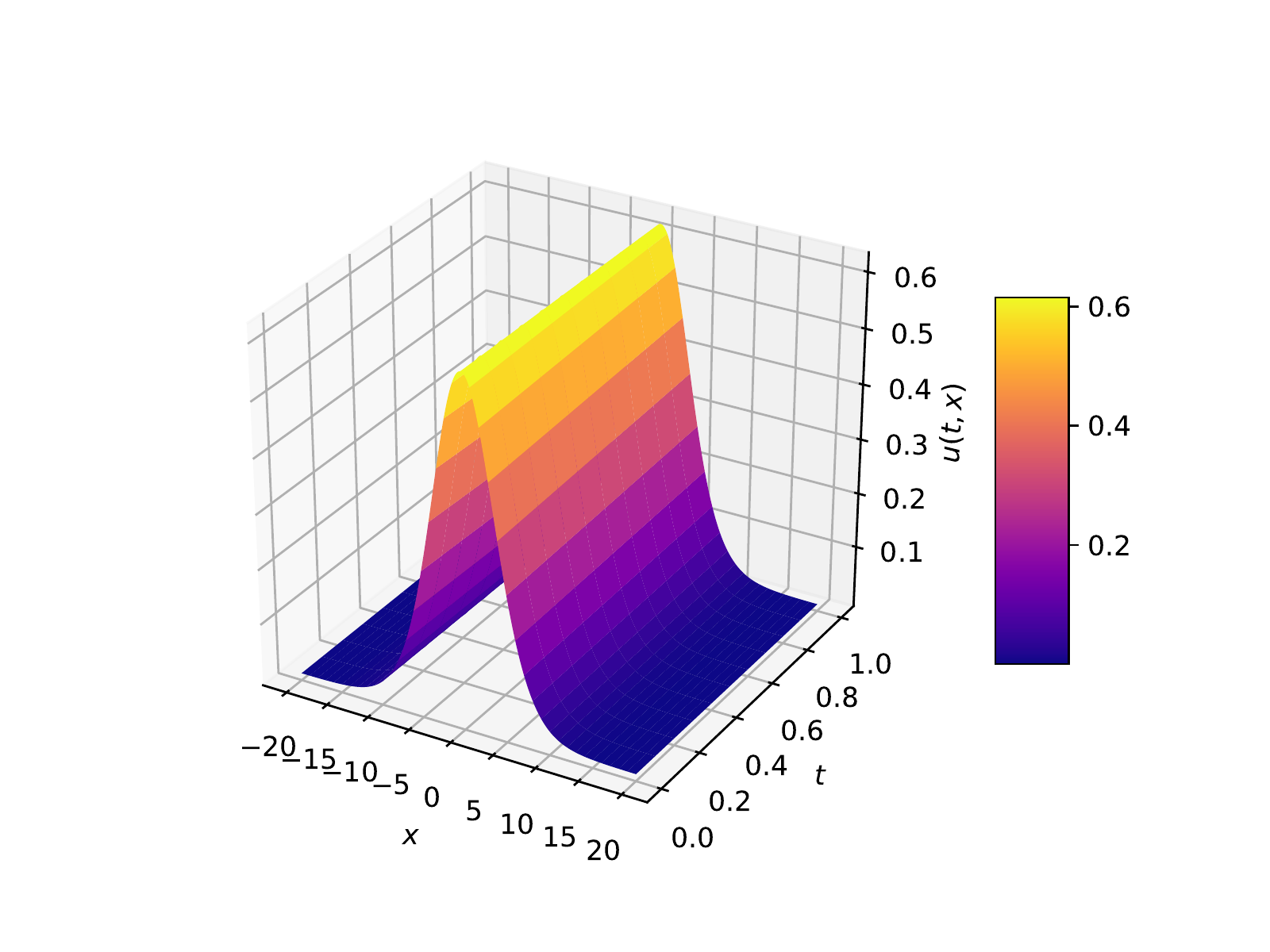}
	\caption{Analytical solution of the Kawahara equation.}
	\label{fig:kaw}
\end{figure}

We consider the Kawahara equation, which is a fifth-order nonlinear  PDE \cite{kawahara1972oscillatory} shown in Table~\ref{tab:eq dis}. This equation is sometimes also referred to as a fifth-order KdV equation or singularly perturbed KdV equation. The fifth-order KdV equation is one of the most well known nonlinear evolution equation which is used in the theory of magneto-acoustic waves in a plasma \cite{kawahara1972oscillatory}, capillary-gravity waves \cite{kawahara1975nonlinear} and  the theory of shallow water waves \cite{hunter1988existence}. This test case is intended to test GEP and STRidge for identifying higher order derivatives from observing data. We use  an analytical solution\cite{sirendaoreji2004new}  which is a traveling wave solution given in Table~\ref{tab:eq dis}. This analytical solution also satisfies the linear wave equation and hence both GEP and STRidge may recover a wave PDE (not shown here) as this is the sparsest model represented by observed data (Fig.~\ref{fig:kaw}). For simplifying the analysis, we remove the potential basis $u_x$ from the feature library \cite{schaeffer2017learning} ($\mathbf{\Theta(U)}$) for STRidge and additionally include $uu_x$ basis in core feature library ($\mathbf{\widetilde \Theta(U)}$) for GEP. 

\begin{figure}[!htpb]
	\centering
	\includegraphics[width=0.35\textwidth]{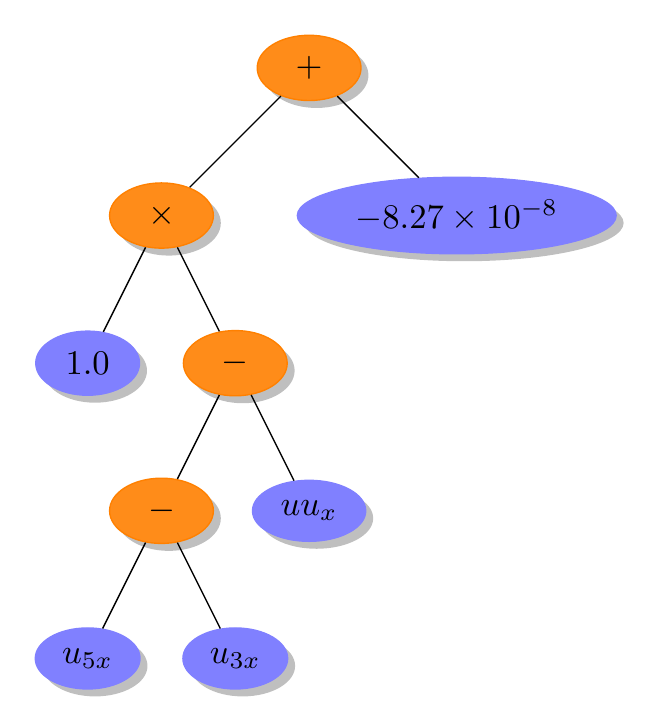}
	\caption{Kawahara equation in terms of ET identified by GEP.}
	\label{fig:kawt}
\end{figure}

\begin{figure}[!htpb]
	\centering
	\includegraphics[width=0.5\textwidth]{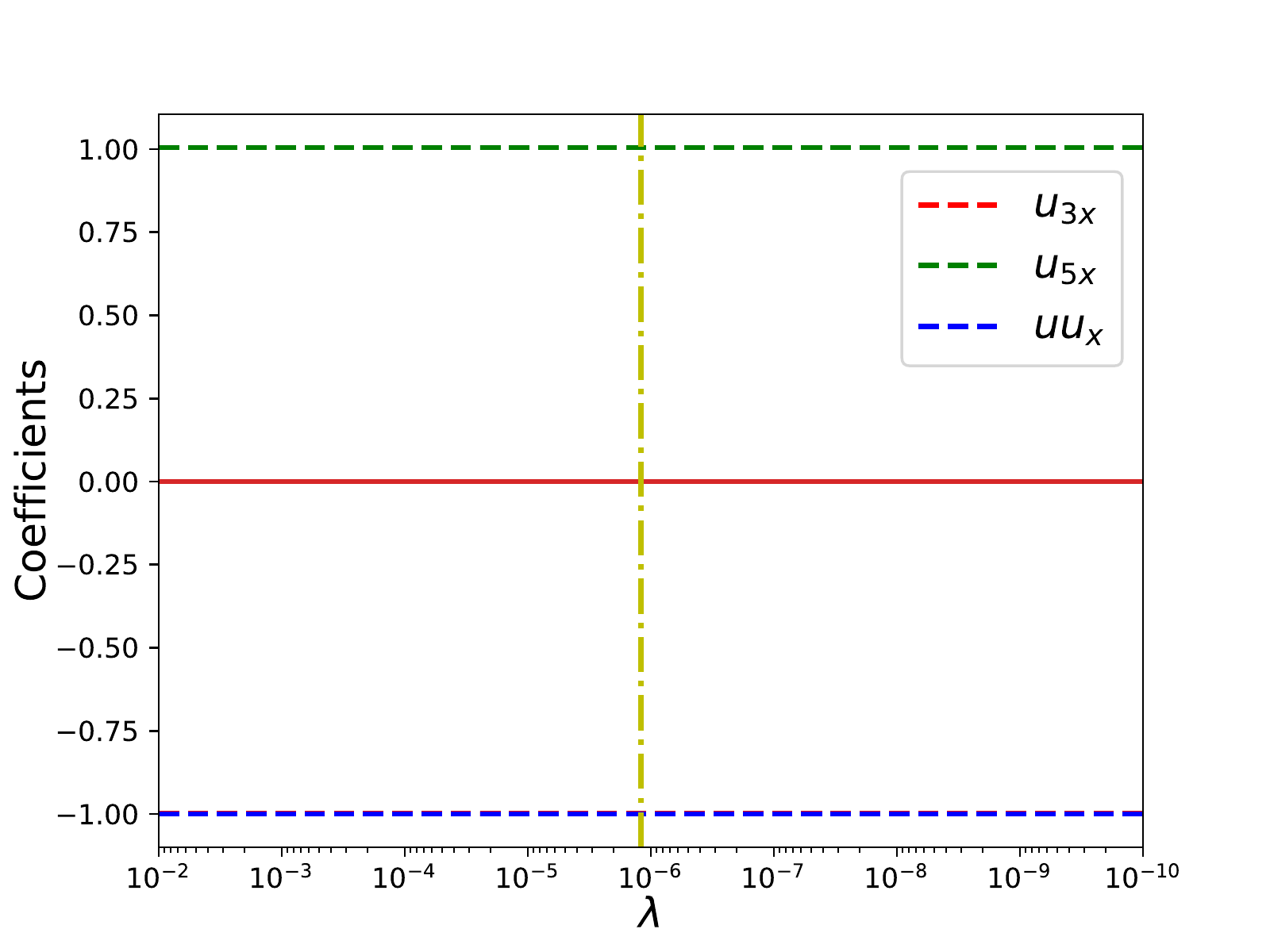}
	\caption{STRidge coefficients as a function of regularization parameter $\lambda$ for the Kawahara equation.}
	\label{fig:kawst}
\end{figure}

Table~\ref{tab:hp2} shows the GEP hyper-parameters used for the identification of the Kawahara equation. Due to simplifying the feature library,  GEP requires smaller head length and single gene. Fig.~\ref{fig:kawt} shows the identified PDE in the form of ET. When ET form is simplified, we can show that the resulting model is the Kawahara equation identified correctly along with its coefficients. For STRidge, the regularization weight ($\lambda$)  is swept across various values as shown in Fig.~\ref{fig:kawst}. The yellow line in Fig.~\ref{fig:kawst} represents the value of $\lambda$ at which the best identified PDE is selected.  Note that STRidge was able to find the Kawahara equation at various values of regularization weights ($\lambda$). Table~\ref{tab:kar} shows the Kawahara equation identified by both GEP and STRidge.

\begin{table*}[!htpb]
\caption{Kawahara equation identified by GEP and STRidge.}
\label{tab:kar}
\bgroup
\def\arraystretch{1.5}
\setlength{\tabcolsep}{1.5em}
\begin{tabular}{@{}llc}
\hline
& \textbf{Recovered} & \textbf{Test error}   \\ \hline
\noalign{\smallskip}
True      &   $u_t  = -1.0~uu_{x} - 1.00~ u_{3x} - 1.0~ u_{5x}$   &   \\
GEP       &   $u_t  = -1.0~uu_{x} - 1.00~ u_{3x} - 1.0~ u_{5x} - 8.27\times10^{-8}$  &  $5.29\times10^{-11}$ \\
STRidge   &   $u_t  = -1.0~ uu_{x} - 0.99~ u_{3x} - 1.0~ u_{5x} $  &  $1.35\times10^{-12}$\\
\hline
\end{tabular}
\egroup
\end{table*}

\subsection{Newell-Whitehead-Segel Equation}
\begin{figure}[!htpb]
	\centering
	\includegraphics[width=0.5\textwidth,trim={1.5cm 0 0 0},clip]{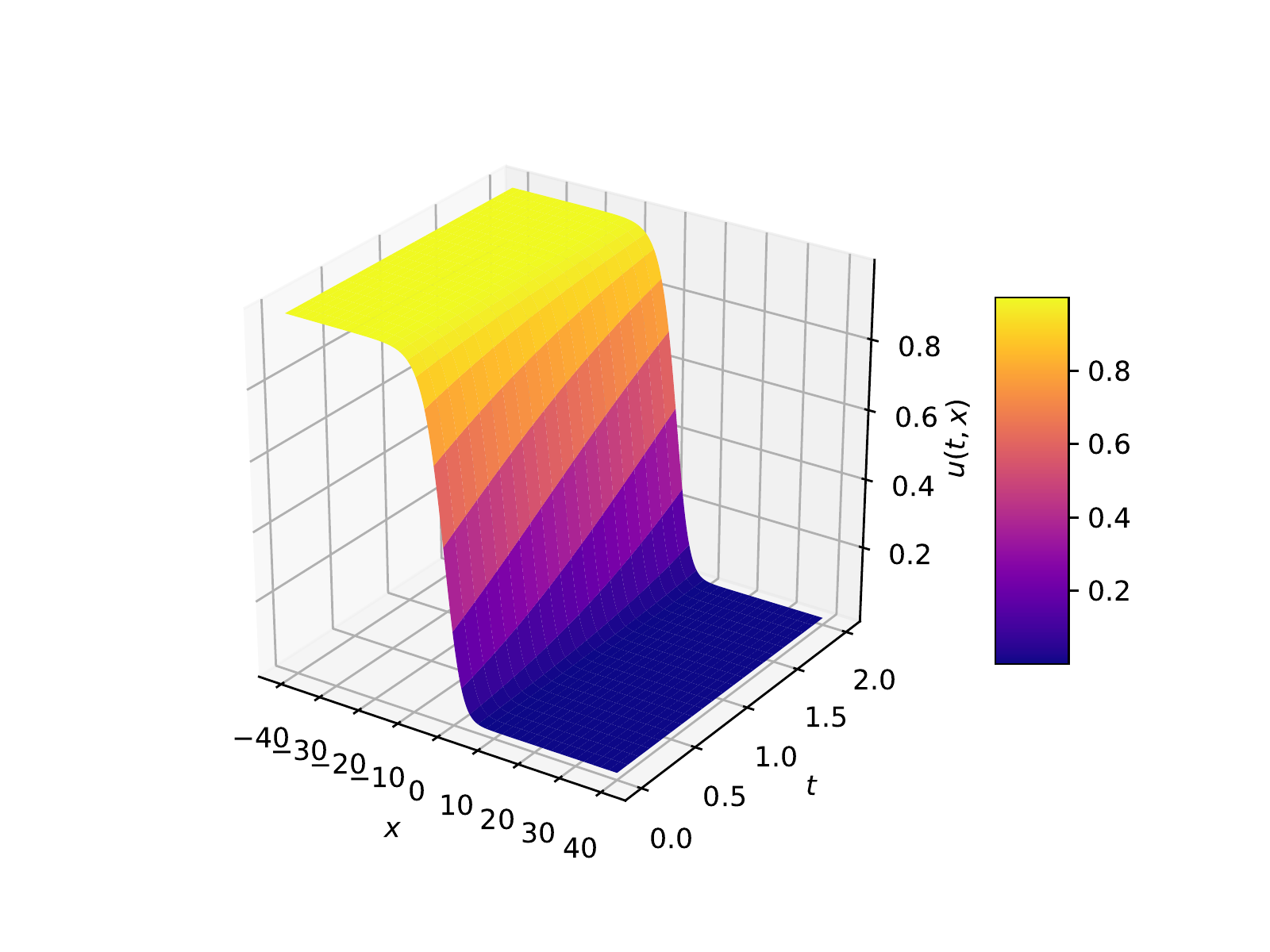}
	\caption{Analytical solution of the NWS equation.}
	\label{fig:new}
\end{figure}

Newell-Whitehead-Segel (NWS) equation is a special case of the Nagumo equation \cite{zhi1992analytic}. Nagumo equation is a nonlinear reaction-diffusion equation that models pulse transmission line simulating a nerve axon \cite{nagumo1962active}, population genetics \cite{aronson1978multidimensional}, and circuit theory \cite{scott1963neuristor}. The NWS equation and its analytical solution are shown in Table~\ref{tab:eq dis}. We use a traveling wave solution \cite{dehghan2011pseudospectral} that satisfies both wave and NWS equations (Fig.~\ref{fig:new}). We carry similar changes to the feature library that was applied to discovering the Kawahara equation.

\begin{figure}[!htpb]
	\centering
	\includegraphics[scale=0.5]{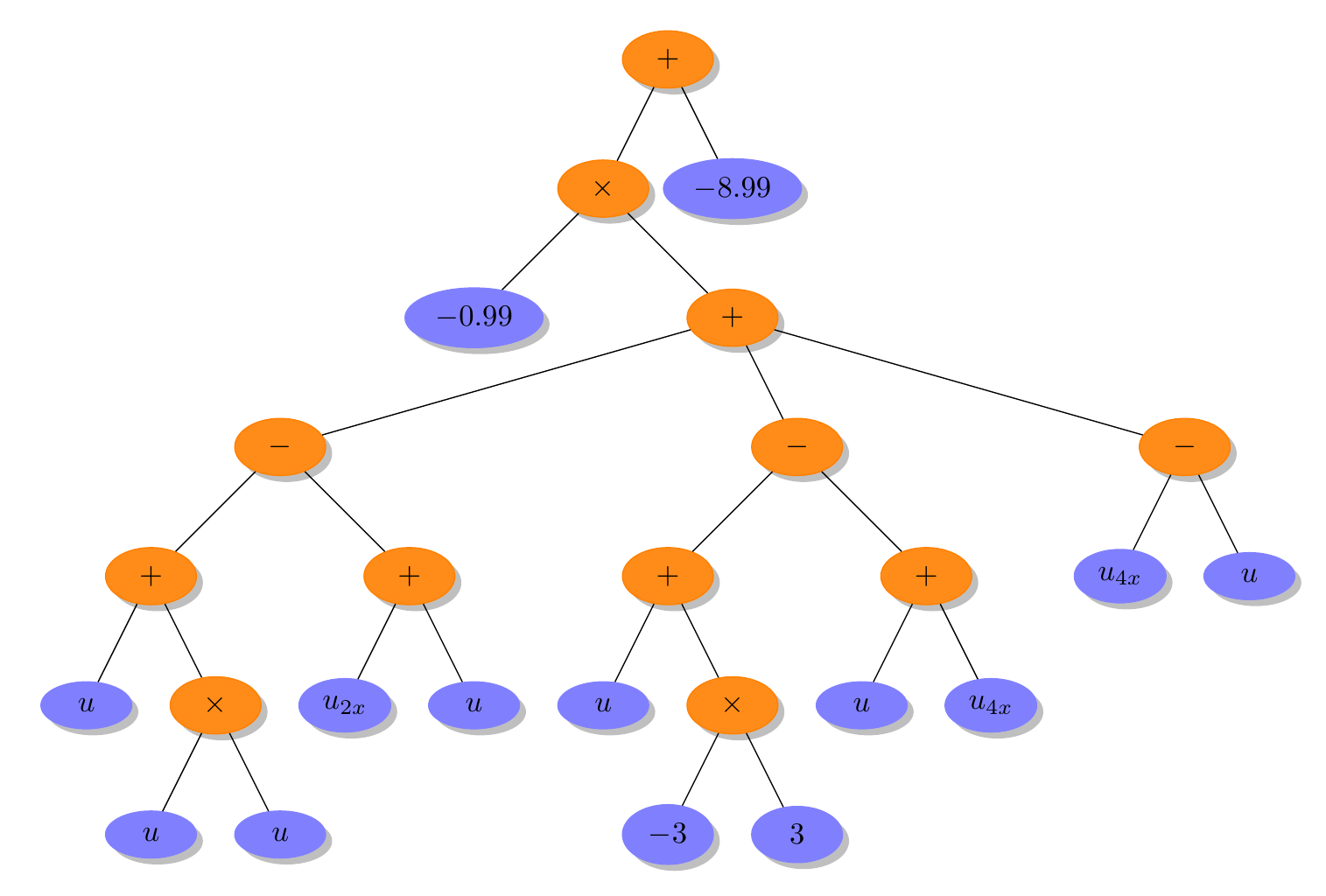}
	\caption{NWS equation in terms of ET identified by GEP.}
	\label{fig:nwst}
\end{figure}

\begin{figure}[!htpb]
	\centering
	\includegraphics[width=0.5\textwidth]{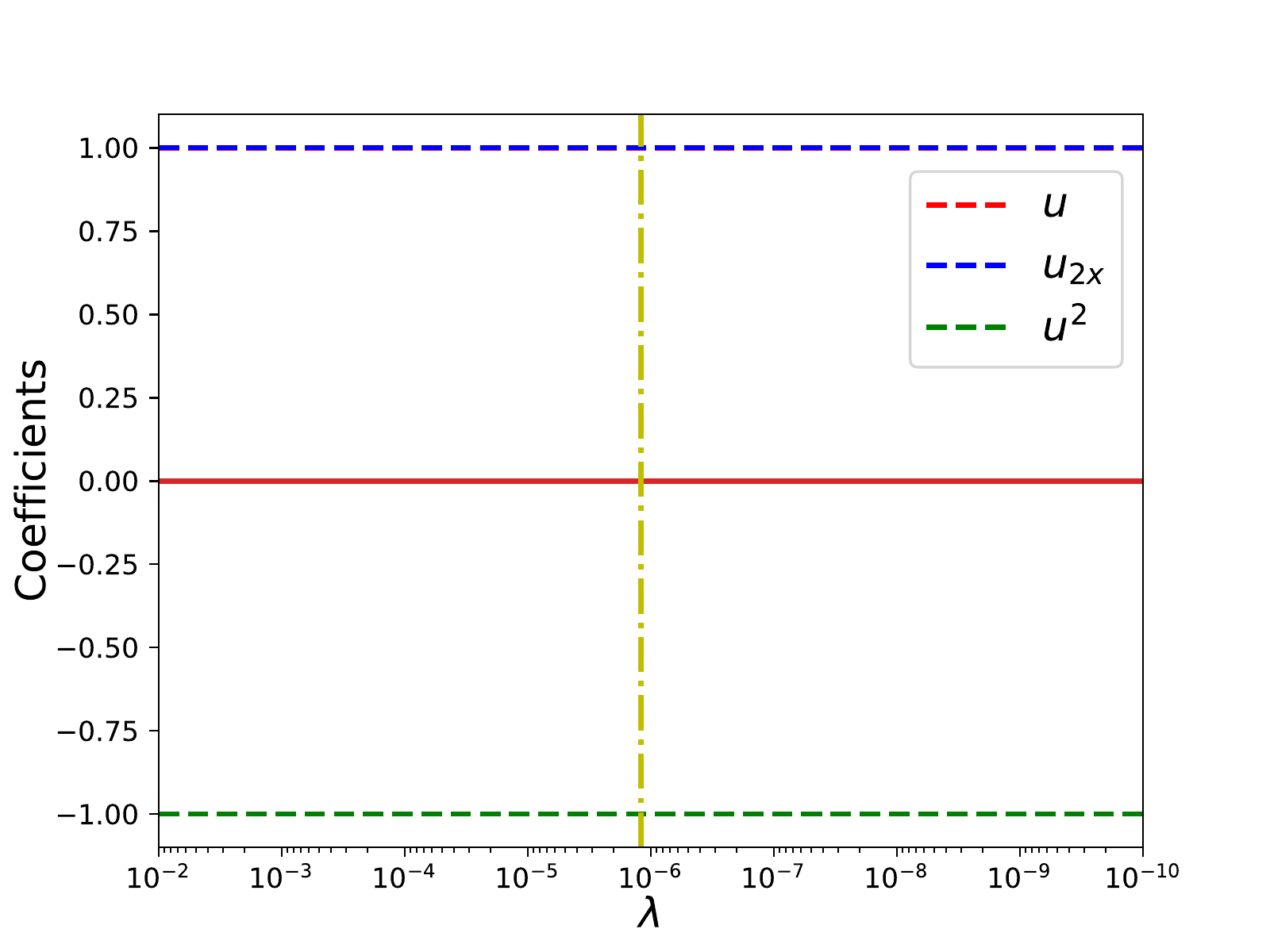}
	\caption{STRidge coefficients as a function of regularization parameter $\lambda$ for the NWS  equation.}
	\label{fig:nwsst}
\end{figure}

Table~\ref{tab:hp2} shows the GEP hyper-parameters used for identification of the NWS equation. However contrast to identifying the Kawahara equation with smaller head length and single gene from simplifying the feature library, for NWS case GEP requires larger head length and more genes for identifying PDE as shown in Table~\ref{tab:hp2}. This is due to the identification of nonlinear interaction feature $u^2$ that appears in the NWS equation.  Fig.~\ref{fig:nwst} shows the identified PDE in the form of ET. When ET form is simplified, we can show that the resulting model is the NWS equation identified along with its coefficients. For STRidge, the  regularization weight ($\lambda$)  is swept across various values as shown Fig.~\ref{fig:nwsst}. The yellow line in Fig.~\ref{fig:nwsst} represents the value of $\lambda$ at which the best identified PDE is selected.  Note that STRidge was able to find the NWS equation at various values of regularization weights ($\lambda$). Table~\ref{tab:nwsr} shows the NWS equation identified by both GEP and STRidge.

\begin{table*}[!htpb]
\caption{ NWS equation identified by GEP and STRidge.}
\label{tab:nwsr}
\bgroup
\def\arraystretch{1.5}
\setlength{\tabcolsep}{1.5em}
\begin{tabular}{@{}llc}
\hline
& \textbf{Recovered} & \textbf{Test error}   \\ \hline
\noalign{\smallskip}
True      & $u_t = 1.00~u_{2x}+  1.00~ u - 1.00~ u^{2}$   &   \\
GEP       & $u_t = 0.99~u_{2x}+  0.99~ u - 0.99~ u^{2} - 8.27\times10^{-8}$  &  $3.02\times10^{-11}$ \\
STRidge   & $u_t = 1.00~u_{2x}+  0.99~ u - 0.99~ u^{2}$  &  $1.36\times10^{-11}$\\
\hline
\end{tabular}
\egroup
\end{table*}

\subsection{Sine-Gordon Equation}
\begin{figure}[!htpb]
	\centering
	\includegraphics[width=0.5\textwidth,trim={1.5cm 0 0 0},clip]{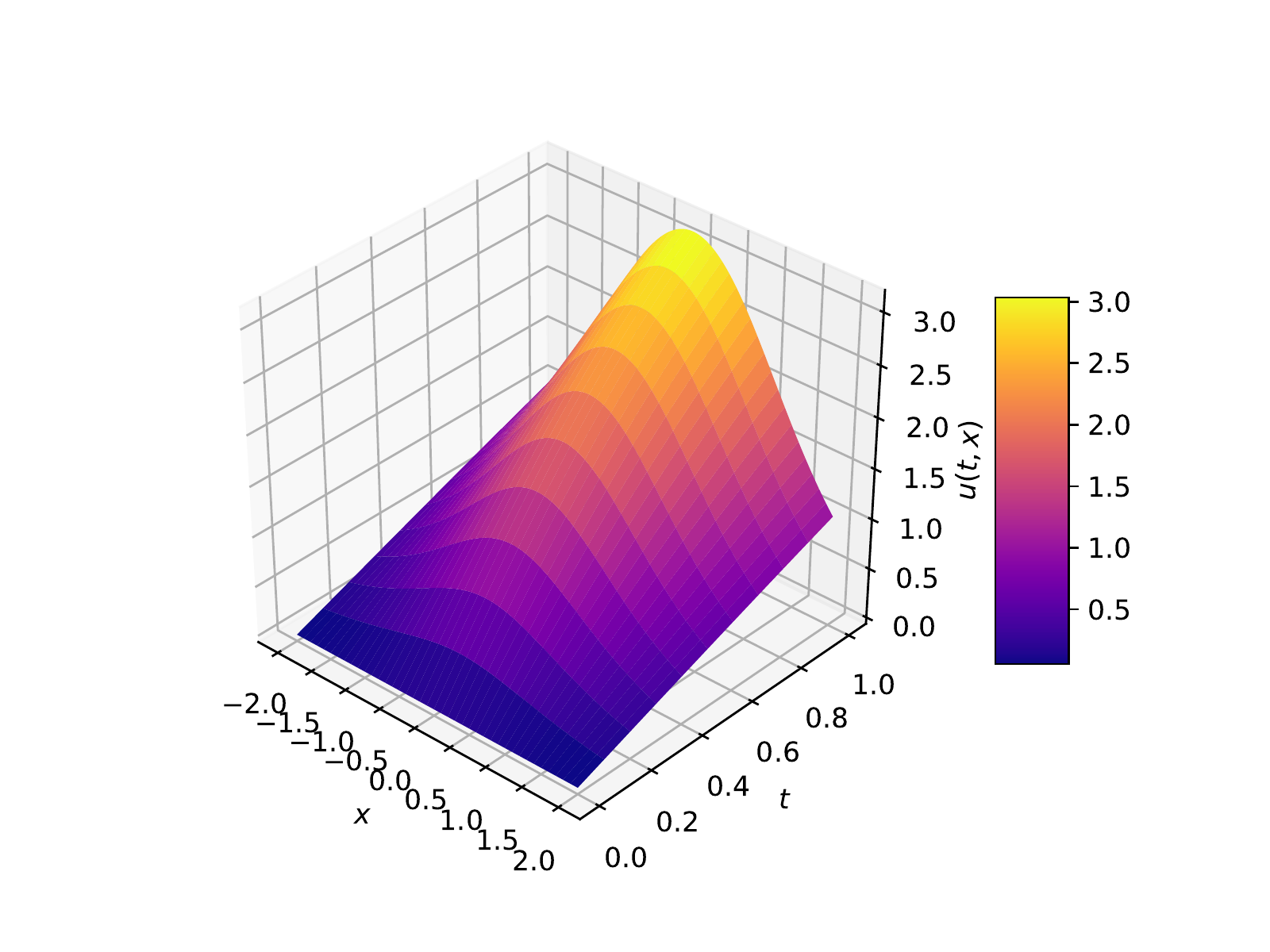}
	\caption{Analytical solution of the Sine-Gordon equation.}
	\label{fig:sg}
\end{figure}

Sine-Gordon equation is a nonlinear PDE that appears in propagating of fluxions in Josephson junctions \cite{barone1971theory}, dislocation in crystals \cite{perring1962model} and nonlinear optics \cite{whitham2011linear}. Sine-Gordon equation has a sine term that needs to be identified by GEP and STRidge by observing data (Fig.~\ref{fig:sg}). This test case is straight forward for GEP as the function set includes trigonometric operators that help to identify the equation. However, the application of STRidge is suitable if features library is limited to basic interactions and does not contain a basis with trigonometric dependencies. STRidge may recover infinite series approximations if higher degree basic feature interactions are included in the feature library \cite{sindy}. Note that the output or target data for the Sine-Gordon equation consists of second order temporal derivative of velocity field $u(t,x)$. Hence, $\textbf{V(t)}$ consists of $u_{2t}$ measurements instead of $u_t$.

\begin{figure}[!ht]
	\centering
	\includegraphics[width=0.4\textwidth]{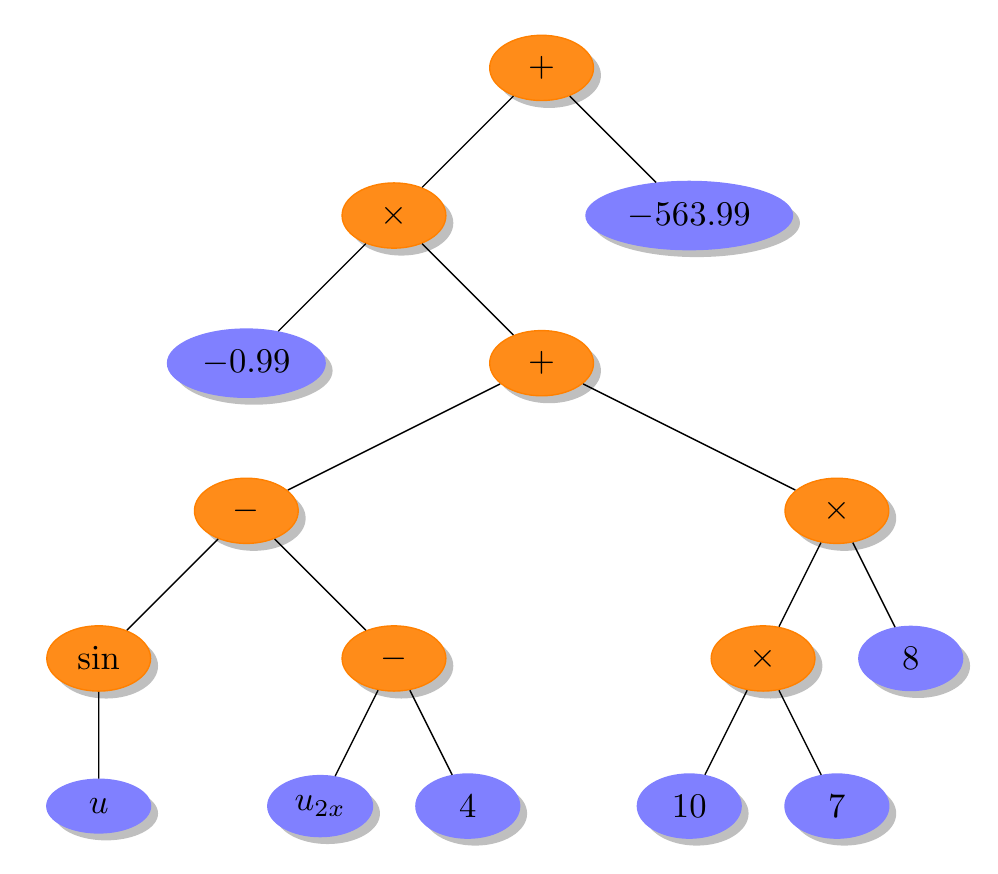}
	\caption{Sine-Gordon equation in terms of ET identified by GEP.}
	\label{fig:sgt}
\end{figure}

Table~\ref{tab:hp2} shows the GEP hyper-parameters used for identifying the Sine-Gordon equation. For our analysis, GEP found the best model when the larger population size used.  Fig.~\ref{fig:sgt} shows the identified PDE in the form of ET. When ET form is simplified, we can show that the resulting model is the Sine-Gordon equation identified along with its coefficients. Table~\ref{tab:sgr} shows the identified equation by GEP. This test case demonstrates the usefulness of GEP in identifying models with complex function composition and limitation of the expressive and predictive power of the feature library in STRidge.

\begin{table}[!htbp]
\caption{Sine-Gordon equation identified by GEP.}
\label{tab:sgr}
\bgroup
\def\arraystretch{1.5}
\setlength{\tabcolsep}{0.35em}
\begin{tabular}{@{}llc}
\hline
& \textbf{Recovered} & \textbf{Test error}   \\ \hline
\noalign{\smallskip}
True      & $u_{2t} = 1.00~u_{2x} - 1.00~\textrm{sin}(u)$   &   \\
GEP       & $u_{2t} = 0.99~u_{2x} - 0.99~\textrm{sin}(u)- 1.82\times10^{-5}$  &  $1.57\times10^{-4}$ \\
\hline
\end{tabular}
\egroup
\end{table}

\section{Truncation Error Analysis}
\label{sec:trunc}
This section deals with constructing a modified differential equation (MDE) for the Burgers equation. We aim at demonstrating both GEP and STRidge techniques as SR tools in the identification of truncation errors resulting from an MDE of the Burgers nonlinear PDE. MDEs provide valuable insights into discretization schemes along with their temporal and spatial truncation errors. Initially, MDE analysis was developed to connect the stability of nonlinear difference equations with the form of the truncation errors\cite{hirt1968heuristic}. In continuation, the symbolic form of MDEs were developed and a key insight was proposed that only the first few terms of the MDE dominate the properties of the numerical discretization\cite{ritchmyer1967difference}. These developments of MDE analysis lead to increasing accuracy by eliminating leading order truncation error terms\cite{klopfer1983nonlinear}, improving stability of schemes by adding artificial viscosity terms \cite{majda1978systematic}, preserving symmetries\cite{ozbenli2017numerical,ozbenli2017high}, and ultimately sparse identification of truncation errors\cite{thaler2019108851}. Therefore, MDE analysis plays a prominent role in implicit large eddy simulations (ILES) \cite{adams2004implicit} as truncation errors are shown to have inherent turbulence modelling capabilities\cite{margolin2002rationale}. Discretization schemes are tuned in the ILES approach as to model the subgrid scale tensor using truncation errors. As the construction of MDEs becomes cumbersome and intractable for complex flow configurations, data driven SR tools such as GEP and STRidge can be exploited for the identification of MDEs by observing the data.

\begin{table*}[!htpb]
\caption{ GEP hyper-parameters selected for identification of truncation error terms of MDEs.}
\label{tab:hp3}
\bgroup
\def\arraystretch{1.5}
\setlength{\tabcolsep}{0.8em}
\begin{tabular}{@{}lcccc}
\hline
\textbf{Hyper-parameters}  &  \textbf{Burgere eq. (i)}  & \textbf{Burgers eq. (ii)}  \\ \hline
\noalign{\smallskip}
Head length              &   $8$     &    $8$      \\
Number of genes          &   $5$     &    $4$      \\
Population size          &   $70$    &    $70$     \\
Generations              &   $1000$   &    $1000$    \\
Length of RNC array      &   $20$    &    $20$      \\
Random constant minimum  &   $1.0\times10^{-6}$     &   $1.0\times10^{-5}$      \\
Random constant maximum  &   $0.01$    &    $0.01$      \\
\hline
\end{tabular}
\egroup
\end{table*}

For demonstration purposes, we begin by constructing an MDE of the Burgers equation,
\begin{align}
\label{eq:tr1}
& u_{t} + uu_{x}= \nu u_{2x}, 
\end{align}
and  discretizing Eq.~(\ref{eq:tr1}) using first order schemes (i.e., forward in time and backward in space approximations for the spatial and temporal derivatives, respectively) and a second order accurate central difference approximation for the second order spatial derivatives. The resulting discretized Burgers PDE is shown below,
\begin{align}
\label{eq:tr2}
& \dfrac{u^{p+1}_{j}-u^{p}_{j}}{dt} + u^p_j\dfrac{u^p_{j}-u^p_{j-1}}{dx}= \nu  \dfrac{u^p_{j+1}-2u^p_j+u^p_{j-1}}{dx^2}, 
\end{align}
where temporal and  spatial steps are given by $dt$ and $dx$, respectively. In Eq.~\ref{eq:tr2}, the spatial location is denoted using  subscript index $j$ and the temporal snapshot using superscript index $p$.  

To derive the modified differential equation (MDE) of the Burgers PDE, we substitute the Taylor approximations for each term,
\begin{equation}\label{eq:tr3}
\left.\begin{aligned}
 u^{p+1}_{j} &= u^{p}_{j} + dt(u_{t})^{p}_{j} + \dfrac{dt^2}{2}(u_{2t})^{p}_{j} +   
                \dfrac{dt^3}{6}(u_{3t})^{p}_{j} + \ldots \\
 u^{p}_{j+1} &= u^{p}_{j} + dx((u_{x}))^{p}_{j} + \dfrac{dx^2}{2}(u_{2x})^{p}_{j} + 
                \dfrac{dx^3}{6}(u_{3x})^{p}_{j} + \ldots \\
 u^{p}_{j-1} &= u^{p}_{j} - dx(u_{x})^{p}_{j} + \dfrac{dx^2}{2}(u_{2x})^{p}_{j} - 
                \dfrac{dx^3}{6}(u_{3x})^{p}_{j} + \ldots \\
\end{aligned}
\right\}
\end{equation}
When we substitute these approximations into Eq.~\ref{eq:tr2}, we obtain the Burgers MDE as follows,
\begin{equation}\label{eq:tr4}
\begin{aligned}
(u_{t} + uu_{x} - \nu u_{2x})^{p}_{j} &=  -R,
 \end{aligned}
\end{equation}
where $R$ represents truncation error terms of the Burgers MDE given as,
\begin{equation}\label{eq:tr5}
\begin{aligned}
R &=  \dfrac{dt}{2}(u_{2t})^{p}_{j} +  \dfrac{dx}{2}(uu_{x})^{p}_{j} -  \dfrac{\nu dx^2}{12}(u_{4x})^{p}_{j} + O(dt^2, dx^4).
 \end{aligned}
\end{equation}
Furthermore, temporal derivative in Eq.~\ref{eq:tr5} is substituted with spatial derivatives resulting in,
\begin{multline}\label{eq:tr6}
R =  dtuu_x^2  - dt\nu u_{x}u_{2x} - dt\nu uu_{3x} \\
- \dfrac{dx}{2}uu_{2x}  + \dfrac{dt}{2}u^2u_{2x}  - \dfrac{\nu dx^2}{12}u_{4x}  + O(dt^2, dx^4).
\end{multline}

The truncation error or residual of discretized equation considering $u(t,x)$ as exact solution to the Burgers PDE is equal to the difference between the numerical scheme (Eq.~\ref{eq:tr2}) and differential equation (Eq.~\ref{eq:tr1})\cite{hirsch2007numerical}. This results in discretized equation with residual as shown below, 
\begin{multline}\label{eq:tr7}
 u^{p+1}_{j}-u^{p}_{j} + u^p_jdt\dfrac{u^p_{j}-u^p_{j-1}}{dx} \\ - \nu dt  \dfrac{u^p_{j+1}-2u^p_j+u^p_{j-1}}{dx^2} = Rdt.
\end{multline}
We follow the same methodology for constructing the output data and feature library as discussed in Section~\ref{sec:meth} for the equation discovery. However, the output or target data $\mathbf{V(t)}$ is stored with the left hand side of Eq.~\ref{eq:tr7} denoted from now as $\mathbf{U_{er}}$. The resulting output and core feature library are shown below,
\begin{equation}\label{eq:tr8}
\left.\begin{aligned}
\textbf{V(t)}  &=  
\left[
  \begin{array}{c}    
     \mathbf{U_{er}}
  \end{array}
\right]\\
\mathbf{\widetilde \Theta(U)}  &=  
\left[
  \begin{array}{cccccccc}    
    \textbf{U}  &  \textbf{U$_x$} & \textbf{U$_{2x}$} & \textbf{U$_{3x}$} & \textbf{U$_{4x}$} 
    \end{array}
\right]
\end{aligned}
\right\}.
\end{equation}

The computation of the output data $\mathbf{V(t)}$ in Eq.~\ref{eq:tr8}  can be obtained using the analytical solution of the Burgers PDE. Furthermore, the derivatives in core feature library $\mathbf{\widetilde \Theta(U)}$ are calculated using the finite difference approximations given by Eq.~\ref{goveq4}. We use both analytical solutions listed in Table~\ref{tab:eq dis} for the Burgers equation (i) and the Burgers equation (ii) to test GEP and STRidge for recovering truncation error terms. 

We use the same extended feature library $\mathbf{\widetilde \Theta(U)}$ as input to STRidge given in Eq.~\ref{goveq7}, but without the fifth order derivative. However, we add additional third degree interaction of features to $\mathbf{\widetilde \Theta(U)}$ to recover the truncation error terms containing third degree nonlinearities. The extra nonlinear features that are added to $\mathbf{\widetilde \Theta(U)}$ are given below,
\[
\begin{split}
[\begin{matrix} \mathbf{U^2U_x}   & \mathbf{U^2U_{2x}} & \mathbf{U^2U_{3x}}  &  \mathbf{U^2U_{4x}} \end{matrix} \\
 &\hspace{-3cm} \begin{matrix}  \mathbf{UU^2_x}   & \mathbf{UU_xU_{2x}} & \mathbf{UU_xU_{3x}}  &  \mathbf{UU_xU_{4x}}  \end{matrix}].
\end{split}
\]
In contrast, GEP uses the core feature $\mathbf{\widetilde \Theta(U)}$ as input as it identifies the higher order nonlinear feature interactions automatically. This test case shows the natural feature extraction capability of GEP and need to modify the feature library to increase the expressive power of STRidge.
\begin{table}[!htpb]
\caption{GEP functional and terminal sets used for truncation error term recovery. `?' is a random constant.}
\label{tab:teset}
\bgroup
\def\arraystretch{1.5}
\setlength{\tabcolsep}{2.1em}
\begin{tabular}{@{}lll}
\hline
\textbf{Parameter} & \textbf{Value}    \\ \hline
\noalign{\smallskip}
Function set     &   $+, -, \times$  \\
Terminal set     &   $\mathbf{\widetilde \Theta(U)}$, $?$ \\
Linking function &   $+$  \\
\hline
\end{tabular}
\egroup
\end{table}

\begin{figure*}[!htpb]
	\centering
	\includegraphics[width=1\textwidth]{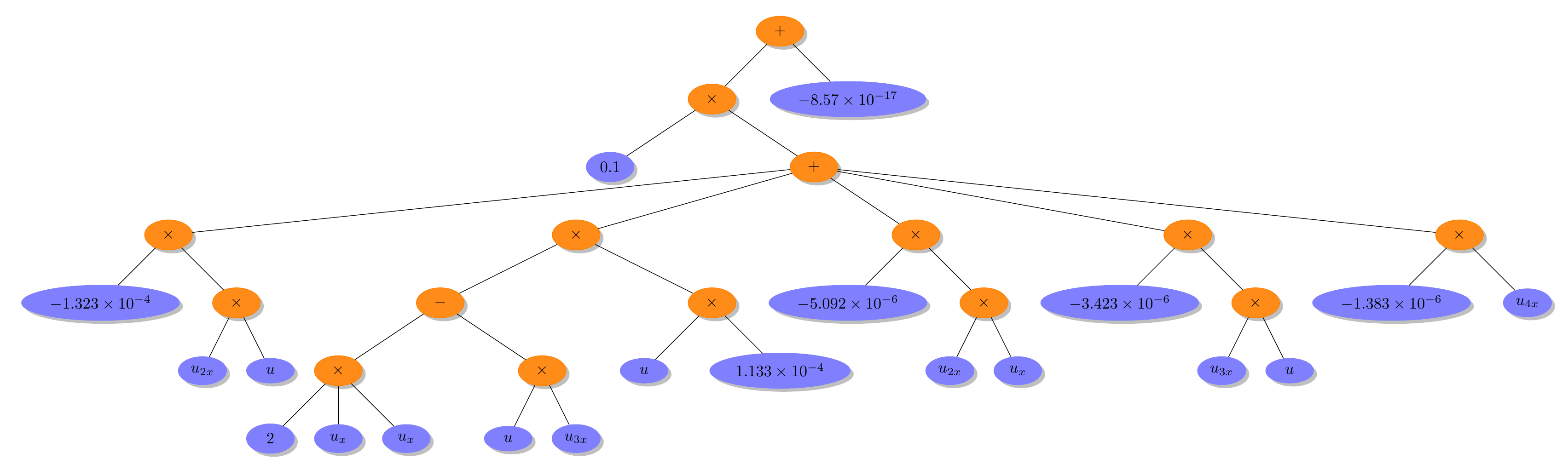}
	\caption{Truncation error of the Burgers MDE using analytical solution of the Burgers equation (i) in terms of ET identified by GEP.}
	\label{fig:mbst}
\end{figure*}

\begin{figure}[!htpb]
	\centering
	\includegraphics[width=0.5\textwidth]{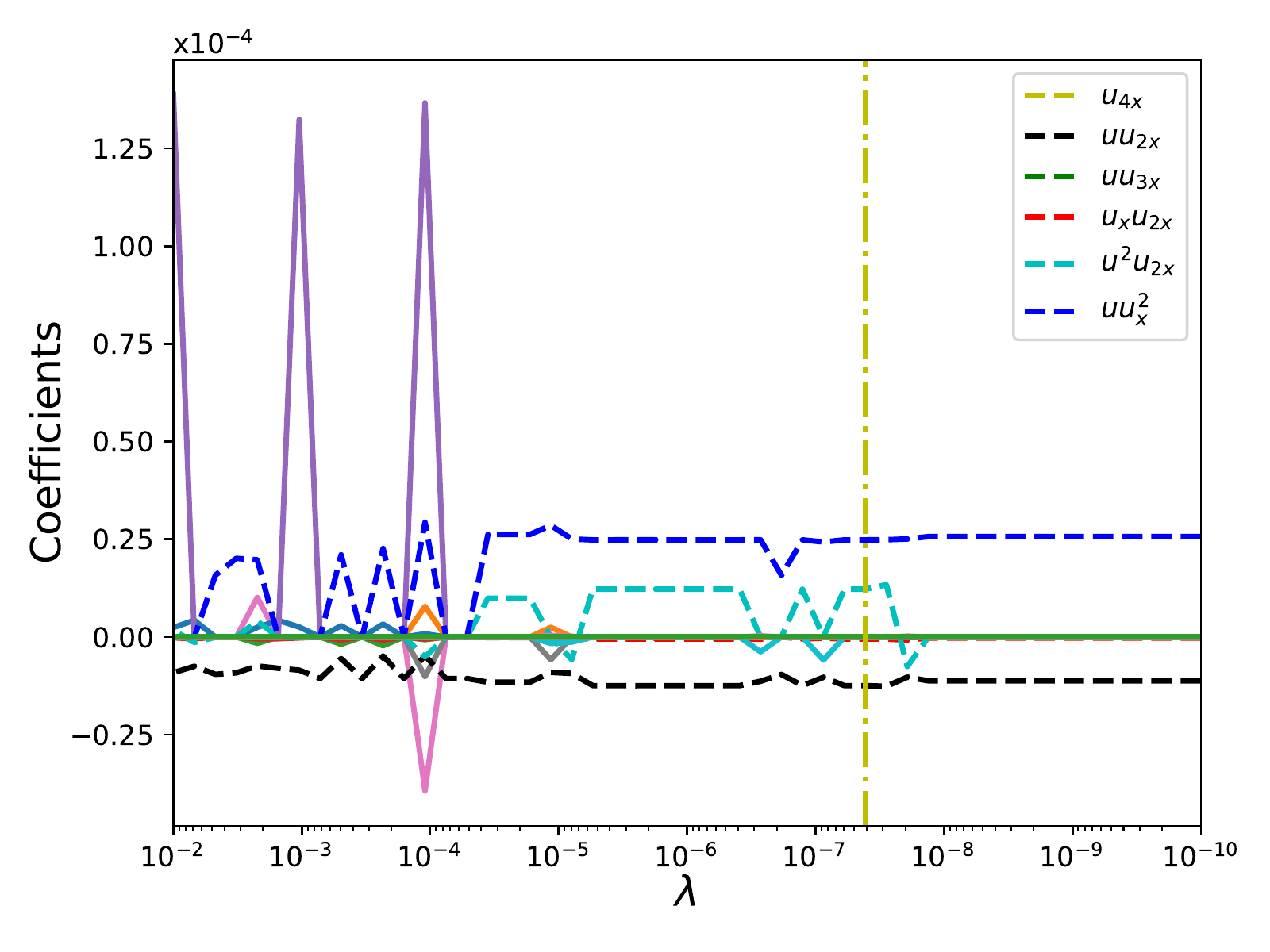}
	\caption{STRidge coefficients as a function of regularization parameter $\lambda$ for truncation error of the Burgers MDE (i).}
	\label{fig:mbsst}
\end{figure}

The functional and terminal sets used for truncation error identification are listed in Table~\ref{tab:teset}.  First, we test the recovery of truncation errors using the analytical solution of the Burgers equation (i) with the same spatial and temporal domain listed in Table~\ref{tab:eq dis}. However, we set spatial discretization to be $dx=0.005$ and temporal discretization to $dt=0.005$ for storing the analytical solution $u(t,x)$. This test case needs a large population size, bigger head length, more genes and more iterations as given in Table~\ref{tab:hp3}, as the truncation error terms consist of nonlinear combinations of features and the coefficients of error terms that are generally difficult for GEP to identify. Fig.~\ref{fig:mbst} shows the ET form of the identified  truncation error terms. The regularization weight $\lambda$ for STRidge is swept across a range of values as shown in Fig.~\ref{fig:mbsst}. The vertical yellow line in  Fig.~\ref{fig:mbsst} is the value of $\lambda$ where STRidge identifies the best truncation error model. Table~\ref{tab:mbst} shows the recovered error terms by GEP and STRidge along with their coefficients. Both GEP and STRidge perform well in identifying the nonlinear spatial error terms with STRidge predicting the error coefficient better than GEP.  

\begin{figure*}[!htpb]
	\centering
	\includegraphics[width=1\textwidth]{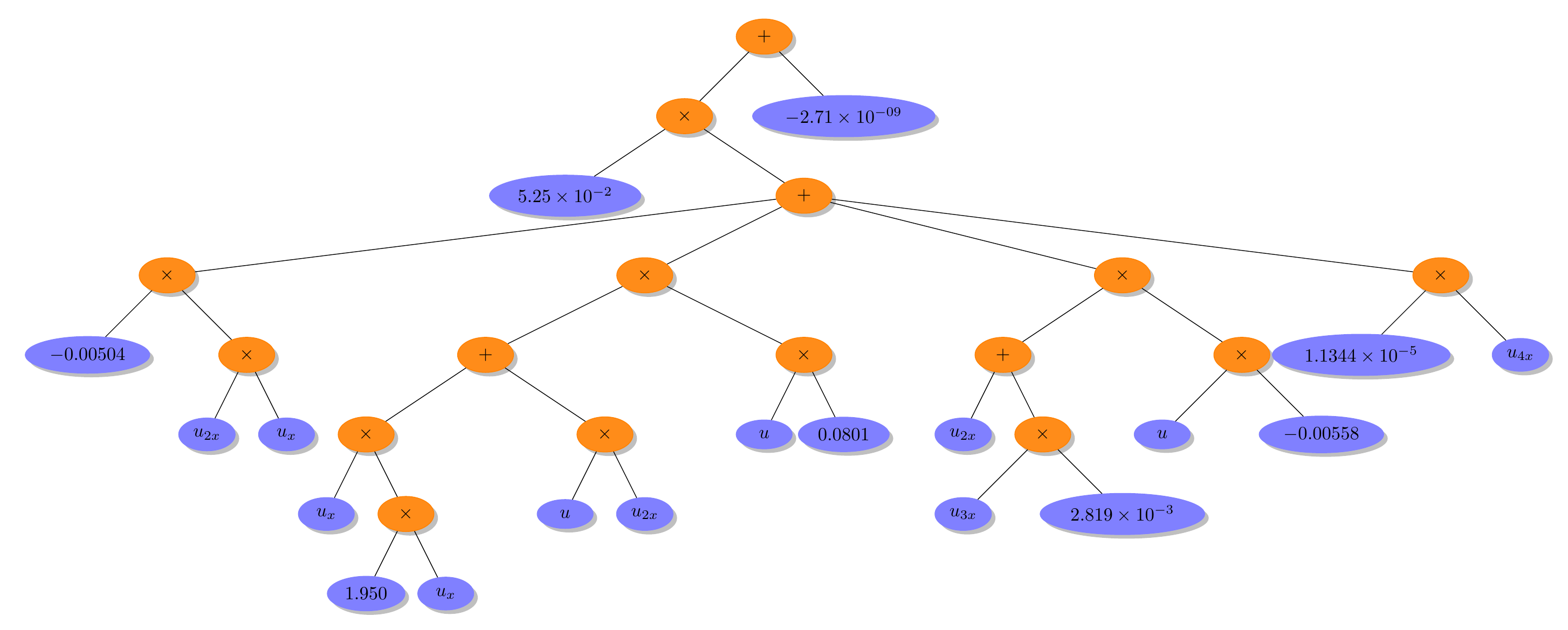}
	\caption{Truncation error term of the  Burgers MDE using analytical solution of the Burgers equation (ii) in terms of ET identified by GEP.}
	\label{fig:mbwt}
\end{figure*}

\begin{figure}[!htpb]
	\centering
	\includegraphics[width=0.5\textwidth]{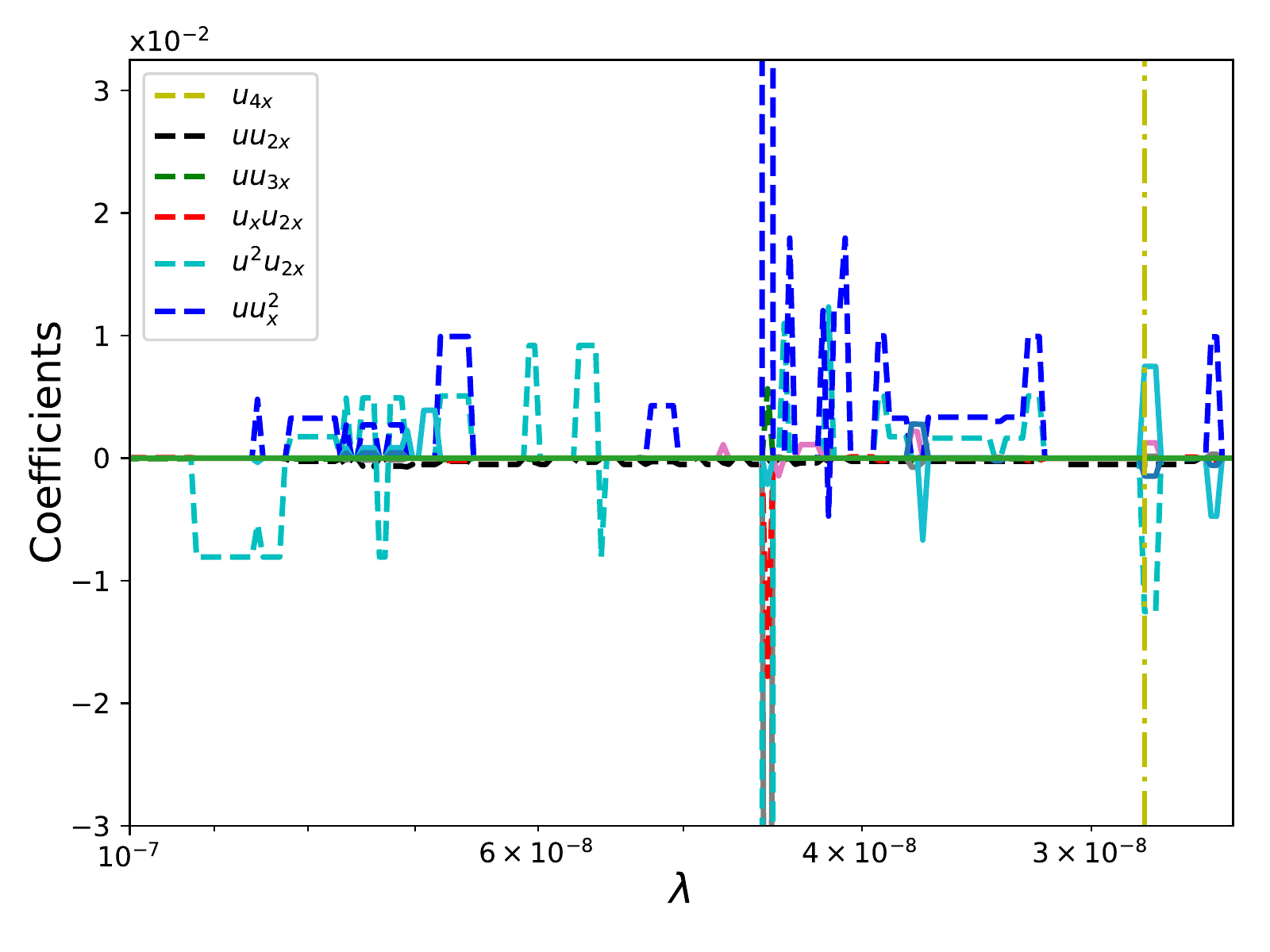}
	\caption{STRidge coefficients as a function of regularization parameter $\lambda$ for truncation error of the Burgers MDE (ii).}
	\label{fig:mbwst}
\end{figure}

In the second case, we test the recovery of truncation errors using an analytical solution of the Burgers eq. (ii)  with the same spatial and temporal domain listed in Table~\ref{tab:eq dis}.  We select the spatial discretization $dx=0.005$ and the temporal discretization $dt=0.1$ for propagating the analytical solution $u(t,x)$. This test case also follows the previous case where a large population size, bigger head length, more genes, and more iterations are needed as shown in Table~\ref{tab:hp3}. Fig.~\ref{fig:mbwt} shows the ET form of identified truncation error terms. The regularization weight $\lambda$ for STRidge is swept across a range of values as shown in Fig.~\ref{fig:mbwst}.  In this test case, the coefficients change rapidly in respect to $\lambda$, and the best model is recovered only at the value of $\lambda$ shown by the vertical yellow line in Fig.~\ref{fig:mbwst}. Table~\ref{tab:mbwt} shows the recovered error terms by GEP and STRidge along with their coefficients. Similar to the previous test case, STRidge predicts the truncation error coefficients better than GEP.

\begin{table*}[!htbp]
\caption{Identified truncation error terms along with  coefficients for the Burgers MDE (i)  by GEP and STRidge.}
\label{tab:mbst}
\bgroup
\def\arraystretch{1.5}
\setlength{\tabcolsep}{1em}
\begin{tabular}{@{}lccccc}
\hline
& \textbf{True} & \textbf{GEP} & \textbf{Relative error (\%)} & \textbf{STRidge} & \textbf{Relative error (\%)}  \\ \hline
\noalign{\smallskip}
$uu_{x}^2$    & $2.5\times10^{-5}$    &     $2.26\times10^{-5}$    & $9.6$   & $2.48\times10^{-5}$  &  $0.8$   \\
$u_{x}u_{2x}$ & $-5.0\times10^{-7}$   &     $-5.09\times10^{-7}$   & $1.8$   & $-5.02\times10^{-7}$ &  $0.4$  \\
$uu_{3x}$     & $-2.5\times10^{-7}$   &     $-3.42\times10^{-7}$   & $36.8$ & $-2.29\times10^{-7}$  &  $8.4$ \\
$u^2u_{2x}$   &  $1.25\times10^{-5}$  &     $1.13\times10^{-5}$    & $9.6$  & $1.22\times10^{-5}$   &  $2.4$\\
$u_{4x}$      &  $1.25\times10^{-9}$  &     $1.38\times10^{-9}$    & $10.4$ & $1.16\times10^{-9}$   &  $7.2$\\
$uu_{2x}$     & $-1.25\times10^{-5}$  &     $-1.33\times10^{-5}$   & $6.4$  & $-1.24\times10^{-5}$  & $0.8$ \\
\hline
\end{tabular}
\egroup
\end{table*}

\begin{table*}[!htbp]
\caption{Identified truncation error terms along with  coefficients for the Burgers MDE (ii)  by GEP and STRidge.}
\label{tab:mbwt}
\bgroup
\def\arraystretch{1.5}
\setlength{\tabcolsep}{1em}
\begin{tabular}{@{}lccccc}
\hline
& \textbf{True} & \textbf{GEP} & \textbf{Relative error (\%)} & \textbf{STRidge} & \textbf{Relative error (\%)}  \\ \hline
\noalign{\smallskip}
$uu_{x}^2$    & $1.0\times10^{-2}$    &     $8.19\times10^{-3}$  & $18.1$ & $9.92\times10^{-3}$   & 0.8  \\
$u_{x}u_{2x}$ & $-2.0\times10^{-4}$   &     $-2.64\times10^{-4}$ & $32.0$  &$-1.99\times10^{-4}$  & 0.5  \\
$uu_{3x}$     & $-1.0\times10^{-4}$   &     $-1.55\times10^{-4}$ & $55.0$ & $-9.91\times10^{-5}$  & 0.9 \\
$u^2u_{2x}$   & $5.0\times10^{-3}$    &     $4.21\times10^{-3}$  & $15.8$ & $5.08\times10^{-3}$   & 1.6 \\
$u_{4x}$      & $5.0\times10^{-7}$    &     $5.65\times10^{-7}$  & $13.0$ & $4.94\times10^{-7}$   & 1.2\\
$uu_{2x}$     & $-2.5\times10^{-4}$   &     $-2.75\times10^{-4}$ & $10$ & $-2.54\times10^{-4}$    & 1.6 \\
\hline
\end{tabular}
\egroup
\end{table*}

\section{Hidden Physics Discovery}
\label{sec:hidden}
In this section, we demonstrate the identification of hidden physical laws from sparse data mimicking sensor measurements using GEP and STRidge. Furthermore, we demonstrate the usefulness of GEP as a natural feature extractor that is capable of identifying complex functional compositions. However, STRidge in its current form is limited by its expressive power which depends on its input feature library. Many governing equations of complex systems in the modern world are only partially known or in some cases still awaiting first principle equations. For example, atmospheric radiation models or chemical reaction models might be not fully known in governing equations of environmental systems\cite{krasnopolsky2006complex, krasnopolsky2006new}. These unknown models are generally manifested in the right hand side of the known governing equations (i.e., dynamical core) behaving as a source or forcing term. The recent explosion of rapid data gathering using smart sensors\cite{dhingra2019internet} has enabled researchers to collect data that  the true physics of complex systems but their governing equations  are only known partially. To this end, SR approaches might be able to recover these unknown physical models when exposed to data representing full physics. 

To demonstrate the proof of concept for identification of unknown physics, we formulate a 1D advection-diffusion PDE and a 2D vortex-merger problem. These problems include a source term that represents the hidden physical law. We generate synthetic data that contains true physics and substitute this data set in to the known governing equations. This results in an unknown physical model left as a residual that must be recovered by GEP when exposed to a target or output containing the known part of the underlying processes. Furthermore, both GEP and STRidge are tested to recover eddy viscosity kernels for the 2D Kraichnan turbulence problem. These eddy viscosity kernels are manifested as source terms in the LES equations that model unresolved small scales. Additionally, the value of the ad-hoc free modelling parameter that controls the dissipation in eddy viscosity models is also recovered using GEP and STRidge. 
\begin{table*}[!htpb]
\caption{ GEP hyper-parameters selected for identifying source terms for the 1D advection-diffusion and the 2D vortex-merger problem. }
\label{tab:hp}
\bgroup
\def\arraystretch{1.5}
\setlength{\tabcolsep}{0.8em}
\begin{tabular}{@{}lcccc}
\hline
\textbf{Hyper-parameters}  &  \textbf{1D advection-diffusion eq.}  & \textbf{2D vortex-merger problem}  \\ \hline
\noalign{\smallskip}
Head length              &   $6$     &    $5$      \\
Number of genes          &   $2$     &    $3$      \\
Population size          &   $50$    &    $50$     \\
Generations              &   $1000$   &    $500$    \\
Length of RNC array      &   $5$      &    $8$      \\
Random constant minimum  &   $\dfrac{\pi}{4}$     &   $-\pi$      \\
Random constant maximum  &   $\pi$    &    $\pi$      \\
\hline
\end{tabular}
\egroup
\end{table*}
\subsection{1D Advection-Diffusion PDE}
In the first test case, we consider a 1D non-homogeneous advection-diffusion PDE which appears in many areas such as fluid dynamics \cite{kumar1983unsteady}, heat transfer \cite{isenberg1973heat}, and mass transfer \cite{guvanasen1983numerical}. The non-homogeneous PDE takes the form,
\begin{equation}\label{eq:ad0}
\begin{aligned}
       u_t + cu_{x} &= \alpha u_{2x} + S(t,x),  \\
 \end{aligned}
\end{equation}
where c = $\dfrac{1}{3\pi}$, $\alpha = \dfrac{1}{4}$ and $S(t,x)$ is the source term. 

\begin{figure}[!htpb]
	\centering
	\includegraphics[scale=0.6,trim={1.5cm 0 0 0},clip]{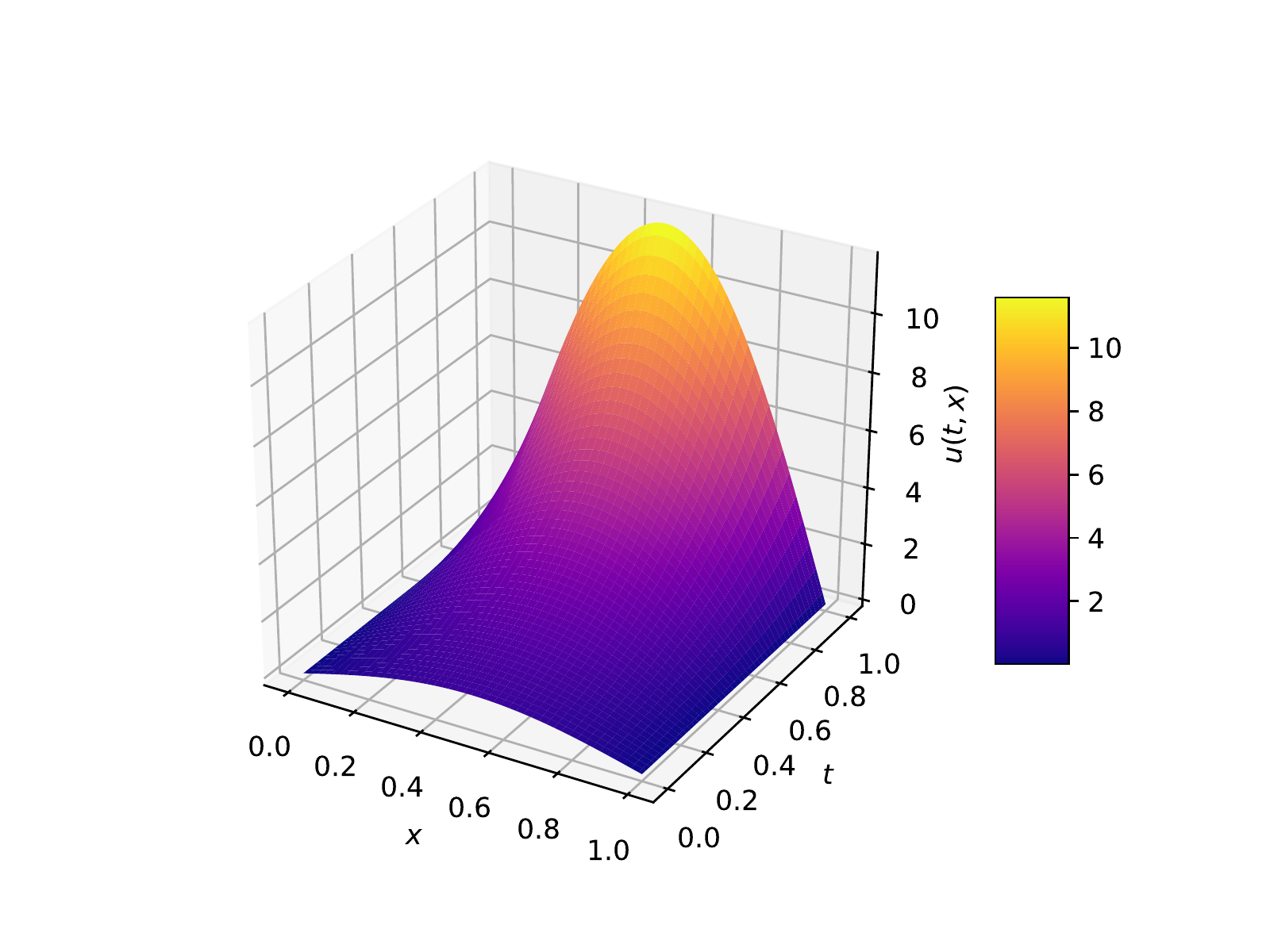}
	\caption{Solution to the 1D advection-diffusion  PDE  with source term. }
	\label{fig:ad}
\end{figure}

We use an analytical solution $u(t,x)$ for solving Eq.~\ref{eq:ad0}. The exact solution for this non-homogeneous PDE is as follows, 
\begin{equation}\label{eq:ad1}
\begin{aligned}
       u(t,x) &= \textrm{exp}\left(\dfrac{\pi^2 t}{4}\right)\textrm{sin}(\pi x), \\
 \end{aligned}
\end{equation}
where the spatial domain $x\in[0,1]$ and the temporal domain $t\in[0,1]$. We discretize the space and time domains with  $n=501$ and $m=1001$, respectively. Fig.~\ref{fig:ad} shows the corresponding analytical solution $u(t,x)$. 

The source term $S(t,x)$, which satisfies Eq.~\ref{eq:ad0} for the analytical solution provided by Eq.~\ref{eq:ad1}, is given as,
\begin{equation}\label{eq:ad2}
\begin{aligned}
       S(t,x) &= \dfrac{\pi^2}{2}\textrm{exp}\left(\dfrac{\pi^2 t}{4}\right)\textrm{sin}(\pi x) + \dfrac{1}{3}\textrm{exp}\left(\dfrac{\pi^2 t}{4}\right)\textrm{cos}(\pi x).\\
 \end{aligned}
\end{equation}

Our goal is to recover this hidden source term once the solution $u(t,x)$ is available either by solving the analytical equation given by Eq.~\ref{eq:ad1} or by sensor measurements in real world applications. Furthermore, we select $64$ random sparse spatial locations to mimic experimental data collection. After the solution $u(t,x)$ is stored at selected sparse spatial locations, we follow the same procedure for constructing output data and feature building as discussed in Section~\ref{sec:meth}. The corresponding output data $\mathbf{V}$ and feature library for recovering source term using GEP are given as,
\begin{equation}\label{eq:ad3}
\left.\begin{aligned}
\textbf{V}  &=  
\left[
  \begin{array}{c}    
     \mathbf{U_{t}} + c\mathbf{U_{x}} - \alpha \mathbf{U_{2x}}
  \end{array}
\right]\\
\mathbf{\widetilde \Theta}  &=  
\left[
  \begin{array}{cc}    
    \textbf{x}  &  \textbf{t} 
    \end{array}
\right]
\end{aligned}
\right\}.
\end{equation}
The derivatives in the output data $\mathbf{V}$ are calculated using Eq.~\ref{goveq4}. Hence, to calculate  spatial derivatives, we also store  additional stencil data $u(t,x)$ around the randomly  selected sparse locations $(u)_j^p$ i.e, $(u)_{j+1}^p$ ,$(u)_{j-1}^p$. Table~\ref{tab:adset} gives the functional and terminal sets used by GEP to recover the source term $S(t,x)$ given in Eq.~\ref{eq:ad2}. 

\begin{table}[!htpb]
\caption{GEP functional and terminal sets used for source term identification. `?' is a random constant.}
\label{tab:adset}
\bgroup
\def\arraystretch{1.5}
\setlength{\tabcolsep}{2.1em}
\begin{tabular}{@{}lll}
\hline
\textbf{Parameter} & \textbf{Value}    \\ \hline
\noalign{\smallskip}
Function set     &   $+, -, \times, / $, exp, sin, cos   \\
Terminal set     &   $\mathbf{\widetilde \Theta}$, $?$ \\
Linking function &   $+$  \\
\hline
\end{tabular}
\egroup
\end{table}

\begin{figure}[!ht]
	\centering
	\includegraphics[width=0.5\textwidth]{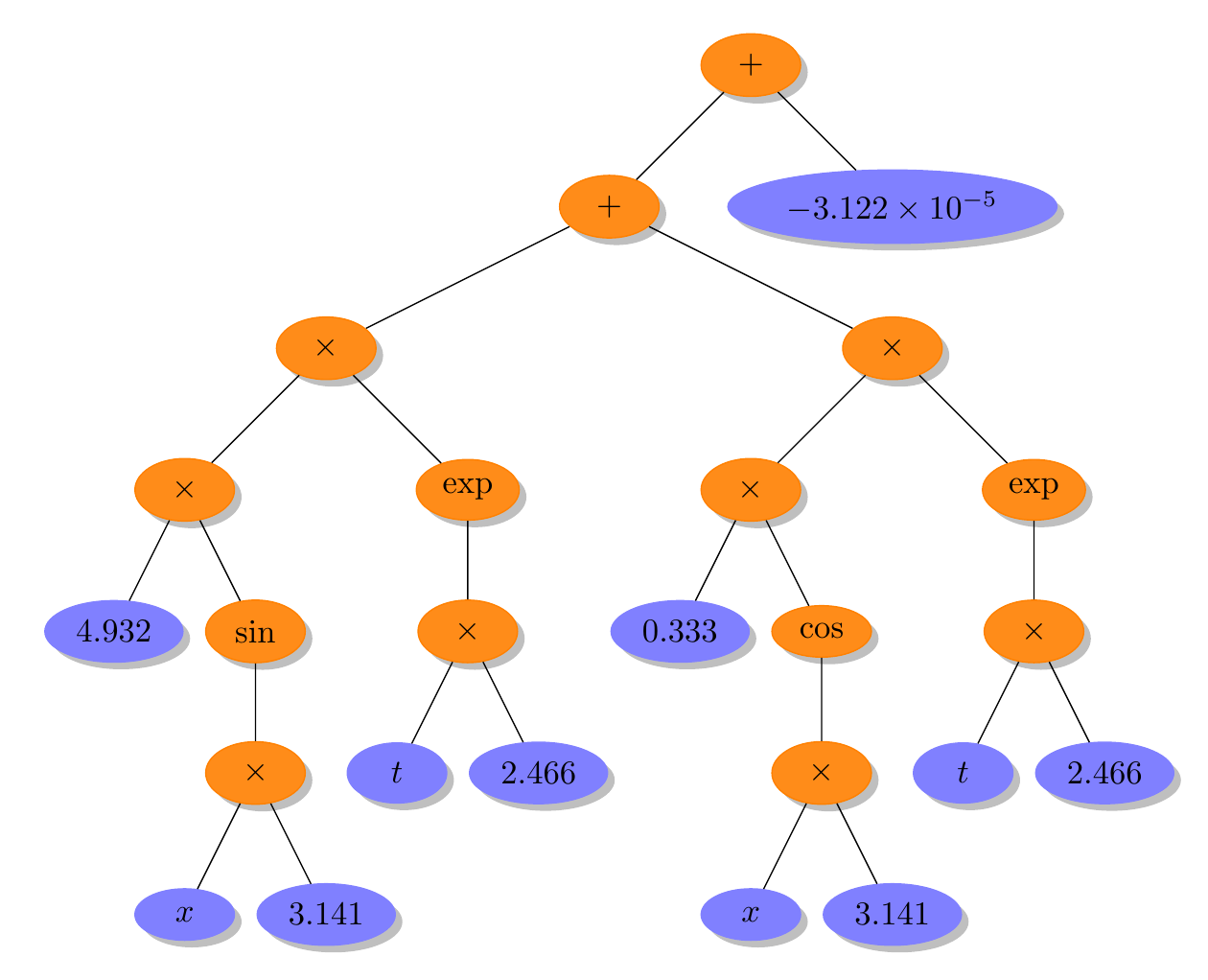}
	\caption{Hidden source term of the 1D advection-diffusion PDE  in terms of ET identified by GEP. }
	\label{fig:adt}
\end{figure}

Table~\ref{tab:hp} lists the hyper-parameters used by GEP for recovering source term of the 1D advection-diffusion equation. As the hidden physical law given in Eq.~\ref{eq:ad2} consists of complex functional compositions, GEP requires a  larger head length, and more generations are required by GEP for identification. The ET form of the source term $S(t,x)$ found by GEP is shown in Fig.~\ref{fig:adt}. The identified source term after simplifying the ET form found by GEP is listed in Table~\ref{tab:adr}. GEP was able to identify the source term $S(t,x)$ given in Eq.~\ref{eq:ad2} from sparse data. 

\begin{table*}[!htbp]
\caption{Hidden source term ($S$) of the 1D advection-diffusion PDE identified by GEP.}
\label{tab:adr}
\bgroup
\def\arraystretch{1.5}
\setlength{\tabcolsep}{1em}
\begin{tabular}{@{}llc}
\hline
& \textbf{Recovered} & \textbf{Test error}   \\ \hline
\noalign{\smallskip}
True      & $S = 4.93~\textrm{exp}(2.47~t)~\textrm{sin}(3.14~x) +0.33~\textrm{exp}(2.47~t)~\textrm{cos}(3.14~x)$   &   \\
GEP       & $S = 4.93~\textrm{exp}(2.46~t)~\textrm{sin}(3.14~x) +0.33~\textrm{exp}(2.46~t)~\textrm{cos}(3.14~x)- 3.12\times10^{-5}$  &  $3.34\times10^{-7}$ \\
\hline
\end{tabular}
\egroup
\end{table*}


\subsection{2D Vortex-Merger Problem}
In this section, we demonstrate the recovery of hidden physical law from the data generated by solving the vortex-merger problem with source terms. The initial two vortices merge to form a single vortex when they are located within a certain critical distance from each other. This two-dimensional process is one of the fundamental processes of fluid motion and it plays a key role in a variety of simulations, such as decaying two-dimensional turbulence\cite{meunier2005physics,san2012high} and mixing layers\cite{san2013coarse}. This phenomenon also occurs in other fields such as astrophysics, meteorology, and geophysics\cite{reinaud2005critical}. The Vortex-merger problem is simulated by using the 2D incompressible Navier-Stokes equations in the domain with periodic boundary conditions. 

We specifically solve the system of PDEs called vorticity-streamfunction formulation. This system of PDEs contains the vorticity transport equation derived from taking the curl of the 2D incompressible Navier-Stokes equations and the Poisson equation representing the kinematic relationship between the streamfunction ($\psi$) and vorticity ($\omega$). The resulting vorticity-streamfunction formulation with source term is given as, 
\begin{equation}\label{eq:vm0}
\left.\begin{aligned}
    \omega_t +J(\omega, \psi) &= \dfrac{1}{\textrm{Re}} \nabla^2\omega + S(t,x,y)\\
    \nabla^2\psi &= -\omega
\end{aligned}
\right\}
\end{equation}
where the Reynolds number is set to $\textrm{Re} = 2000$. In Eq.~\ref{eq:vm0}, $S(t,x,y)$ is the source term and $J(\omega, \psi)$ is the Jacobian term given as $\psi_y\omega_x - \psi_x\omega_y$. We use the Cartesian domain  $(x,y)\in[0,2\pi]\times[0,2\pi]$ with a spatial resolution of $128\times128$. The initial vorticity field consisting of a co-rotating vortex pair is generated using the superposition of two Gaussian-distributed vortices given by,
\begin{multline}\label{eq:vm1}
  \omega(0,x,y) = \Gamma_1 \textrm{exp}\left(-\rho\left[(x-x_1)^2+(y-y_1)^2\right]\right)\\
  + \Gamma_2 \textrm{exp}\left(-\rho\left[(x-x_2)^2+(y-y_2)^2\right]\right),
\end{multline}
where the circulation $\Gamma_1 = \Gamma_2 =1$, the interacting constant $\rho=\pi$ and the intial vortex centers are located near each other with coordinates $(x_1,y_1)=(\frac{3\pi}{4}, \pi)$ and $(x_2,y_2)=(\frac{5\pi}{4}, \pi)$. We choose the source term $S(t,x)$ as,
\begin{equation}\label{eq:vm2}
\begin{aligned}
  S(t,x,y) &= \Gamma_0 \textrm{sin}(x)\textrm{cos}(y)\textrm{exp}\left(\frac{-4\pi^2}{\textrm{Re}}t\right),\\
 \end{aligned}
\end{equation}
where the magnitude of the source term is set to $\Gamma_0 = 0.01$.

The vorticity  field $\omega$ and streamfunction  field$\psi$ are obtained by solving the Eq.~\ref{eq:vm0} numerically. We use a third-order Runge-Kutta scheme for the time integration, and a second order Arakawa scheme \cite{arakawa1966computational} for the discretization of the Jacobian term $J(\omega, \psi)$. As we have a periodic domain, we use a fast Fourier transform (FFT) for solving the Poisson equation in  Eq.~\ref{eq:vm0} to obtain the streamfunction at every time step. Numerical details for solving the vortex-merger problem can be found in San et al\cite{san2013coarse,pawar2019cfd}. We integrate the solution from time $t=0$ to $t=20$ with a temporal step $dt=0.01$. 

\begin{figure*}[!htpb]
	\centering
	\includegraphics[width=0.9\textwidth]{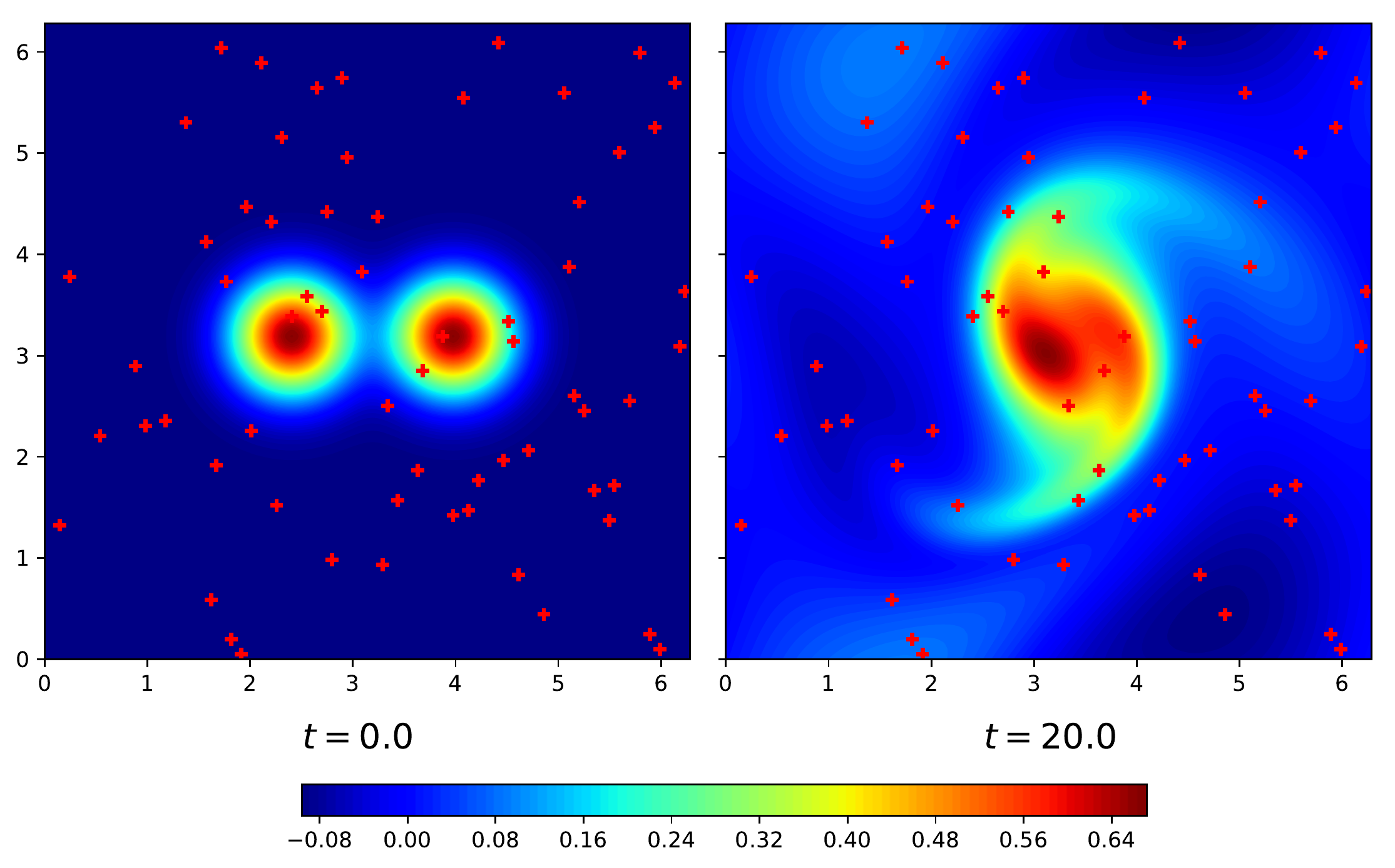}
	\caption{The 2D vortex-merger problem  with source term at time $t=0.0$ and $t=20.0$. The red markers shows $64$ random sensor locations used to collect vorticity ($\omega$) and streamfunction ($\psi$) data for recovering source term $S(t,x,y)$. }
	\label{fig:vm1}
\end{figure*}

Fig.~\ref{fig:vm1} shows the merging process of two vortices at the initial and final times. The red markers in  Fig.~\ref{fig:vm1} are $64$ randomly selected sparse locations to collect both 
streamfunction $\psi$ and vorticity $\omega$ data. Once the streamfunction  and vorticity  data at sparse locations are available, we can construct the target data $\textbf{V}$ and feature library $\mathbf{\widetilde \Theta}$ as discussed in Section~\ref{sec:meth}. The resulting input-response data is given as,

\begin{equation}\label{eq:ad3}
\left.\begin{aligned}
\textbf{V}  &=  
\left[
  \begin{array}{c}    
     \boldsymbol{\omega_{t}} + \boldsymbol{J(\omega, \psi)} - \dfrac{1}{\textrm{Re}}\boldsymbol{\nabla^2\omega}
  \end{array}
\right]\\
\mathbf{\widetilde \Theta}  &=  
\left[
  \begin{array}{ccc}    
    \textbf{x}  & \textbf{y} & \textbf{t} 
    \end{array}
\right]
\end{aligned}
\right\}.
\end{equation}

The derivatives in the output data  $\textbf{V(t)}$ are calculated using finite difference approximations similar to  Eq.~\ref{goveq4}. As streamfunction $(\psi)_{i,j}^p$ and vorticity $(\omega)_{i,j}^p$ data are selected only at sparse spatial locations, we also store the surrounding stencil, i.e., $(\psi)_{i+1,j}^p$, $(\psi)_{i-1,j}^p$, $(\psi)_{i,j+1}^p$, $(\psi)_{i,j-1}^p$, and $(\omega)_{i+1,j}^p$, $(\omega)_{i-1,j}^p$, $(\omega)_{i,j+1}^p$, $(\omega)_{i,j-1}^p$ in order to calculate the derivatives. The index $i$ represents spatial location in $x$ direction, and $j$ represents spatial location in $y$ direction. 

\begin{figure}[!htpb]
	\centering
	\includegraphics[width=0.4\textwidth]{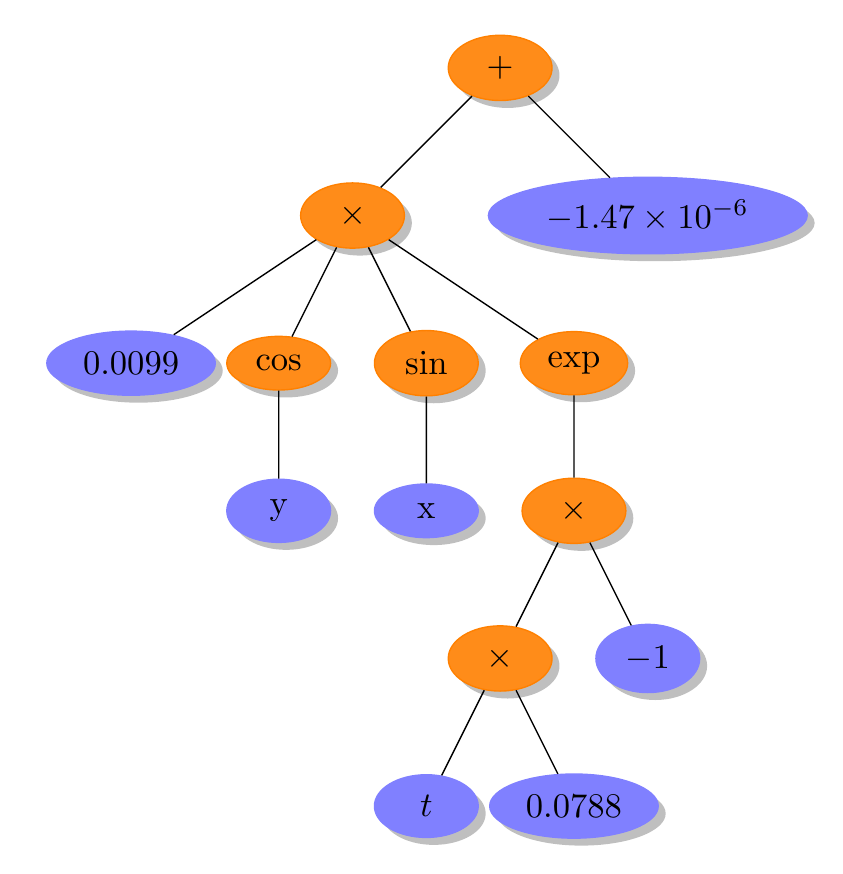}
	\caption{ Hidden source term of the 2D vortex-merger problem in terms of ET identified by GEP. }
	\label{fig:vmt}
\end{figure}

In this test case, we demonstrate the identification of hidden physics which is the source term $S(t,x,y)$ given by Eq.~\ref{eq:vm2} from the data obtained at sparse spatial locations using GEP. Table~\ref{tab:hp} lists the hyper-parameters used by GEP to recover the hidden physical law. We use the same function and terminal sets as shown in Table~\ref{tab:adset} but $\times$ is used as a linking function. Fig.~\ref{fig:vmt} shows the ET form of hidden physical law (source term) obtained by GEP. Simplification of the ET form shows the identified source term which is close to true source term as shown in Table~\ref{tab:vmr}. 

\begin{table*}[!htbp]
\caption{Hidden source term ($S$) of the 2D vortex-merger problem identified by GEP.}
\label{tab:vmr}
\bgroup
\def\arraystretch{1.5}
\setlength{\tabcolsep}{1em}
\begin{tabular}{@{}llc}
\hline
& \textbf{Recovered} & \textbf{Test error}   \\ \hline
\noalign{\smallskip}
True      & $S = 0.0100~\textrm{sin}(x)~\textrm{cos}(y)~\textrm{exp}(-0.078~t)$   &   \\
GEP       & $S = 0.0099~\textrm{sin}(x)~\textrm{cos}(y)~\textrm{exp}(-0.078~t)- 1.47\times10^{-6}$  &  $1.35\times10^{-8}$ \\
\hline
\end{tabular}
\egroup
\end{table*}

The 1D advection-diffusion and 2D vortex-merger problem demonstrate the usefulness of GEP in recovering hidden physics, i.e., a source term that composed of complex functions using randomly selected sparse data. The expressive power of the feature library limits the applications of STRidge for identifying complex composition models. However, STRidge might be able to identify the infinite series approximations of these nonlinear functions \cite{sindy}. In the next test case, we use both STRdige and GEP to identify eddy viscosity kernels along with their free modelling coefficient that controls the dissipation of these kernels.


\subsection{2D Kraichnan Turbulence}
\label{sec:kt}
The concept of two-dimensional turbulence helps in understanding many complex physical phenomenon such as geophysical and astrophysical flows\cite{boffetta2010evidence, boffetta2012two}. The equations of two-dimensional turbulence can model idealized flow configurations restricted to two-dimensions such as flows in rapidly rotating systems and in thin films over rigid bodies. The physical mechanism associated with the two-dimensional turbulence is explained by the Kraichnan-Batchelor-Leith (KBL) theory\cite{kraichnan1967inertial, batchelor1969computation, leith1971atmospheric}.  Generally, large eddy simulation (LES) is performed for both two and three dimensional flows to avoid the fine resolution and thereby computational requirements of direct numerical simulation (DNS)    \cite{piomelli1999large, meneveau2000scale}. In LES, the flow variables are decomposed into resolved low wavenumber (or large scale) and unresolved high wavenumber (or small scale). This is achieved by the application of a low pass spatial filter to the flow variables. By arresting high wavenumber content (small scales), we can reduce the high resolution requirement of DNS, and hence faster simulations and reduced storage requirements. However, the procedure of introducing a low pass filtering results in an unclosed term for the LES governing equations representing the finer scale effects in the form of a source term.

Thus the quality of LES depends on the modeling approach used to close the spatial filtered governing equations to capture the effects of the unresolved finer scales\cite{sagaut2006large}.  This model also called the subgrid scale model is a critical part of LES computations. A functional or eddy viscosity approach is one of the popular approaches to model this closure term. These approaches propose an artificial viscosity to mimic the dissipative effect of  the fine scales. Some of the popular functional models are the Smagorinsky \cite{smagorinsky1963general}, Leith \cite{leith1968diffusion}, Balwin-Lomax\cite{baldwin1978thin} and Cebeci-smith models\cite{smith1967numerical}.  All these models require the specification of a model constant that controls the quantity of dissipation in the simulation, and its value is often set based on the nature of the particular flow being simulated.  In this section, we demonstrate the identification of an eddy viscosity kernel (model) along with its ad-hoc model constant from observing the source term of the LES equation using both GEP and STRidge as robust SR tools. To this end, we use the vorticity-streamfunction formulation for two-dimensional fluid flows given in  Eq.~\ref{eq:vm0}. We derive the LES equations for the two dimensional Kraichnan turbulence by applying a low pass spatial filter to the vorticity-streamfunction PDE given in Eq.~\ref{eq:vm0}.  The resulting filtered equation is given as,
\begin{equation}\label{eq:kt1}
\begin{aligned}
\overbar{\omega}_t + \overbar{J(\psi, \omega)} = \frac{1}{\textrm{Re}}\nabla^2\overbar{\omega},\\
 \end{aligned}
\end{equation}
where Re is the Reynolds number of the flow and $J(\omega, \psi)$ is the Jacobian term given as $\psi_y\omega_x - \psi_x\omega_y$. Furthermore the  Eq.~\ref{eq:kt1} is rearranged as,
\begin{equation}\label{eq:kt2}
\begin{aligned}
\overbar{\omega}_t + J(\overbar{\psi}, \overbar{\omega}), = \frac{1}{\textrm{Re}}\nabla^2\overbar{\omega} + \Pi,\\
 \end{aligned}
\end{equation}
where the LES source term $\Pi$ is given as,
\begin{equation}\label{eq:kt3}
\begin{aligned}
\Pi =  J(\overbar{\psi}, \overbar{\omega}) -\overbar{J(\psi, \omega)}. \\
 \end{aligned}
\end{equation}

The source term $\Pi$ in Eq.~\ref{eq:kt3} represents the influence of the subgrid scales on larger resolved scales. The term  $\overbar{J(\psi, \omega)}$ is not available, which necessitates the use of a closure modelling approach. In functional or eddy viscosity models, the source term of LES equations is represented as,
\begin{equation}\label{eq:kt4}
\begin{aligned}
\Pi =  \nu_e\nabla^2\overbar{\omega}.\\
 \end{aligned}
\end{equation}
where eddy viscosity $\nu_e$  is given by, but not limited to, the Smagorinsky, Leith, Baldwin-Lomax, and Cebeci-Smith kernels.  The choice of these eddy viscosity kernels essentially implies the choice of a certain function of local field variables such as the strain rate or gradient of vorticity as a control parameter for the magnitude of $\nu_e$.
\begin{table}[!htpb]
\caption{GEP functional and terminal sets used for identifying eddy viscosity kernel. `?' is a random constant.}
\label{tab:krset}
\bgroup
\def\arraystretch{1.5}
\setlength{\tabcolsep}{2.1em}
\begin{tabular}{@{}lll}
\hline
\textbf{Parameter} & \textbf{Value}    \\ \hline
\noalign{\smallskip}
Function set     &   $+, -, \times, / $   \\
Terminal set     &   $\mathbf{\widetilde \Theta}$, $?$ \\
Linking function &   $+$  \\
\hline
\end{tabular}
\egroup
\end{table}

In Smagorisnky model, the eddy viscosity kernel is given by,
\begin{equation}\label{eq:kt5}
\begin{aligned}
\nu_e =  (c_s\delta)^2|\overbar{S}|,\\
 \end{aligned}
\end{equation}
where $c_s$ is  a free modelling constant that controls the magnitude of the dissipation and $\delta$ is a characteristic grid length scale given by the square root of the product of the cell sizes in each direction. The $|\overbar{S}|$  is based on the second invariant of the filtered field deformation, and given by,
\begin{equation}\label{eq:kt6}
\begin{aligned}
|\overbar{S}|=\sqrt{4\overbar{\psi}_{xy}^2 + (\overbar{\psi}_{2x} - \overbar{\psi}_{2y} )^2},\\
 \end{aligned}
\end{equation}

The Leith model proposes that eddy viscosity kernel is a function of vorticity and given as,
\begin{equation}\label{eq:kt7}
\begin{aligned}
\nu_e =  (c_s\delta)^3|\nabla\overbar{\omega}|,\\
 \end{aligned}
\end{equation}
where $|\nabla\overbar{\omega}|$ controls the dissipative character of the eddy viscosity as against the resolved strain rate used in the Smagorinsky model. The magnitude of the gradient of vorticity is defined as,
\begin{equation}\label{eq:kt8}
\begin{aligned}
|\nabla\overbar{\omega}|=\sqrt{\overbar{\omega}_x^2 + \overbar{\omega}_{y}^2}.\\
 \end{aligned}
\end{equation}
\begin{table}[!htpb]
\caption{ GEP hyper-parameters selected for identification of the eddy viscosity kernel for the Kraichnan turbulence.}
\label{tab:les}
\bgroup
\def\arraystretch{1.5}
\setlength{\tabcolsep}{0.2em}
\begin{tabular}{@{}lcccc}
\hline
\textbf{Hyper-parameters}  &  \textbf{Kraichnan turbulence}  \\ \hline
\noalign{\smallskip}
Head length              &   $2$         \\
Number of genes          &   $2$         \\
Population size          &   $20$        \\
Generations              &   $500$       \\
Length of RNC array      &   $3$          \\
Random constant minimum  &   $-1$         \\
Random constant maximum  &   $1$          \\
\hline
\end{tabular}
\egroup
\end{table}

The Baldwin-Lomax is an alternative approach that models the eddy viscosity kernel as,
\begin{equation}\label{eq:kt9}
\begin{aligned}
\nu_e =  (c_s\delta)^2|\overbar{\omega}|,\\
 \end{aligned}
\end{equation}
where $|\overbar{\omega}|$ is the absolute value of the vorticity considered as a measure of the local energy content of the flow at a grid point and also a measure of the dissipation required at that location.
\begin{figure}[!ht]
	\centering
	\includegraphics[width=0.4\textwidth]{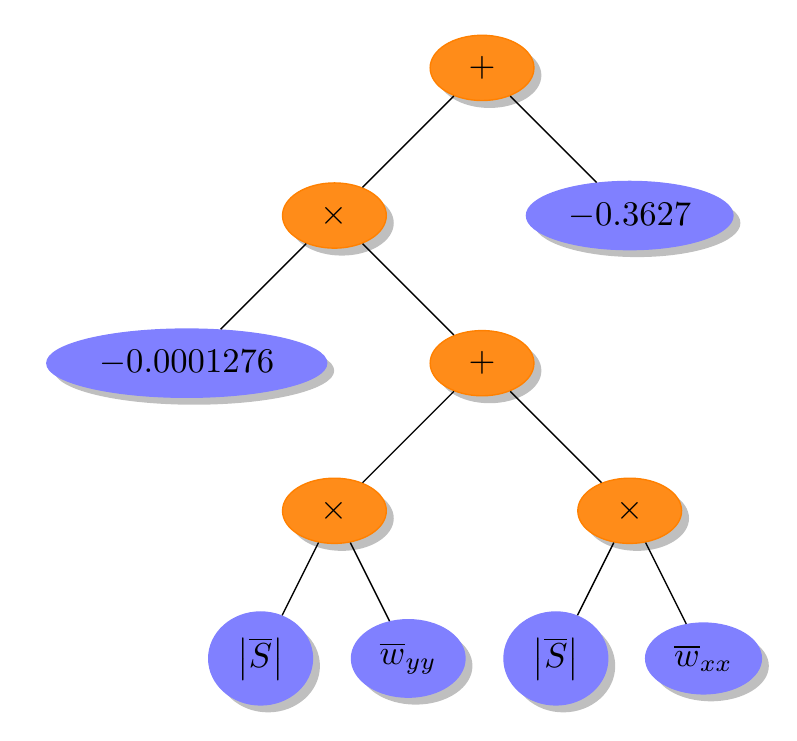}
	\caption{Samgorisnsky kernel in terms of ET identified for the two-dimensional Kraichnan turbulence problem  by GEP.}
	\label{fig:pit}
\end{figure}

The Cebeci-Smith model was devised for the Reynolds Averaged Navier-Stokes (RANS) applications. The model is modified for LES setting, and is given as,
\begin{equation}\label{eq:kt10}
\begin{aligned}
\nu_e =  (c_s\delta)^2|\overbar{\Omega}|,\\
 \end{aligned}
\end{equation}
where $|\overbar{\Omega}|$ is given as,
\begin{equation}\label{eq:kt11}
\begin{aligned}
|\overbar{\Omega}|=\sqrt{\overbar{\psi}_{2x}^2 + \overbar{\psi}_{2y}^2}.\\
 \end{aligned}
\end{equation}
\begin{table*}[!htbp]
\caption{LES source term ($\Pi$) for two-dimensional Kraichnan turbulence problem identified by GEP and STRidge.}
\label{tab:2dkr}
\bgroup
\def\arraystretch{1.5}
\setlength{\tabcolsep}{1em}
\begin{tabular}{@{}llc}
\hline
& \textbf{Recovered}    \\ \hline
\noalign{\smallskip}
GEP       & $\Pi = 0.000128~\left|S\right| ~w_{2x} + 0.000128~\left|S\right|~w_{2y} - 0.362$    \\  
STRidge   & $\Pi = 0.000132~\left|S\right|~w_{2x} + 0.000129~\left|S\right|~w_{2y}$   \\
\hline
\end{tabular}
\egroup
\end{table*}

High fidelity DNS simulations are performed for Eq.~\ref{eq:kt1}. We use a square domain of length $2\pi$  with periodic boundary conditions in both directions. We  simulate homogeneous isotropic decaying turbulence which may be specified by an initial energy spectrum that decays through time. High fidelity DNS simulations are carried out for $\textrm{Re}=4000$ with $1024\times1024$ resolution from time $t=0$ to $t=4.0$ with time step $0.001$. The filtered flow quantities and LES source term $\Pi$ in Eq.~\ref{eq:kt3} are obtained from coarsening the DNS quantities to obtain quantities with a $64\times64$ resolution. The further details of solver and coarsening can be found in San and Staples\cite{san2012high}. Once the LES source term $\Pi$ in Eq.~\ref{eq:kt3} and filtered flow quantities are obtained, we build the feature library and output data similar to the discussion in Section~\ref{sec:meth}. The resulting input-response data is given as,
\begin{equation}\label{eq:ktr0}
\left.\begin{aligned}
\textbf{V}  &=  
\left[
  \begin{array}{c}    
     \boldsymbol{\Pi}
  \end{array}
\right]\\
\mathbf{\widetilde \Theta}  &=  
\left[
  \begin{array}{cccccccc}    
    \boldsymbol{\overbar{\omega}_{2x}} &  \boldsymbol{\overbar{\omega}_{2y}}  & \boldsymbol{|\overbar{S}|}  &  \boldsymbol{|\nabla\overbar{\omega}|} & \boldsymbol{|\overbar{\omega}|} & \boldsymbol{|\overbar{\Omega}|} 
\end{array}
\right]
\end{aligned}
\right\}.
\end{equation}

GEP uses the output and feature library given in Eq.~\ref{eq:ktr0} to automatically extract the best eddy viscosity kernel for decaying turbulence problems along with the model's ad-hoc coefficient.
\begin{figure}[!htpb]
	\centering
	\includegraphics[width=0.5\textwidth]{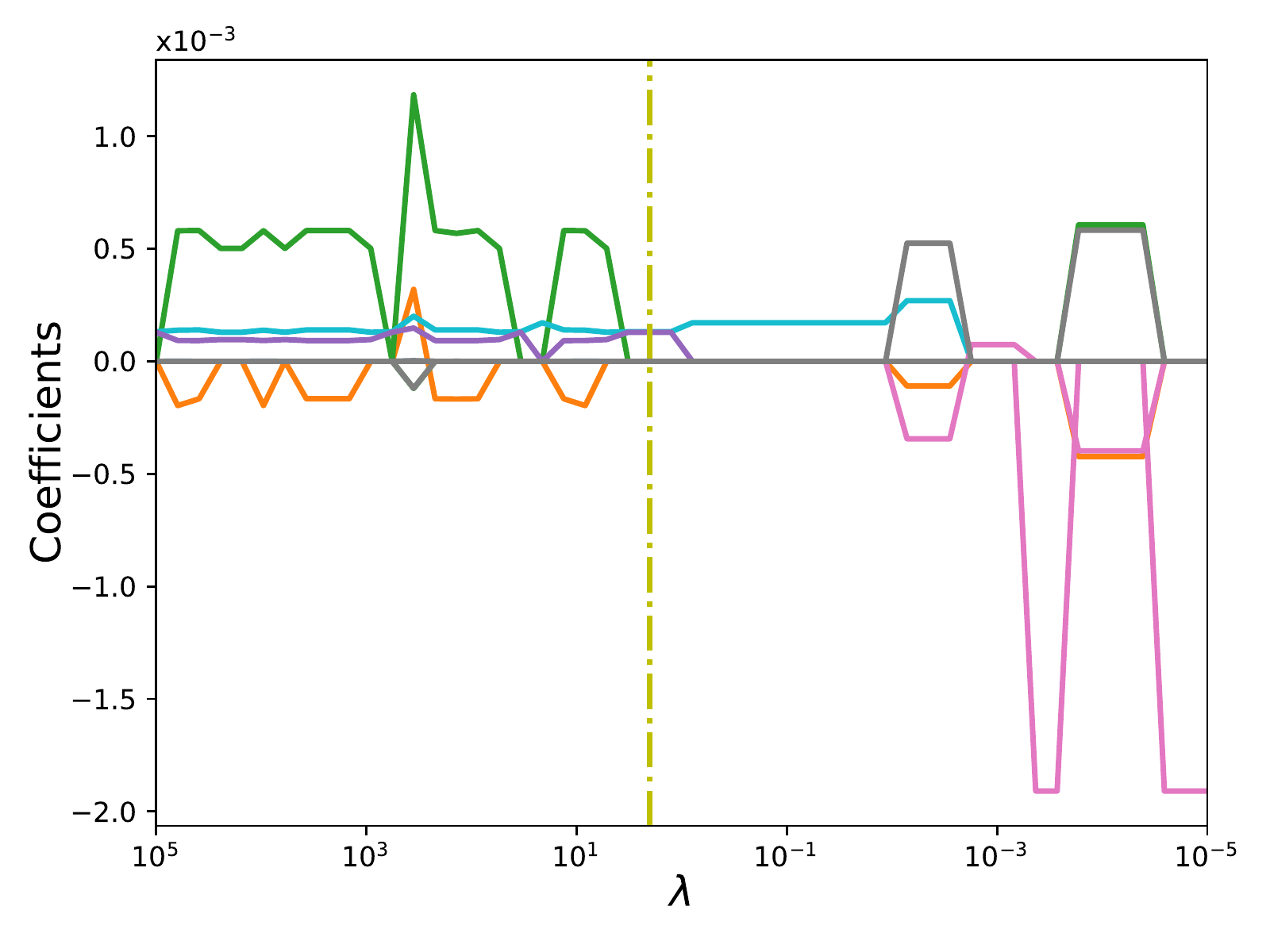}
	\caption{STRidge coefficients as a function of regularization parameter $\lambda$ for the two-dimensional Kraichnan turbulence problem.}
	\label{fig:kttr}
\end{figure}
\begin{figure}[!htpb]
	\centering
	\includegraphics[width=0.5\textwidth]{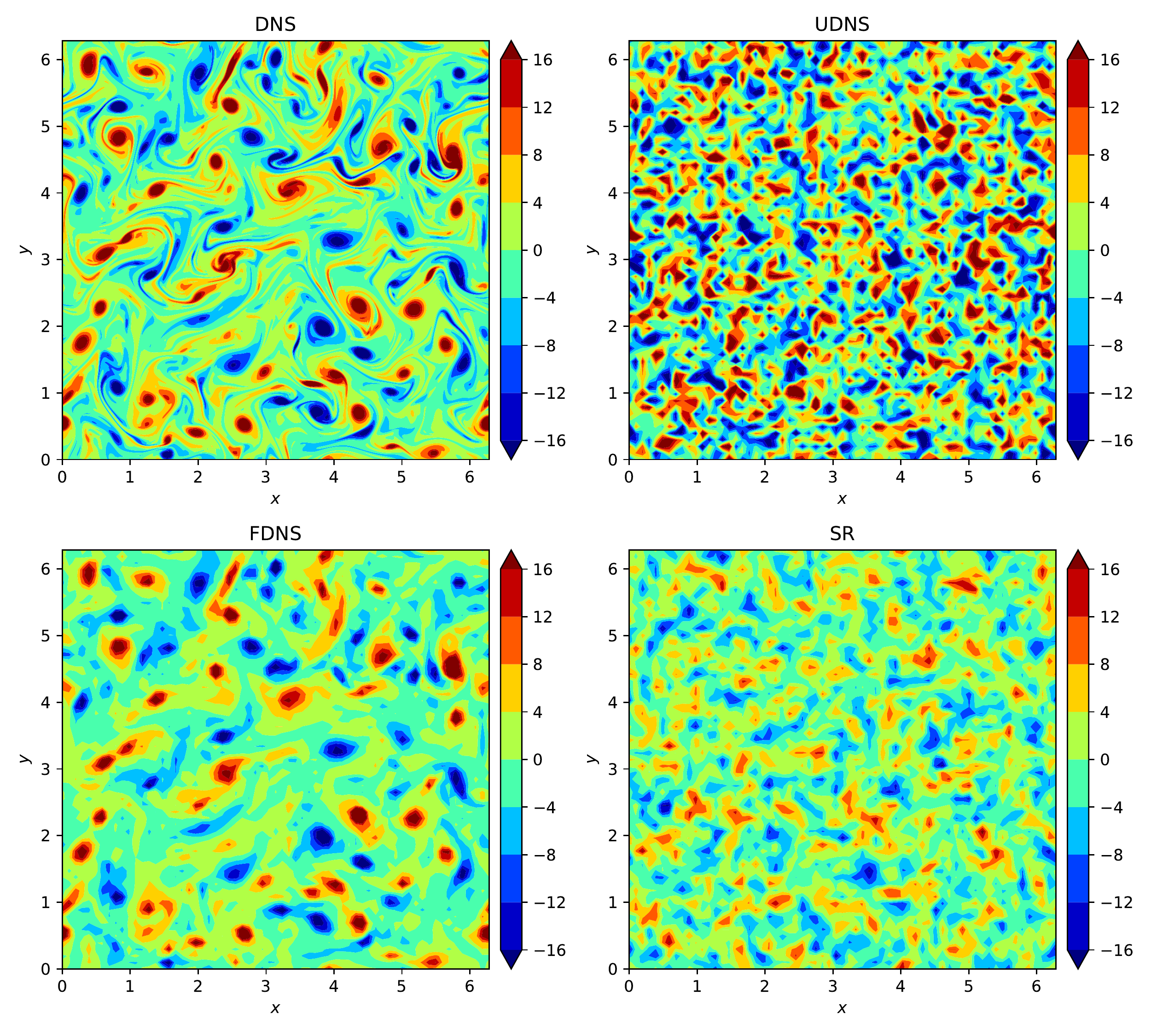}
	\caption{Controur plots for the two-dimensional Kraichnan turbulence problem at $t=4$. SR refers to the identified model of the Smagorinsky kernel with $c_s=0.12$. UDNS and FDNS refer to the no-model and filtered DNS simulations, respectively.}
	\label{fig:ktfield}
\end{figure}

The extended feature library is constructed to include nonlinear interactions up to the quadratic degree to expand the expressive power for the STRidge algorithm. The resulting extended feature library is given as,
\begin{equation}\label{eq:ktr1}
\mathbf{\Theta}  =  
\left[
  \begin{array}{cccccccc}    
    \textbf{1}   &  \boldsymbol{\overbar{\omega}_{2x}} &  \boldsymbol{\overbar{\omega}_{2x}}^2  & \boldsymbol{\overbar{\omega}_{2y}} & \boldsymbol{\overbar{\omega}_{2x}}\boldsymbol{\overbar{\omega}_{2y}} & \boldsymbol{\overbar{\omega}_{2y}}^2& \ldots  & \boldsymbol{|\overbar{\Omega}|^2}
  \end{array}
\right].
\end{equation}

The function and terminal sets used for identification of eddy viscosity kernel by GEP are listed in Table~\ref{tab:krset}. Furthermore, the hyper-parameters of GEP are listed in Table~\ref{tab:les}. Both GEP and STRidge identify the Smagorinsky kernel with approximately the same coefficients as shown in Table~\ref{tab:2dkr}. The ET form of the Smagorinsky kernel found by GEP is shown in Fig.~\ref{fig:pit}. The regularization weight $\lambda$ is varied  to recover multiple models of different complexity as shown in Fig.~\ref{fig:kttr}. The yellow line in Fig.~\ref{fig:kttr} corresponds to the value of $\lambda$ where STRidge identifies the Smagorinsky kernel. We can take the average coefficient from both SR tools and derive the value of the free modelling constant identified by SR approaches. The average model of both approaches is given by,
\begin{equation}\label{eq:kt9}
\begin{aligned}
 \Pi = 0.000129~(\left|S\right| ~w_{2x} + \left|S\right|~w_{2y}).\\
 \end{aligned}
\end{equation}
By comparing with Eq.~\ref{eq:kt4} and Eq.~\ref{eq:kt5} and using the spatial cell size $\delta=\frac{2\pi}{64}$, the value of the free  modelling constant is retrieved as $c_s=0.12$.

\begin{figure}[!htpb]
	\centering
	\includegraphics[width=0.5\textwidth]{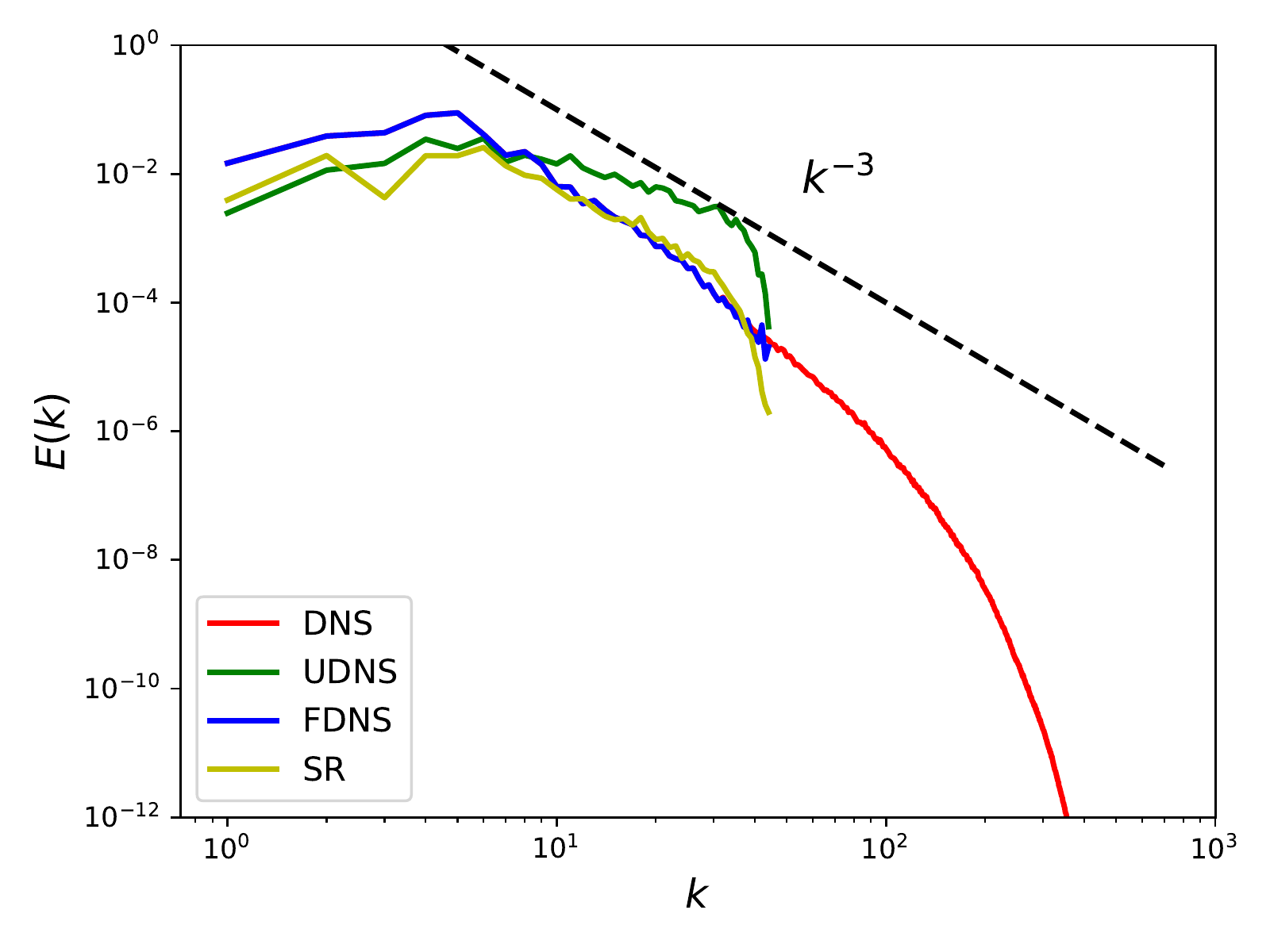}
	\caption{Energy spectra for the two-dimensional Kraichnan turbulence problem at $t=4$. SR refers to the identified model of the Smagorinsky kernel with $c_s=0.12$. UDNS and FDNS refer to the no-model and filtered DNS simulations, respectively.}
	\label{fig:es}
\end{figure}

The SR identified Smagorinsky kernel with $c_s=0.12$ is plugged into the LES source term $\Pi$ in Eq.~\ref{eq:kt2} and a forward LES simulation is run for the 2D decaying turbulence problem. 
Fig.~\ref{fig:ktfield} shows the vorticity fields at time $t=4.0$ for the DNS, under-resolved no-model simulation (UDNS), filtered DNS (FDNS), and LES with SR retrieved  Smagorinsky kernel. Energy spectra at time $t=4.0$ are showed in Fig.~\ref{fig:es}. We can observe that SR approaches satisfactorily identify the value of the modelling constant $c_s$, which controls reasonably well the right amount of dissipation needed to account for the unresolved small scales. \textcolor{rev1}{We also highlight that several deep learning frameworks such as ANNs have been exploited for subgrid scale modelling for 2D Kraichnan turbulence\cite{maulik2019subgrid,maulik2018data,maulik2019sub}. The importance of feature selection can be seen in these works where different invariant kernels, like those listed in the feature library given in Eq.~\ref{eq:ktr0}, are used as inputs to improve the ANN's predictive performance. The authors compared a posteriori results with different free modelling coefficients of the Smagorinsky and Leith models. Furthermore, it is evident from the energy spectrum comparisons in their studies that the appropriate addition of dissipation with the right tuning of the free modelling coefficient can lead to better predictions of the energy spectrum. To this end, SR approaches automatically distill traditional models along with the right values for the ad-hoc free modelling coefficients. Although the present study establishes a modular regression approach for discovering the relevant free parameters in LES models, we highlight that it can be extended easily to a dynamic closure modelling framework reconstructed automatically by sparse data on the fly based on the flow evolution, a topic we would like to address in future.}

\section{Conclusion}
\label{sec:conc}
Data driven symbolic regression tools can be extremely useful for researchers for inferring complex models from sensor data when the underlying physics is partially or completely unknown. Sparse optimization techniques are envisioned as an SR tool that is capable of recovering hidden physical laws in a highly efficient computational manner. Popular sparse optimization techniques such as LASSO, ridge, and elastic-net are also known as feature selection methods in machine learning. These techniques are regularized variants of least squares regression adapted to reduce overfitting and promote sparsity. The model prediction ability of sparse regression methods is primarily dependent on the expressive power of its feature library which contains exhaustive combinations of nonlinear basis functions that might represent the unknown physical law. This limits the identification of physical models that are represented by complex functional compositions. GEP is an evolutionary optimization algorithm widely adapted for the SR approach. This genotype-phenotype algorithm takes advantage of the simple chromosome representations of GA and the free expansion of complex chromosomes of GP. GEP is a natural feature extractor that may not need a priori information of nonlinear bases other than the basic features as a terminal set. Generally, with enough computational time, GEP may recover unknown physical models that are represented by complex functional compositions by observing the input-response data.  

In this paper, we demonstrate that the sparse regression technique STRidge and the evolutionary optimization algorithm GEP are effective SR tools for identifying hidden physical laws from observed data. We first identify various canonical PDEs using both STRidge and GEP. We demonstrate that STRidge is limited by its feature library for identifying the Sine-Gordon PDE. Following equation discovery, we demonstrate the power of both algorithms in identifying the leading truncation error terms for the Burgers MDE.  While both algorithms find the truncation terms, coefficients found by STRidge were more accurate than coefficients found by GEP. We note that, when the feature library is capable of expressing the underlying physical model, the application of STRidge is suitable due to its fewer hyper-parameters and lower computational overhead. Next, we illustrate the recovery of hidden physics that is supplied as the source or forcing term of a PDE. We use randomly selected sparse measurements that mimic real world data collection. STRdige is not applied in this setting as the feature library was limited to represent the unknown physical model that consists of complex functional compositions. GEP was able to identify the source term for both 1D advection-diffusion PDE and 2D vortex-merger problem using sparse measurements. Finally, both STRdige and GEP were applied to discover the eddy viscosity kernel along with its ad-hoc modelling coefficient as a subgrid scale model for the LES equations simulating the 2D Kraichnan turbulence problem. This particular example demonstrates the capability of inverse modelling or parametric estimation for turbulence closure models using SR approaches. Future studies will focus on identifying LES closure models that augment the known closure models by accounting for the various nonlinear physical process. Furthermore, various SR tools are being investigated for the identification of nonlinear truncation error terms of MDEs for implicit LES approaches that can be exploited for modelling turbulent flows without the need for explicit subgrid scale models. 

\begin{acknowledgements}
This material is based upon work supported by the U.S. Department of Energy, Office of Science, Office of Advanced Scientific Computing Research under Award Number DE-SC0019290. O.S. gratefully acknowledges their support. 
Disclaimer: This report was prepared as an account of work sponsored by an agency of the United States Government. Neither the United States Government nor any agency thereof, nor any of their employees, makes any warranty, express or implied, or assumes any legal liability or responsibility for the accuracy, completeness, or usefulness of any information, apparatus, product, or process disclosed, or represents that its use would not infringe privately owned rights. Reference herein to any specific commercial product, process, or service by trade name, trademark, manufacturer, or otherwise does not necessarily constitute or imply its endorsement, recommendation, or favoring by the United States Government or any agency thereof. The views and opinions of authors expressed herein do not necessarily state or reflect those of the United States Government or any agency thereof.
\end{acknowledgements}

\bibliography{main}
\end{document}